\providecommand{\martin}[1]{#1}
\providecommand{\martinc}[1]{}

\providecommand{\tilf}{\tilde{f}}

\documentclass[preprint,12pt,authoryear]{elsarticle}

\usepackage[left=1.0in,right=1.0in]{geometry}
\usepackage{amssymb}
\usepackage{amsmath}
\usepackage{graphicx}
\usepackage{subcaption}   
\usepackage{xcolor}
\definecolor{BurntOrange}{rgb}{0.8,0.33,0.0}
\usepackage[colorlinks=true,
            linkcolor=blue,
            citecolor=blue]{hyperref}

\usepackage{cleveref}
\crefname{equation}{Eq.}{Eqs.}
 \usepackage{booktabs}
\usepackage{natbib}
\usepackage{float}
\usepackage{longtable}
\usepackage{placeins}
\usepackage{cleveref}
\usepackage[colorlinks=true,citecolor=blue]{hyperref}

\usepackage{units}
\usepackage{color}
\definecolor{red}{rgb}{0.8,0,0}
\definecolor{green}{rgb}{0,0.8,0}
\definecolor{blue}{rgb}{0,0,1}

\journal{Transportation Research Part B: Methodological}

\providecommand{\sub}[1]{dummy} 
\providecommand{\sup}[1]{dummy} 

\renewcommand{\sup}[1]{^{\text{#1}}}
\providecommand{\sub}[1]{_{\text{#1}}}

\providecommand{\ablpart}[2]{\frac{\partial #1}{\partial #2}}

\begin{document}

\begin{frontmatter}



\title{Second-Order Continuum Model for Disordered Traffic \martinc{be specific!}} 

\author[1]{Shashank Rajput}
\author[1]{Venkatesan Kanagaraj}
\author[2]{Martin Treiber}
\author[3]{Gowri Asaithambi\corref{cor1}}
\ead{gowri@iittp.ac.in}
\author[2]{Ostap Okhrin}

\cortext[cor1]{Corresponding author}

\affiliation[1]{
    organization={Department of Civil Engineering, Indian Institute of Technology Kanpur},
    city={Kanpur},
    country={India}
}

\affiliation[2]{
    organization={Institute of Transport and Economics, Technische Universität Dresden},
    city={Dresden},
    country={Germany}
}

\affiliation[3]{
    organization={Department of Civil and Environmental Engineering, Indian Institute of Technology Tirupati},
    city={Tirupati},
    country={India}
}



\begin{abstract}
In this study, a two-dimensional second-order macroscopic traffic flow model is proposed to capture the complex dynamics of disordered traffic by explicitly incorporating coupled longitudinal and lateral interactions. The framework extends the classical continuity equation to two spatial dimensions and introduces acceleration equations consisting of self-driven, interaction-induced, and road-boundary-related components. The macroscopic longitudinal acceleration is formulated based on the Full Velocity Difference Model (FVDM), while lateral dynamics are governed by interaction principles consistent with the Optimal Velocity Model (OVM). The explicit inclusion of road-boundary effects enables the representation of vehicle confinement and space-sharing behaviour that are essential in lane-free and weakly lane-disciplined traffic systems. The model is examined through a series of numerical experiments in which longitudinal and lateral dynamics are analysed individually as well as simultaneously within a coupled two-dimensional setting under a range of initial conditions, including transitions between free-flow, medium congestion, and heavy congestion, and different lateral density configurations. The simulations demonstrate the model’s ability to reproduce key traffic features such as shockwave propagation, lateral dispersion, vehicle rearrangement, and stable density evolution across the road width. The numerical scheme, based on upwind and Lax--Friedrichs discretisations, ensures stable solutions of the coupled partial differential equations. Overall, the proposed framework provides a robust macroscopic description of disordered traffic and offers a consistent basis for analysing two-dimensional vehicular flow dynamics.
\end{abstract}




\end{frontmatter}



\section{Introduction}
\label{sec1}
\noindent Modelling disordered traffic has always been a challenging task due to its complexity, nonlinearity, and higher vehicle interactions. In contrast to ordered traffic, where vehicles adhere to strict lane discipline, disordered traffic is characterized by highly dynamic, heterogeneous, and irregular vehicle movements. These characteristics of unpredictable movement in disordered traffic make it difficult for researchers to develop a traffic model that is applicable and accurately reproduces real-world conditions across all scenarios.  Traffic flow modelling can be categorised based on the level of aggregation into microscopic and macroscopic models. Microscopic models describe the behaviour of individual vehicles and their interactions with surrounding vehicles, capturing driving processes such as car-following, lane-changing, overtaking, and gap acceptance. They provide a detailed representation of traffic dynamics but are computationally intensive and not practical for large-scale traffic simulations. Macroscopic models assume traffic as a fluid flow, where the behaviour of individual vehicles is not explicitly considered. Traffic is described using aggregate variables such as flow, density, and speed, whose evolution over time helps explain large-scale traffic phenomena such as congestion, shockwaves, queuing, and stop-and-go waves. These models of traffic flow are computationally efficient and can be used in the analysis of large-scale traffic, but tend to oversimplify vehicle interactions.  Macroscopic traffic flow models can be classified into two types: First-order models (kinematic wave models) and second-order models, \martin{both of which} are explained in the subsections below.

\subsection{First-Order Models}
\noindent In all continuous macroscopic models, the three basic quantities density $\rho$, speed $v$, and flow $Q$ are connected by the hydrodynamic relation $Q=\rho v$ reflecting the definitions of these quantities and by the continuity equation, a partial differential equation reflecting vehicle conservation. First-order models complete these relations by the fundamental diagram, a static relation between flow and density. Models of this class are also called LWR models in recognition of \citealp{lighthill1955kinematic} and (\citealp{richards1956shock}) who independently proposed them first. 
\vspace{0.2cm}

\noindent 
Different studies have proposed various functional forms for the fundamental diagram with a linear relationship between speed and density by \citealp{greenshields1935study}, a Logarithmic speed-density relationship by \citealp{greenberg1959analysis}, and an exponential relationship by \citealp{underwood1960speed}. \citealp{newell1961nonlinear} introduced a triangular fundamental diagram, which is a piecewise linear approximation of the traffic states, making it particularly useful for first-order macroscopic models. Based on this foundation, the following sections will discuss modelling concepts of the first-order modelling approaches for both ordered (lane-disciplined) and disordered traffic conditions.\vspace{0.2cm}

\noindent Within this modelling framework, several researchers have attempted to incorporate lane-changing behaviour at an aggregate level. One of the earliest efforts in this direction was undertaken by \citealp{gazis1962density}, who introduced a macroscopic treatment of lane-changing phenomena. This work was later extended by \citealp{munjal1971propagation}, who assumed that lane-changing manoeuvres occur as a mechanism to achieve a more uniform density distribution along a road segment. \martinc{Michalopoulos shifted to the second-order models.}These models were developed in a continuous space–time framework and faced practical limitations due to the absence of suitable numerical discretisation techniques at the time, which hindered their effective implementation. Subsequently, \citealp{laval2006lane} proposed a hybrid modelling approach in which lane-changing vehicles were treated as discrete particles with bounded acceleration capabilities. However, a key limitation of this class of models lies in the aggregation of traffic flow variables across lanes, an assumption that does not accurately reflect lane-specific dynamics observed in real traffic. To address this limitation, \citealp{shiomi2015multilane} and \citealp{roncoli2015traffic} developed first-order multilane macroscopic models in which lane-changing behaviour was explicitly modelled through transfer rates dependent on local traffic conditions.
\vspace{0.2cm}

\noindent While these multilane formulations improve the representation of lane-changing dynamics in lane-disciplined traffic, they remain inadequate for traffic systems where lane boundaries are weak or effectively absent. In such disordered traffic environments, especially those characterised by different vehicle compositions, smaller vehicles often exploit available gaps and road space in a manner that cannot be described solely through lane-based interactions. To capture these behaviours, several studies such as those by \citealp{nair2011porous,chunchu2014analysis,fan2015heterogeneous,agarwal2016modeling,mohan2017heterogeneous,nagalur2019first} have introduced alternative conceptual frameworks, including porous flow, lane sharing, creeping, seepage, and gap-filling mechanisms, which describe how smaller vehicles manoeuvre through traffic by utilising inter-vehicular spaces rather than adhering to fixed lanes.
\vspace{0.2cm}

\noindent Several studies have extended the classical LWR model to account for multiple vehicle classes and lane-specific interactions, resulting in multi-class macroscopic models for heterogeneous flow (\citealp{wong2002multi,benzoni2003populations,logghe2008multi,ngoduy2007multiclass,tang2009new,sreekumar2022multi}). Although these approaches distinguish between different vehicle classes based on attributes such as vehicle length and desired speed, they often rely on simplified representations of the fundamental diagram. In a multi-class extension of the LWR framework, \citealp{wong2002multi} proposed a model in which driver heterogeneity is captured through variations in free-flow speed, with vehicle speeds in the isotropic formulation expressed as functions of the total traffic density.
\vspace{0.2cm}

\noindent \citealp{ngoduy2007multiclass} introduced an alternative multi-class fundamental diagram in which each vehicle class is assigned a distinct desired speed under free-flow conditions, while a single common speed is assumed for all classes in congested regimes. Building on this formulation, \citealp{ngoduy2011multiclass} developed a stochastic variant, in which roadway capacity was treated as a random variable to account for uncertainty and variability in traffic conditions. \citealp{logghe2008multi} employed PCUs as scaling factors to map class-specific fundamental diagrams onto a common reference class. In their formulation, interactions among vehicle classes are treated as non-cooperative, such that slower vehicles function as moving bottlenecks that constrain faster vehicles, while the dynamics of slower vehicles remain largely unaffected by the presence of faster traffic.
\vspace{0.2cm}

\noindent While multi-class and multilane extensions of the LWR framework have improved the representation of disordered traffic conditions, their underlying assumptions continue to rely on a one-dimensional flow structure and well-defined lane boundaries. Consequently, these models remain limited in their ability to describe traffic systems where lateral motion is continuous, and lane demarcations are weak or absent. Recognising these limitations, recent research efforts have shifted toward developing two-dimensional macroscopic formulations that explicitly account for lateral vehicle interactions and lane-free driving behaviour. In this context, \citealp{agrawal2023two} proposed a first-order two-dimensional macroscopic framework for lane-free traffic by extending the classical continuity equation to a two-dimensional domain. Their formulation incorporates lateral diffusion effects and repulsive forces from road boundaries, enabling a directed representation of vehicle movement under disordered traffic conditions. \citealp{hadadi2025extended} introduced an extended LWR-based model that explicitly accounts for behavioral characteristics associated with non-lane-disciplined driving, including lateral separation distances and time headway effects. This extension allows the model to better represent driver interactions beyond purely longitudinal motion. In a related development, \citealp{goswami2025solution} presented a two-dimensional local fractional LWR model aimed at addressing non-differentiable traffic variables. Their approach integrates lateral dynamics influenced by diffusion and boundary repulsion, providing an alternative mathematical framework for modeling lane-free traffic behavior. These variations aim to enhance the realism of traffic flow representation, yet they still rely on the assumption of a steady-state relationship between flow and density.
\vspace{0.2cm}

\subsection{Second-Order Models}
\noindent A second-order traffic flow model is a continuum (macroscopic) traffic model governed by two coupled partial differential equations: a continuity equation that ensures conservation of vehicles and a momentum (or velocity) equation that describes the temporal and spatial evolution of local velocity. By modelling velocity as an independent dynamic variable rather than a direct function of density, second-order models are able to represent non-equilibrium traffic behaviour, including finite driver reaction times, traffic instabilities, stop-and-go waves, and non-equilibrium traffic states.
\martin{Specifically, second-order models replace the fundamental diagram of the LWR models by an acceleration equation as a function of density, speed, and gradients/nonlocalities thereof.} 
\vspace{0.2cm}

\noindent Some second-order models are phenomenological, with their main foundations coming directly from the principles of fluid dynamics. Other models are derived from car-following models using the dynamics between vehicles in terms of relative speeds and gaps, or through gas-kinetic theory, which uses statistical mechanics to represent a collective behavior for vehicles in a traffic stream. 
In the \martin{following}, the review of second-order macroscopic traffic flow models is presented in a model-based manner, tracing the systematic development of continuum formulations and highlighting their underlying assumptions, key contributions, and inherent limitations.
\vspace{0.2cm}

\noindent \citealp{payne1971model} proposed a one-dimensional second-order continuum model by 
deriving the dynamic speed equation from the Optimal-Velocity Model (OVM). The acceleration from the driver's perspective consists of two parts: The anticipation term comes from the fact that the driver responds to the leader, i.e., to the density ahead while the relaxation term is a direct reformulation of the OVM acceleration and tends to revert the speed to the steady-state speed given by the fundamental diagram implicitly contained in the OVM. 
\citeauthor{witham1974linear} independently proposed a similar traffic flow model based on fluid-dynamics principles, commonly referred to as the Payne–Whitham (PW) model. This model assumes that all vehicles exhibit uniform behavior (\citealp{witham1974linear}). However, a key limitation of Payne's model is the presence of a characteristic speed greater than the vehicle velocity, implying that traffic waves can propagate faster than the vehicles themselves, which is not consistent with real-world traffic behavior (\citealp{daganzo1995requiem}).
\vspace{0.2cm}

\noindent \citealp{del1994reaction} enhanced the PW model by integrating anticipation and reaction time to account for small variations in density and velocity. \citealp{berg2000continuum} introduced a diffusion term to the PW model to reduce unrealistic acceleration and deceleration. However, this modification was unable to effectively capture abrupt changes in density, and the PW model still faced challenges in representing detailed interactions between vehicles on the road (\citealp{ambarwati2014empirical}). \citealp{zhang1998theory} proposed a non-equilibrium model to address the challenges associated with defining second-order continuum models and ensuring realistic characteristic propagation speeds. However, \citeauthor{zhang1998theory} model assumes that drivers adjust to traffic density instantly, overlooking the physiological limitations of human reaction time. In the 1990s, several other second-order continuum models were developed, including Papageorgiou's high-order model (\citealp{papageorgiou1989macroscopic}), Michalopoulos' viscous model (\citealp{michalopoulos1991continuum}), and the semi-viscous model (\citealp{michalopoulos1992development}). Earlier contributions to this domain, such as the works of (\citealp{phillips1979new,kuhne1984macroscopic,ross1988traffic}), also offered significant advancements. While these models addressed specific limitations of Payne's model, the issue of characteristic speeds exceeding macroscopic flow velocity remains unresolved.
\vspace{0.2cm}

\noindent \citealp{aw2000resurrection} introduced a pressure term as an increasing function of density $P(\rho)$ to ensure the anisotropic property of traffic and to enhance the Payne model. While this approach successfully addressed several limitations, it could lead to significant variations in acceleration and deceleration in high-density regions (\citealp{richardson2012refined}). \citealp{zhang2002non} also proposed a model that is a special case of the Aw-Rascle model, derived from a microscopic car-following framework. Together, these two models are commonly referred to as the ARZ model. ARZ model may become unstable at low-density regions, hence a bi-phase extension of \martin{using the LWR model at low densities} should be used in these regions (\citealp{goatin2006aw}). {\citealp{jiang2002new} proposed a modified continuum traffic from the microscopic traffic model in which the density gradient term is replaced by a speed gradient term, allowing the model to incorporate the anisotropic nature of traffic flow without requiring bi-directional information. Further efforts have been made to integrate bi-directional information in macroscopic modelling while preserving the anisotropic nature of traffic and ensuring physically meaningful behaviour (\citealp{zheng2015anisotropic}). \citealp{zheng2015anisotropic} model considers the influence of a single leader and follower in the driving strategy, providing an insightful approximation. To further generalise this approach, \citealp{ngoduy2017multi} extended the model to account for both backwards-looking and multiple forward-looking interactions, incorporating information from multiple neighbouring vehicles. Several other studies have also focused on maintaining the anisotropic nature of traffic flow while ensuring that characteristic speeds remain within realistic limits, aligning with macroscopic flow speeds (\citealp{gupta2006new,zhang2009conserved,zheng2017anisotropic,khan2019macroscopic}).
\vspace{0.2cm}

\noindent \citealp{colombo20022} introduced a phase-transition model to enhance the ARZ model by incorporating maximum density considerations and improving the representation of phase transitions. \citealp{lebacque2007generic} further advanced traffic modelling by proposing the GSOM, a unified framework that generalises models such as ARZ and Colombo’s, allowing for a more comprehensive representation of traffic dynamics. GSOM effectively captures complex traffic phenomena, including scattering and hysteretic phase transitions. \citealp{goatin2005modeling} conducted an in-depth mathematical analysis of phase transitions within GSOM, contributing to a better understanding of its theoretical properties. \citealp{fan2017collapsed} refined GSOM by calibrating it with empirical traffic data, enhancing its accuracy in representing real-world traffic conditions. \citealp{seo2017traffic} successfully applied GSOM to large-scale traffic networks, demonstrating its effectiveness in traffic state estimation and control. \citealp{yu2019traffic} further explored its practical applications by leveraging GSOM for designing control strategies aimed at mitigating traffic congestion. These studies collectively highlight GSOM’s adaptability and robustness in capturing intricate traffic behaviors, making it a valuable tool for both theoretical research and practical traffic management solutions.
\vspace{0.2cm}

\noindent Second-order macroscopic models can also be derived from gas-kinetic considerations initially introduced by \citealp{prigogine1961boltzmann} for traffic flow, providing a strong theoretical foundation for capturing detailed traffic dynamics. These gas-kinetic based traffic flow (GKT-based) models explicitly incorporate velocity variance as a key factor influencing speed adaptation. Unlike models that rely on a predefined FD, GKT-based models derive the FD from the steady-state definition of traffic flow. Some GKT-based models include a third dynamic equation for velocity variance, acknowledging that velocity distribution may be skewed (\citealp{phillips1979kinetic,helbing1995high,helbing1995improved,helbing1996gas,hoogendoorn2001platoon}). Alternatively, some models present a functional form for velocity variance to simplify the framework while maintaining essential traffic characteristics. For instance, \citealp{treiber1999derivation} proposed a GKT-based model where velocity variance follows a Gaussian-like function. This non-local model primarily determines acceleration based on traffic conditions ahead. However, in contrast to the second-order models mentioned above, it describes anticipation not in terms of gradients but by explicit nonlocalities. GKT-based models (\citealp{helbing1998gas,treiber1999macroscopic}) have contributed significantly to understanding complex traffic phenomena, including hysteresis and scattering in the flow-density relationship. 
\martinc{Following sentence shifted from first-order models to here.}\citealp{michalopoulos1984multilane} extended earlier first-order macroscopic formulations by proposing a second-order traffic flow model with lane changing driven by momentum differences. Further advancements include the work of \citealp{ngoduy2006continuum}, who introduced a multilane continuum model incorporating on- and off-ramps to describe traffic dynamics near bottlenecks. This model was based on the GKT approach and was derived from microscopic principles using the gap-acceptance model and renewal theory, enhancing the understanding of multilane traffic behaviour. Building on this approach, \citealp{ngoduy2014multianticipative} developed a model for multianticipative driving behaviour, where vehicles react to multiple vehicles ahead, offering insights valuable for real-time traffic prediction and control. Several studies have derived continuum traffic models from car-following models by incorporating drivers' memory (\citealp{gupta2005analyses,ge2006density,sun2020new}), introducing viscosity parameters into the macroscopic traffic flow equations, reflecting the influence of historical traffic states on current dynamics. 
\vspace{0.2cm}

\noindent Another class of model known as Stochastic second-order continuum traffic models incorporates randomness and variability into the framework of traditional deterministic traffic models to better represent the inherent uncertainty in real-world traffic systems. These models extend the deterministic macroscopic equations by introducing stochastic terms that account for fluctuations in driver behaviour, environmental conditions, and other external factors influencing traffic flow. One common approach to introducing stochasticity is by adding noise terms to the governing equations, such as the continuity equation or the momentum equation. These noise terms can represent random variations in traffic density, speed, or flow caused by unpredictable driver responses or variations in vehicle performance. \citealp{ngoduy2021noise} examined a class of second-order deterministic macroscopic models and incorporated stochastic perturbations in the acceleration and deceleration processes by introducing Langevin noise terms, showing the formation of a stop-go wave at low speed.
\vspace{0.2cm}

\noindent The models discussed above are primarily designed for homogeneous traffic flows, assuming all vehicles have identical characteristics and driving behaviors. However, this assumption does not hold in real-world traffic scenarios, where highways and urban roads often feature a mix of vehicle types with diverse performance capabilities, such as varying maximum speeds, accelerations, and braking capacities. Accurately modelling such mixed traffic is essential to better capture the impacts of mixed traffic compositions and to design effective traffic management and control strategies. Few studies have developed a continuum model considering mixed traffic conditions by adding multiple vehicle types (\citealp{hoogendoorn1999multiclass,tampere2003gas,jiang2004extended,ngoduy2004comparison,ngoduy2006macroscopic,tang2007mixed,tang2009new,nair2011porous,ngoduy2012application,gupta2014analyses,mohan2017heterogeneous,fosu2021multilane,zhang2021extended,wen2025novel})
These continuum models, which incorporate continuity and momentum equations, can simulate various traffic phenomena on freeways. However, to accurately represent disordered traffic, it is crucial to extend continuum traffic dynamics into two dimensions, effectively capturing lateral movements.
\vspace{0.2cm}

\noindent Most of the models proposed in the literature are either one-dimensional second-order macroscopic models or two-dimensional first-order macroscopic models (\citealp{herty2017two,balzotti2020two,agrawal2023two}), both of which fail to comprehensively address the complexities of disordered traffic. A two-dimensional second-order macroscopic model is necessary to account for the interplay between longitudinal and lateral dynamics, providing a more realistic representation of traffic flow. Only a limited number of studies have proposed two-dimensional models that incorporate both lateral and longitudinal dynamics (\citealp{herty2018macroscopic,vikram2022stabilized}). \citealp{vikram2022stabilized} considered the longitudinal dynamics as a function of density and the negative density gradient along the longitudinal direction, while the lateral dynamics were governed solely by the negative density gradient in the lateral direction and ignoring the influence from road boundary.
\vspace{0.2cm}

\subsection{Paper Outline}
\noindent The literature review above underlines a clear gap in the existing studies on \martin{two-dimensional} second-order macroscopic traffic flow models. Although recent contributions have indeed explored the understanding of either second-order longitudinal behaviour or two-dimensional first-order dynamics, a unified framework that captures second-order effects in both the longitudinal and lateral directions remains largely unexplored. The explicit modelling of lateral interactions and road boundary effects is still lacking; current approaches can hardly represent key features of disordered traffic, such as space-sharing behaviour, density redistribution, and boundary-driven positioning.
\vspace{0.2cm}

\noindent This study addresses these limitations by proposing a two-dimensional second-order macroscopic traffic flow model that consistently couples longitudinal and lateral dynamics within a unified continuum framework. The formulation introduces momentum equations in both directions, enabling a realistic description of interaction-driven motion and road boundary forces in disordered traffic environments. Vehicle–vehicle interactions are explicitly represented to capture longitudinal speed adaptation and lateral redistribution arising from local density variations. In addition, the model incorporates road-boundary influences, allowing the representation of boundary-induced confinement, space-sharing behaviour, and density redistribution across the road width. By integrating coupled longitudinal–lateral dynamics, interaction-driven mechanisms, and boundary influences, the proposed model provides a physically consistent and numerically stable tool for analysing disordered and lane-free traffic systems, thereby addressing a critical gap in current macroscopic traffic flow modelling.
\vspace{0.2cm}

\noindent The manuscript is organized into five sections. The next section presents the formulation of the proposed macroscopic model. Section 3 describes the numerical scheme and its associated stability conditions. Section 4 reports the results of numerical experiments conducted on isolated longitudinal and lateral dynamics, as well as fully coupled two-dimensional dynamics. Finally, Section 5 summarizes the main findings and concludes the paper.

\section{Model Formulation}
\noindent In disordered traffic, modelling both longitudinal and lateral dynamics is crucial to accurately capture the movement of vehicles, as they not only travel forward but also manoeuvre laterally to utilise available road space. Unlike traditional macroscopic models, which primarily focus on one-dimensional dynamics, a two-dimensional macroscopic model provides a more comprehensive representation by including a continuity equation and acceleration equations for both directions. This approach allows for the simulation of behaviours such as lane-changing, overtaking, and lateral gap utilisation, which are common in disordered traffic. Additionally, incorporating lateral dynamics enables the model to account for the interactions between different vehicle classes and the resulting effects on overall traffic flow efficiency. This two-dimensional framework is particularly well-suited for studying mixed traffic conditions and complex road environments. Hence, in this study, the one-dimensional continuity equation is first extended to a two-dimensional form to incorporate lateral vehicle movements. Subsequently, acceleration equations are proposed separately in the longitudinal and lateral directions to fully describe the two-dimensional traffic dynamics. The continuity equation has been extended to 2D to account for the lateral movements as \autoref{2dcontinuty}	
\begin{equation}
\frac{\partial \rho}{\partial t} + \frac{\partial (\rho u_x)}{\partial x} + \frac{\partial (\rho u_y)}{\partial y} = 0
\label{2dcontinuty}
\end{equation}
where $\rho(x,y,t)$ is traffic density represented in $\text{veh/m}^2$, while $u_x$ and $u_y$ denote the velocity components (m/s) in the longitudinal and lateral directions, respectively. The flow density (flow per cross section, unit $\unit{veh (ms)^{-1}}$) in these directions follow the hydrodynamic relationships $Q_x = \rho u_x$ and $Q_y = \rho u_y$.
\vspace{0.2cm}

\subsection{Longitudinal dynamics}
\noindent Macroscopic second-order traffic flow models can be formulated using different theoretical frameworks. As already mentioned, their acceleration equation can be derived from microscopic models through systematic averaging and upscaling procedures. Other approaches include a derivation from kinetic theory, where macroscopic variables are obtained as moments of an underlying vehicle distribution function, while phenomenological approaches directly postulate macroscopic evolution equations based on empirical observations. In this study, we adopt the first approach and derive the longitudinal acceleration equation from the underlying microscopic dynamics. The proposed model consists of three terms: The first two terms are derived from a modified version of the Full Velocity Difference Model (FVDM) by \citealp{jiang2001full} while the third term describing a deceleration effect near a road boundary has no car-following equivalent. The FVDM is a widely used car-following model that captures the fundamental aspects of vehicular acceleration and deceleration behavior. Its acceleration equation is given by 

\begin{equation}
\frac{dv_i}{dt} = \frac{v_{\text{opt}}(s) - v_i}{\tau} + \gamma (v_{i-1} - v_i),
\label{FVDM}
\end{equation}
where $v_i$ is the current speed of the $i$-th vehicle, $v_{i-1}$ is the speed of the leading vehicle, $v_{\text{opt}}(s)$ denotes the optimum speed that the driver aims to achieve as a function of the gap $s$, and $\gamma$ represents the sensitivity to relative speed. The FVDM has certain limitations, particularly in its inability to capture all traffic situations, as its relative-speed term does not depend on the inter-vehicle gap. Consequently, in scenarios where the leading vehicle is slow and far away, the relative-speed term prevents the subject vehicle to reach its desired speed, which is unrealistic in real-world driving behavior \citep{treiber2025traffic}. To address this limitation, we make the relative-speed term proportional to the gap sensitivity $v_{\text{opt}}'(s)$ of the optimal-velocity (OV) function. This interaction of the relative-speed sensitivity with the OV function ensures a plausible reaction whenever the OV function itself is plausible, i.e., $v_{\text{opt}}'(s) \to 0$ for sufficiently large gaps $s$. Finally, a road boundary component is incorporated to account for the tendency of drivers to slow down if very near the edges. We assume this decelerating component to be proportional to the speed (no slowing down if the speed is already zero) and to be exponentially increasing when approaching a boundary. In summary, for a road of width $2b$, the microscopic model underlying the longitudinal acceleration component of our proposed second-order model is expressed as
\begin{equation}
\frac{dv_i}{dt} = f_x^{\text{OVM}} + f_x^{\text{antic}} + f_x^{\text{road}}
\label{eq:long_gen}
\end{equation}
with
\begin{equation}
\label{eq:long_gen_forces}
f_x^{\text{OVM}}=\frac{v_{\text{opt}}(s) - v_i}{\tau},\quad
f_x^{\text{antic}} = \delta\, v'_{\text{opt}}(s)\,\bigl(v_{i-1} - v_i\bigr),\quad
f_x^{\text{road}} = \martin{-\frac{\beta_b\sup{long}}{\tau}} v_i\, e^{-\frac{b}{s_y}} \cosh\left(\frac{y}{s_y}\right),
\end{equation}
where $\tau$, $\delta$, \martin{$\beta_b\sup{long}$}, and $s_y$ are model parameters to be described below.
\vspace{0.2cm}

\noindent We now proceed to derive the longitudinal acceleration equation of our second-order model from \autoref{eq:long_gen} with \autoref{eq:long_gen_forces} by transforming the microscopic quantities to their macroscopic equivalents. Specifically, we transform the car-following speeds $v_i$ and $v_{i-1}$, the bumper-to bumper gap $s_i$, the distance $\Delta x=x_{i-1}-x_i-l$ (where $l$ is the effective vehicle length) to the macroscopic quantities local density $\rho$, local longitudinal speed $u_x$, and macroscopic steady-state speed $U_e(\rho)$ as follows:


\begin{subequations}
\label{eq:micro_to_macro}
\begin{align}
s_i &\;\rightarrow\;\frac{1}{\rho^{1/2}}-l,
\qquad
\Delta x \;\rightarrow\;\frac{1}{\rho^{1/2}}, 
\label{eq:s}
\\[4pt]
v_i &\;\rightarrow\; \martin{u_x(x,y,t), }
\qquad
v_{i-1} \;\rightarrow\; u_x(x+\Delta x,y,t),
\label{eq:v}
\\[4pt]
v_{\text{opt}}(s) 
&\;\rightarrow\; U_e(\rho(x_a)),
\label{eq:vopt}
\\[4pt]
v'_{\text{opt}}(s) 
&\;\rightarrow\; \frac{d U_e(\rho(s))}{d s}
\;\rightarrow\; -2\rho^{3/2} U_e'(\rho),
\label{eq:vopts}
\\[4pt]
 u_x(x+\Delta x,\martin{y},t) - u_x(x,y,t)
&\;\rightarrow\; \frac{\partial u_x}{\partial x}\Delta x
\;\rightarrow\; \frac{1}{\rho^{1/2}}\,\frac{\partial u_x}{\partial x}.
\label{eq:dv}
\end{align}
\end{subequations}

\noindent The first term $f_x^{\text{OVM}}$ of \autoref{eq:long_gen} represents the driver’s effort to attain the optimal speed for the actual gap. In contrast to the derivation of Payne's model where the density $\rho$ is evaluated at an intermediate location $(x+\Delta x/2,y)$ between the follower's and the leader's position separated by $\Delta x$, we assume a general anticipation point  I have moved up the definition of $x_a$ to its first occurrence.
\begin{equation}
\label{xa}
    x_a = x+s_a=x + l_{\rm eff}+ T u_x
\end{equation}
denotes the anticipated point for which the density is calculated. Here, $l_{\rm eff}$ denotes the minimum anticipation distance which can be interpreted as the average longitudinal distance headway for zero speed (it is larger than $(\rho_{\rm max})^{-1/2}$ due to longitudinal-lateral anisotropy) and the parameter $T$ can be interpreted as an anticipation time typically in the range \unit[1]{s} -- \unit[1.5]{s}, cf. \citealp{treiber2025traffic}. With \autoref{eq:vopt}, this results in
\begin{equation}
v_{\text{opt}}(s) \;\rightarrow\; U_e\!\left(\rho(x_a,\martin{y},t)\right),
\label{eq:opt_to_macro}
\end{equation}
Using \autoref{eq:opt_to_macro} and performing the appropriate change of variables, the macroscopic form of the OVM force \autoref{eq:long_gen_forces} is obtained as
\begin{equation}
\tilde{f}_x^{\text{OVM}} = \frac{U_e\!\left(\rho(x_a,t)\right) - u_x}{\tau},
\label{eq:self_force_macro}
\end{equation}
where the parameter $\tau$ denotes the speed adaptation time. The inclusion of anticipation allows the model to account for downstream traffic conditions, enabling drivers to adapt their behavior by reducing acceleration or increasing braking deceleration when necessary. This anticipation mechanism introduces a stabilizing effect, confining traffic flow instabilities to realistic density regimes under congested conditions. Notice that, in 1d, a Taylor expansion of $U_e(.)$ around $x$ with $x_a=x+\Delta x/2=x+2/\rho$ would lead to Payne's model but it turned out that an explicit nonlocal term increased the numerical stability so we do not apply this expansion.
\vspace{0.2cm}

\noindent The anticipation force $f_x^\text{anti}$ of \autoref{eq:long_gen_forces} acts as \martin{an additional} mechanism to prevent critical situations or potential collisions in the presence of slow leaders/speed gradients. When the leading vehicle is sufficiently far ahead, the interaction force vanishes since $v'_{\text{opt}}\to 0$ for any sensible OV function. The prefactor $\delta$ is a model parameter denoting the degree of speed anticipation.
This force can be translated into macroscopic quantities using the transformation given in \autoref{eq:micro_to_macro}, resulting in
\begin{equation}
\tilde{f}_x^{\text{antic}}
=
-\martin{\frac{\beta\sup{anti}}{\tau}}\,\rho\,U_e'(\rho)\,
\frac{\partial u_x}{\partial x},
\label{eq:interaction_macro}
\end{equation}
\martin{where $\beta\sup{anti}=2\delta\tau$ has been defined as the new model parameter giving the weighting of the anticipation relative to the weighting of the deviation from the steady-state speed.} Finally, the decelerating boundary force is directly transformed from the microscopic model leading to
\begin{equation}
\tilde{f}_x^{\text{road}}
=
-\frac{\beta_b\sup{long}}{\tau} u_x \, g(y)
\label{eq:road_boundary_force}
\end{equation}
\martin{with
\begin{equation}
    \label{eq:gy}
    g(y)=e^{-\frac{b}{s_y}} \cosh\left(\frac{y}{s_y}\right),
\end{equation}
where $y$ is the lateral distance measured from the road centerline, $b$ is equal to half the road width, and $s_y$ denotes the characteristic attenuation length. Since $s_y \ll b$ holds in most situations, the symmetric boundary interaction function
is essentially zero near the road center and takes on its maxima $g(\pm b)\approx 0.5$ at the road boundaries.
}
\vspace{0.2cm}

\noindent Since the car-following models \autoref{eq:long_gen} is formulated in the comoving (Lagrangian) frame of reference (unit mass, Lagrangian acceleration $du_x/dt=\tilde{f}_x$) while macroscopic models are generally formulated in the fixed (Eulerian) frame, we apply a transformation to the latter resulting into the final longitudinal acceleration equation
\begin{equation}
\frac{d u_x}{d t} = \frac{\partial u_x}{\partial t} + u_x \frac{\partial u_x}{\partial x}
+ u_y \frac{\partial u_x}{\partial y} = \tilde{f}_x.
\label{eq:long}
\end{equation}

\noindent An explicit expression for the longitudinal force is obtained by combining \autoref{eq:self_force_macro} - \autoref{eq:road_boundary_force} leading to
\martin{
\begin{equation}
   \tilde{f}_x=\frac{1}{\tau}\left(
    U_e (\rho(x_a, y,t)) - u_x 
    -\beta\sup{anti}\rho U_e^{\prime} (\rho) \frac{\partial u_x}{\partial x} 
    -\beta_b\sup{long} u_x\, g(y)
    \right),
  \label{eq:fx}
\end{equation}
}
where $x_a=x+ u_x T$ is the anticipation point and $\tau$, $T$, \martin{$\beta\sup{anti}=2\delta\tau$, $\beta_b\sup{long}$}, and $s_y$ are model parameters.

\subsection{Lateral Dynamics}
\noindent The lateral dynamics in the proposed model consists of three key components: self-driven force, traffic interaction force, and road-boundary force, 

\begin{equation}
\frac{du_y}{dt} = \tilf_y^{\text{self}} + \tilf_y^{\text{traffic}} + \tilf_y^{\text{road}}.
\label{eq:lat-gen}
\end{equation}  
The self-driven force $\tilde{f}_y\sup{self}$  originates from the necessity of \martin{shifting sideways for tactical reasons}, for example, to prepare for a turn or an exit. \martin{When simulating  mid-block road sections, this force can be set to zero.} The traffic force is motivated by a \martin{driver’s desire to improve driving conditions}, such as increasing speed, obtaining a better line of sight, or maintaining a larger following distance. Finally, \martin{the road boundary force  ensures} that vehicles remain within the permissible road boundaries and adapt to geometric and infrastructural constraints, thereby enhancing the model’s applicability to real-world traffic scenarios with varying lane configurations and obstacles.
\vspace{0.2cm}

\noindent We assume that the \martin{lateral dynamics due to tactical considerations and traffic is captured by a lateral OVM},
\begin{equation}
    \tilf_y^{\text{self}} + \tilf_y^{\text{traffic}} = \frac{u_y^{\text{self}} + u_y^{\text{traffic}} - u_y}{\tau_y}.
    \label{eq:OVM_Macro}
\end{equation}  
\martin{We set the traffic related desired lateral speed $u_y^{\text{traffic}}$ reflecting the
incentive to move sideways proportional to the optimal-velocity gradient in the lateral direction,
\begin{equation}
    \label{eq:uytraffic}
    \martin{u_y^{\text{traffic}} \propto \frac{\partial U_e(\rho)}{\partial y}.}
\end{equation}  
Notice that we do \emph{not} use the lateral gradient of the full OVM acceleration $\propto(U_e(\rho)-u_x)$ since inclusion of the latter would mean that, at the new lateral position $y+dy$, the longitudinal force is calculated with the local speed $u_x+\partial u_x/\partial_y \, dy$ while the vehicles at $y$ have their local speed $u_x$ as decision base which is consistent with the reasoning of the microscopic lane-changing model MOBIL (\citealp{kesting2007general})}.
\vspace{0.2cm}

\noindent \martin{Since drivers become more sensitive at higher speed (and also to avoid pure sidewards "sliding" at zero $u_x$) we set the lateral sensitivity $\beta\sup{lat} u_x$ proportional to the longitudinal speed and obtain
\begin{equation}
     \tilde{f}_y^{\text{self}}+ \tilde{f}_y^{\text{traffic}} 
     = \frac{u_y\sup{self}+\beta\sup{lat} u_x \frac{\partial U_e(\rho)}{\partial y} - u_y}{\tau_y}\, .
     \label{eq:lat_traffic}
\end{equation}  
}

\noindent \noexpand The road-boundary force \( (\tilf_y^{\text{road}}) \) ensures that the influence of road edges is incorporated into the model, preventing vehicles from deviating beyond the road boundaries. This force is derived as being directly proportional to the gradient of the boundary interaction function $g(y)$ with respect to the lateral direction, maintaining realistic vehicle positioning. \martin{With the lateral boundary repulsion parameter $\beta_b\sup{lat}$, the lateral road boundary force is given by } 

\begin{equation}
     \tilde{f}_y^{\text{road}}
     = -\frac{\beta_b\sup{lat}}{\tau_y} u_x g'(y)
     =-\frac{\beta_b\sup{lat}}{\tau_y} u_x
     \frac{\partial}{\partial y}
     \left(e^{-\frac{b}{s_y}}\cosh \left( \frac{y}{s_y} \right)\right).
     \label{eq:lat_road}
\end{equation}

\noindent Going again from the Lagrangian to the Euler coordinates, setting $\tilde{f}_y^\text{self}=0$ and inserting the expressions \autoref{eq:lat_traffic} and \autoref{eq:lat_road} into \autoref{eq:lat-gen}, we finally obtain
\begin{equation}
\frac{du_y}{dt} = \frac{\partial u_y}{\partial t}  
 + u_x \frac{\partial u_y}{\partial x} 
 + u_y \frac{\partial u_y}{\partial y} =\tilde{f}_y
  \label{eq:lat}
\end{equation}
with
\martin{
\begin{equation}
    \tilde{f}_y = \frac{1}{\tau_y}\left(
    u_y\sup{self}
    +\beta\sup{lat} u_x \frac{\partial U_e(\rho)}{\partial y} - u_y
    -
    \beta_b\sup{lat} u_x g'(y)\right), \quad
    g(y)=e^{-\frac{b}{s_y}}\cosh \left( \frac{y}{s_y} \right).
  \label{eq:fy}
\end{equation}
}
\noindent This specification contains three model parameters, the lateral speed adaptation time $\tau_y$, \martin{the sensitivity $\beta\sup{lat}$ to go laterally if there are more favorable conditions, and the sensitivity $\beta_b\sup{lat}$  to the repulsive boundary}. To this comes the scale $s_y$ of the boundary effects taken from the longitudinal acceleration equation.
To capture the dynamics of disordered traffic, it is necessary to solve \autoref{2dcontinuty}, \autoref{eq:fx} and \autoref{eq:fy}. Since these equations form a coupled system of nonlinear second-order partial differential equations, obtaining an analytical solution is not possible for all but the simplest configurations. Therefore, a numerical scheme must be developed.
The choice of the numerical method plays a critical role in ensuring stability and accuracy while capturing the intricate interactions within disordered traffic dynamics.

\section{Numerical Scheme}
\noindent 
Numerical schemes are broadly classified into explicit and implicit methods. Explicit methods calculate the system's future state based on its current state, while implicit methods solve equations involving both current and future states simultaneously. For realistic traffic simulations, where data is continuously updated to reflect varying boundary conditions, explicit methods are often preferred due to their simplicity and computational efficiency. We employ an explicit finite difference scheme to solve the longitudinal and lateral dynamics of the traffic model. This method involves discretizing the road space into finite segments of equal length and dividing time into intervals of consistent duration. The longitudinal and lateral dynamics are then approximated as finite-difference equations, which can be iteratively solved to compute traffic flow variables such as density and speed at each spatial and temporal grid point. A separate finite difference scheme has been used for longitudinal and lateral dynamics as explained in the subsections below.

\martinc{I would drop the isolated longitudinal nd transversal dynamics completely because these special cases follow in a very straightforward way from the general 2d case.}

\martin{\subsection{Flow-Conservative Formulation}}
\noindent Generally, conservation laws as our macroscopic model are most efficiently numerically analyzed in the flow-conservative form (\citealp{leveque1992numerical}), i.e., in terms of the quantities that are conserved without source terms. Here, these quantities are the density (conservation of the vehicle number) and the flow vector $\vec{Q}$ (momentum conservation).  By multiplying \autoref{eq:long} and \autoref{eq:lat} with $\rho$, eliminating the macroscopic local speed components \( u_x \) using the hydrodynamic relations \(Q_x = \rho u_x\) and $Q_y=\rho u_y$ and the continuity equation to get rid of the resulting time derivatives of the density, we obtain the flow-conservative form suitable for numerics as follows.

\begin{equation}
\frac{\partial \rho}{\partial t} + \frac{\partial Q_x}{\partial x} + \frac{\partial Q_y}{\partial y} = 0
\label{full_continuty}
\end{equation}
\begin{equation}
    \frac{\partial Q_x}{\partial t} + \frac{\partial}{\partial x} \left(\frac{Q_x^2}{\rho} \right) + \frac{\partial}{\partial y} \left(\frac{Q_x Q_y}{\rho} \right) =  \rho\tilde{f}_x
    \label{Qx-full}
\end{equation}
\begin{equation}
    \frac{\partial Q_y}{\partial t} + \frac{\partial}{\partial y} \left(\frac{Q_y^2}{\rho} \right) + \frac{\partial}{\partial x} \left(\frac{Q_x Q_y}{\rho} \right) =  \rho\tilde{f}_y
    \label{Qy-full}
\end{equation}
where \autoref{eq:fx} and \autoref{eq:fy} are reformulated as
\begin{eqnarray}
\label{eq:fxcons}
   \rho\tilde{f}_x &=& \frac{1}{\tau}\left[
    \rho U_e (\rho(x_a, y,t)) - Q_x 
    -\beta\sup{anti}U'_e(\rho)\rho^2\ablpart{}{x}\left(\frac{Q_x}{\rho}\right) 
    -\beta_b\sup{long} Q_x\, g(y)
    \right],\\
  \label{eq:fycons}
  \rho\tilde{f}_y &=& \frac{1}{\tau_y}\left[
    \rho u_y\sup{self}
    +\beta\sup{lat} Q_x \ablpart{U_e}{y}
    - Q_y
    -\beta_b\sup{lat} Q_x g'(y)\right].
\end{eqnarray}
\autoref{full_continuty} - \autoref{Qy-full} can be formulated \martin{as a vector conservation law} with source or sink terms as 
\begin{equation}
 \frac{\partial \vec{k}}{\partial t} + \frac{\partial \vec{f}(\vec{k})}{\partial x} +\frac{\partial \vec{g}(\vec{k})}{\partial y}= \vec{s}(\vec{k}) 
 \label{conservative}
\end{equation}   
where,
\begin{equation}
\vec{k} = \begin{bmatrix} \rho \\ Q_x \\ Q_y\end{bmatrix}, \quad
\vec{f}(\vec{k}) = \begin{bmatrix} Q_x \\  \frac{Q_x^2}{\rho} \\ \frac{Q_x Q_y}{\rho} \end{bmatrix}, \quad
\vec{g}(\vec{k}) = \begin{bmatrix} Q_y \\ \frac{Q_x Q_y}{\rho} \\ \frac{Q_y^2}{\rho} \end{bmatrix}, \quad
\vec{s}(\vec{k}) = \begin{bmatrix}
0 \\ 
\rho\tilde{f}_x\\
\rho\tilde{f}_y
\end{bmatrix}.
\end{equation}

\subsection{Numerical Scheme for the Longitudinal Dynamics}
\noindent The first-order upwind scheme has been employed to numerically solve the longitudinal dynamics using asymmetric upwind differences, which effectively capture information from the upstream direction. This approach is particularly advantageous for anticipative traffic models, such as the GKT model by \citealp{helbing1999numerical}, as it inherently incorporates nonlocal effects that allow downstream traffic conditions to influence upstream dynamics. This capability is crucial for accurately modelling congested traffic scenarios, where the propagation of disturbances—such as stop-and-go waves—plays a significant role in traffic flow behaviour \citep{treiber2025traffic}. The upwind scheme has widely been used in the literature for solving second-order macroscopic traffic models (\citealp{helbing1999numerical, jiang2002new, delis2014high, fosu2020two, ngoduy2021noise, zhang2021extended}) due to its robustness, simplicity, and effectiveness in handling discontinuities, such as shock waves and rarefaction waves, that commonly arise in traffic flow. Moreover, its computational efficiency makes it well-suited for large-scale traffic simulations, where real-time or near-real-time processing of complex traffic dynamics is required. Additionally, its straightforward implementation helps in easier integration with boundary conditions and empirical traffic data, making it a practical choice for real-world applications. 
\vspace{0.2cm}

\noindent With the anticipation point $x_a=x+u_xT$ containing the anticipation time $T$.  Since $x_a$ generally does not coincide with a grid point, the anticipated quantities (e.g., $\rho_a = \rho(x_a,t)$ and $Q_a = Q(x_a,t)$) are obtained using piecewise linear interpolation. Specifically, if the anticipation distance is $s_a = x_a - x$ at grid location $x = j\Delta x$, the interpolated value of a generic state variable $\vec{k}$  is omputed as
\[
\vec{k}^{a}_{j} = \vec{k}_{j+m} + \left(\vec{k}_{j+m+1} - \vec{k}_{j+m}\right)\left(\frac{s_a}{\Delta x} - m\right),
\quad
m = \left\lfloor \frac{s_a}{\Delta x} \right\rfloor.
\]
Here, $m$ denotes the number of grid cells between the current position and the upstream interpolation point, determined by the integer part (floor function) of the ratio $s_a/\Delta x$. In many practical cases, the anticipation distance is smaller than the cell size, resulting in $m=0$.  
\vspace{0.2cm}

\noindent In summary, dividing the road section into a cell size of \( \Delta x \), a time interval of \( \Delta t \), and denoting the traffic states \( \vec{k}(x,t) \) as \( \vec{k}_j^n \) at location \( j\Delta x \) and time \( n\Delta t \), the discretized form of the longitudinal part of \autoref{conservative} using the upwind scheme can be written as
\begin{equation}
    \vec{k}_j^{n+1} = \vec{k}_j^n - \frac{\Delta t}{\Delta x} (\vec{f}_j^n - \vec{f}_{j-1}^n) + \Delta t \vec{s}_j^n.
    \label{long_discretize}
\end{equation}

\subsection{Numerical Scheme for the Lateral Dynamics}
\noindent The numerical scheme for the lateral dynamics must be designed to allow the flux to propagate in either the left or right direction, depending on the sign of the lateral velocity. To achieve this, we employ the Lax-Friedrichs scheme, which is first-order accurate in space and time. This scheme is particularly advantageous for handling numerical fluxes that can propagate in multiple directions, making it well-suited for capturing the lateral dynamics of disordered traffic. 
\vspace{0.2cm}

\noindent Dividing the lateral road dimension into cells of size \( \Delta y \), a time interval of \( \Delta t \), and denoting the traffic states \( \vec{k}(y,t) \) as \( \vec{k}_l^n \) at location \( y=-b+l\Delta y \) and time \( t=n\Delta t \), the discretized form of the lateral dimension of \autoref{conservative} using the Lax-Friedrichs scheme can be written as
\begin{equation}
    \vec{k}_l^{n+1} = \frac{\vec{k}_{l-1}^n + \vec{k}_{l+1}^n}{2} - \frac{\Delta t}{2 \Delta y} (\vec{f}_{l+1}^n - \vec{f}_{l-1}^n) + \Delta t \vec{s}_l^n 
    \label{lat_discretize}
\end{equation}
When simulating the full two-dimensional dynamics, the discretized state variables are given by $\vec{k}_{lj}^n$ and longitudinal and lateral derivatives are discretized using \autoref{long_discretize} and \autoref{lat_discretize}, respectively.
\noindent \martin{Note that, in the full 2d case, the state vector $\vec{k}_{jl}^n$  will get two subscripts (spatial indices).}

\subsection{\martin{Numerical Stability Conditions}}
\noindent The Courant--Friedrichs--Lewy (CFL) conditions are fundamental stability criterion in the numerical solution of partial differential equations, ensuring that the computed solution remains stable, accurate, and physically meaningful. It establishes a relationship between the time step $\Delta t$, the spatial resolutions $\Delta x$ and $\Delta y$, and the maximum characteristic wave speed of the system, thereby preventing violations of causality and the emergence of non-physical oscillations or numerical distortions. The First-order CFL condition, associated with convective \martin{numerical} instability arising from first-order spatial derivatives, \martin{states that signals may not travel more than one cell per time step. Since the maximum signal speed is essentially the free-flow speed $v_f$, it reads for the longitudinal dynamics}
\begin{equation}
\Delta t < \frac{\Delta x}{v_f}.
\label{eq:cfl_convective}
\end{equation}
\martin{Since the lateral dynamics involves much slower speeds, the CFL condition is not critical for this dimension. }
\martinc{Check! You have much smaller lateral cell width!!}
\vspace{0.2cm}

\noindent Furthermore, since the proposed model does not contain diffusion terms, there are no second-order CFL conditions to take care of. In addition to the CFL condition, the explicit time integration of the source terms imposes a relaxation stability restriction on the time step. Since the anticipation term is associated with the convective dynamics, only the local relaxation and road-boundary source terms are considered. Accordingly, the longitudinal acceleration equation reduces to

\begin{equation}
\frac{du_x}{dt}
=
\frac{1}{\tau}
\left(
U_e-u_x-\beta_b^{\mathrm{long}}u_xg(y)
\right).
\label{eq:relax1}
\end{equation}

\noindent The source term is integrated using the explicit Euler method,

\begin{equation}
u_x^{n+1}
=
u_x^{n}
+
\frac{\Delta t}{\tau}
\left(
U_e-u_x^{n}
-\beta_b^{\mathrm{long}}u_x^{n}g(y)
\right).
\label{eq:relax2}
\end{equation}

\noindent The steady-state solution is obtained by setting the source term equal to zero,

\begin{equation}
u_x^{e}
=
\frac{U_e}
{1+\beta_b^{\mathrm{long}}g(y)}.
\label{eq:relax3}
\end{equation}

\noindent To analyze the stability of the numerical integration, a small perturbation is introduced about the steady state,

\begin{equation}
u_x=u_x^{e}+\epsilon .
\label{eq:relax4}
\end{equation}

\noindent  Substituting Eq.~(\ref{eq:relax4}) into Eq.~(\ref{eq:relax2}) and using the steady-state relation Eq.~(\ref{eq:relax3}) gives

\begin{equation}
\epsilon^{\,n+1}
=
\left[
1-
\frac{\left(1+\beta_b^{\mathrm{long}}g(y)\right)\Delta t}{\tau}
\right]
\epsilon^{\,n}.
\label{eq:relax5}
\end{equation}

\noindent Therefore, the amplification factor is

\begin{equation}
G
=
1-
\frac{\left(1+\beta_b^{\mathrm{long}}g(y)\right)\Delta t}{\tau}.
\label{eq:relax6}
\end{equation}

\noindent For the perturbation to decay monotonically, the amplification factor must satisfy
$0<G<1$, which gives

\begin{equation}
\Delta t<
\frac{\tau}
{1+\beta_b^{\mathrm{long}}g(y)}.
\label{eq:relax7}
\end{equation}

\noindent Since the strongest boundary influence occurs at the road boundary, where $g(y)$ attains its maximum value $g_{\max}$, the most restrictive longitudinal relaxation stability condition becomes

\begin{equation}
\Delta t<
\frac{\tau}
{1+\beta_b^{\mathrm{long}}g_{\max}}.
\label{eq:relax8}
\end{equation}

\noindent A similar analysis can be carried out for the lateral momentum equation, leading to the corresponding relaxation stability condition

\begin{equation}
\Delta t<\tau_y.
\label{eq:relax9}
\end{equation}

\noindent Here, $\tau$ and $\tau_y$ denote the longitudinal and lateral relaxation times, respectively. Since the simulation time step must satisfy both the convective (CFL) and relaxation stability constraints, the upper bound on the time step is given by
\begin{equation}
\Delta t
=
\min\!\left(
\frac{\Delta x}{v_f},
\;
\frac{\tau}{1+\beta_b^{\mathrm{long}}g_{\max}},
\;
\tau_y
\right).
\label{eq:dt_final}
\end{equation}
In the following section, numerical experiments are conducted on \autoref{long_discretize} and \autoref{lat_discretize} under various traffic conditions to investigate the lateral and longitudinal dynamics of the proposed model.

\section{Numerical Experiments}
\noindent To evaluate the effectiveness of the proposed model in capturing critical traffic dynamics, we conduct a series of numerical simulations. These simulations are performed on a uniform road segment of length \( L \) and width \( 2b \), defined within the spatial domain \([0, L] \times [-b, b]\). The setup assumes a mid-block road section with no lateral access or egress of vehicles. 
Consequently, the source term in the continuity equation \autoref{2dcontinuty} is set to zero
ensuring the conservation of vehicle density within the system. The \martin{complete set of simulation parameters, including model parameters, parameters of the used triangular fundamental diagram,} and discretization values, are listed in \autoref{tab:simulationparams}. Furthermore, details regarding the initial conditions, as well as an in-depth discussion of the model's behaviour under various traffic scenarios, are presented in the subsequent subsections. \martin{We emphasize that we have chosen a very fine discretisation to show the details of the lateral dynamics. However, we tested the numerics also with the much coarser discretisation $\Delta x=\unit[50]{m}$, $\delta y=\unit[0.5]{m}$, and $\Delta t=\unit[0.1]{s}$ with good results (cf. Fig.~\ref{fig:density1d_3})} \martinc{Fig. 3 needs to be improved; see there}

\renewcommand{\arraystretch}{1.11}
\begin{table}[H]
\centering
\fontsize{10}{12}\selectfont
\caption{Simulation Parameters}
\label{tab:simulationparams}
\begin{tabular}{@{}ll@{}}
\toprule
Parameter & Value \\
\midrule
Road Length, $L$ & 5000 m \\
Road Width, $2b$ & 10 m \\
Longitudinal step size, $\Delta x$ & \martin{\unit[50]{m}} \\
Lateral step size, $\Delta y$ & \unit[0.05]{m} \\
Grid size, $N_x \times N_y$ & $100 \times 201$ \\
Timestep, $\Delta t$ & 0.001 s \\
Fundamental diagram: maximum density, $\rho_{\rm max}$ & 0.16 veh/m$^2$ \\
Fundamental diagram: free-flow speed, $v_f$ & 30.55 m/s \\
Fundamental diagram: wave velocity, $w$ & $-\unit[5.2]{m/s}$ \\
Anticipation time, $T$ & \unit[1.5]{s} \\
Minimum anticipation distance, $l_{\rm eff}$ & \unit[4]{m} \\
Scale of boundary effects, $s_y$ & 0.5 m \\
Longitudinal speed adaptation time, $\tau$ & 0.65 s \\
\martin{Longitudinal anticipation factor, $\beta\sup{anti}$} & \unit[0.65]{s} \\
Longitudinal boundary sensitivity, $\beta_b\sup{long}$ & 2 \\
Lateral speed adaptation time, $\tau_y$ & 1.5 s \\
Lateral sensitivity, $\beta\sup{lat}$ & \martin{\unit[0.1]{s}} \\
Lateral boundary sensitivity, $\beta_b\sup{lat}$ & \unit[1.2]{m} \\
\bottomrule
\end{tabular}
\end{table}

\subsection{Longitudinal Dynamics}
\noindent In the first set of simulations, we only consider the longitudinal component ignoring all lateral movements and their coupling to the longitudinal dynamics. This simplification allows the investigation of traffic phenomena that mainly evolve along the longitudinal direction of the roadway, such as shockwave propagation, queue formation, and other dynamic behaviors occurring along the length of the road. It also allows to compare the longitudinal model -- which is a new model on its own -- with other one-dimensional second-order models such as Payne's model (\citealp{payne1971model}) or the GKT model (\citealp{treiber1999derivation}). In each simulation scenario, we consider two cases, (i) complete lateral homogeneity, $\frac{\partial}{\partial y} = 0$, providing a direct link to the one-dimensional models, and (ii) lateral dependence of $\rho$ and $u_x$ via the boundary conditions but without considering relaxation due to lateral velocity components that would arise in the full model (lane synchronization in the lane-based case). For the equilibrium flow, a triangular fundamental diagram is adopted due to its simplicity and its ability to effectively capture the essential characteristics of traffic flow. It represents traffic behavior as follows:
\begin{itemize}
    \item In the free-flow regime, the traffic wave speed is equal to the desired speed $v_f$ of the vehicles.
    \item In the congested regime, traffic waves travel backwards at a constant velocity $w$.
    \item At the maximum density, the flow reaches zero. This also defines the capacity density (maximum flow per meter cross section) as (\citealp{treiber2025traffic})
    \[
    Q_\text{max}=\rho_\text{max}\left(\frac{1}{v_f}-\frac{1}{w}\right)^{-1}
    \]
\end{itemize}
The triangular fundamental diagram can be expressed as 
\begin{equation}
 Q_e(\rho) = 
\begin{cases}
v_f \rho & \text{if } \rho \leq \rho_c = \frac{Q_\text{max}}{v_f}=\frac{\rho_\text{max}}{1-\frac{v_f}{w}}, \ \text{(free traffic),}\\
Q_\text{max}\left(1 - \frac{w}{v_f}\right) + w\rho & \text{if } \rho_c < \rho \leq \rho_{\text{max}} \ \text{(congested traffic).}
\end{cases}
\label{eq:triangularFD}
\end{equation}
The parameters $v_f$, $w$, and $\rho_\text{max}$  are summarized in \autoref{tab:simulationparams}. 
\vspace{0.2cm}

\noindent The initial flow condition is specified as 
$Q(x,0) = Q_e\big(\rho(x,0)\big)$ and is consistently implemented for all scenarios, including cases with boundary effects. The longitudinal dynamics are examined through three distinct scenarios, each designed to capture different traffic flow conditions and behaviors. These cases are outlined as follows:

\subsubsection*{Case 1: Free-Flow Region}

\noindent In this case, the propagation of shockwaves is examined when the traffic density lies entirely within the free-flow regime. Under the assumption of lateral homogeneity, the analysis is simplified by focusing exclusively on the longitudinal dynamics, including shockwave propagation and vehicle interactions. The \martin{initial conditions are} prescribed as \martinc{Check the speed initial condition!}
\begin{equation}
\rho(x\martin{,0}) =
\begin{cases}
0.004~\mathrm{veh/m^2}, & 0 \leq x < \martin{\unit[1\,650]{m}},\\
0.015~\mathrm{veh/m^2}, & \unit[1\,650]{m} \leq x < \unit[3\,300]{m},\\
0.004~\mathrm{veh/m^2}, & \unit[3\,300]{m} \leq x < \unit[5\,000]{m},
\end{cases} \quad
\martin{u_x(x,0)=U_e(\rho(x,0))=v_f}
\label{eq:init_case1}
\end{equation}
which represents a sharp discontinuity between two low-density states, both corresponding to conditions where inter-vehicle spacing is large and longitudinal interactions remain weak ($\rho < \rho_c=\unit[0.023]{veh/m^2}$). Under these conditions, vehicle motion is dominated by the self-driven acceleration toward the equilibrium speed, while interaction forces play a negligible role. As shown in \autoref{fig:density1d_1_lines}, the density wave front propagates downstream at approximately the free-flow speed $v_f$, which is also evident in the corresponding speed profile in \autoref{fig:speed1d_1_steady}.
When road-boundary forces are included, vehicles traveling near the boundaries experience additional resistance, leading to a reduced propagation speed of the wave front. This effect is clearly visible in the speed distribution shown in \autoref{speed_with_boundary} \martin{and, via the relaxation to the steady state, has also implications for the density \autoref{density_with_bundary}.}

\begin{figure}[H]
    \centering
    
    \begin{subfigure}{0.48\textwidth}
        \centering
        \includegraphics[width=\linewidth]{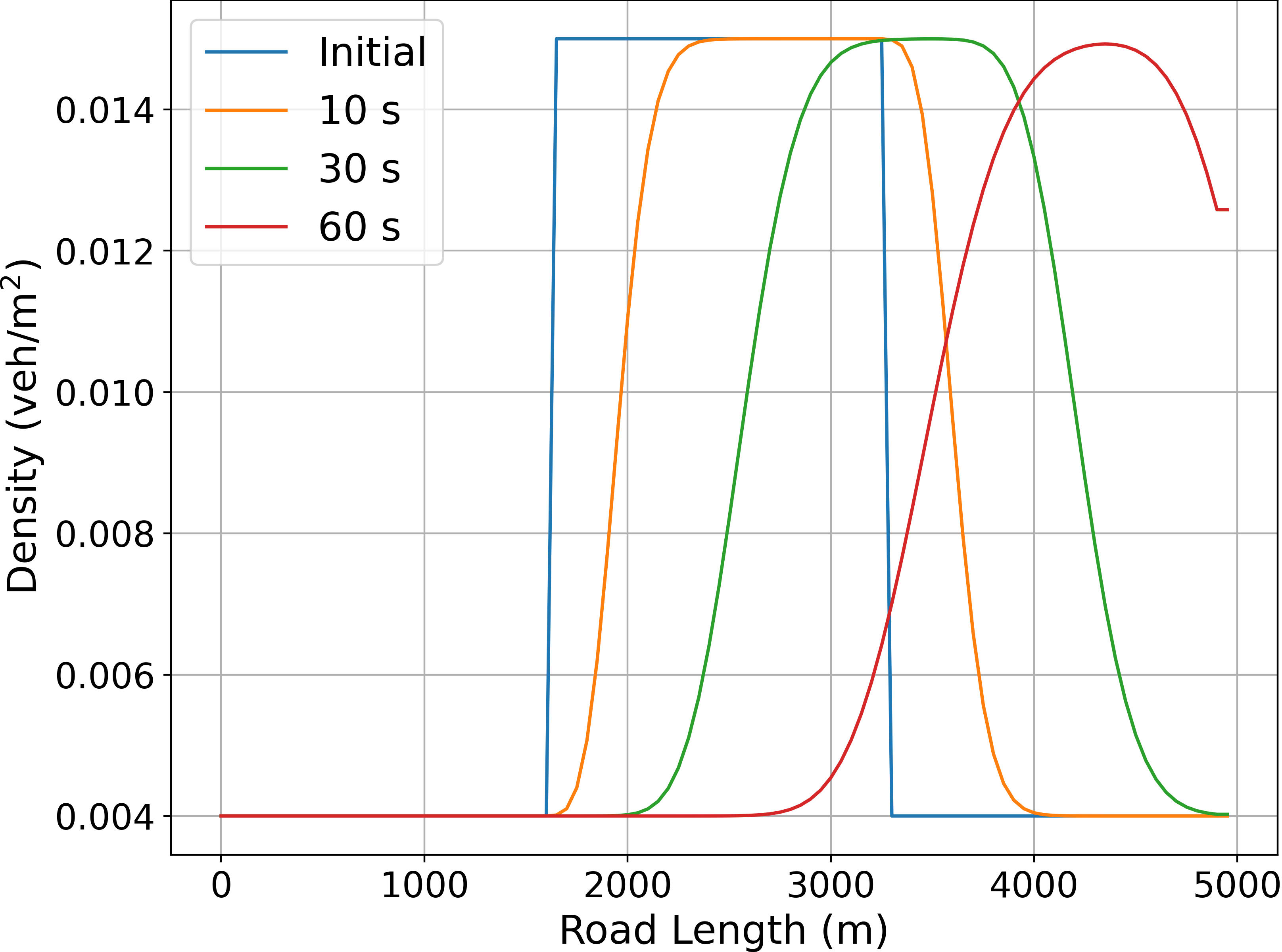}
        \caption{Without Boundary Forces}
    \end{subfigure}
    \hspace{0.02\textwidth}
    \begin{subfigure}{0.48\textwidth}
        \centering
        \includegraphics[width=\linewidth]{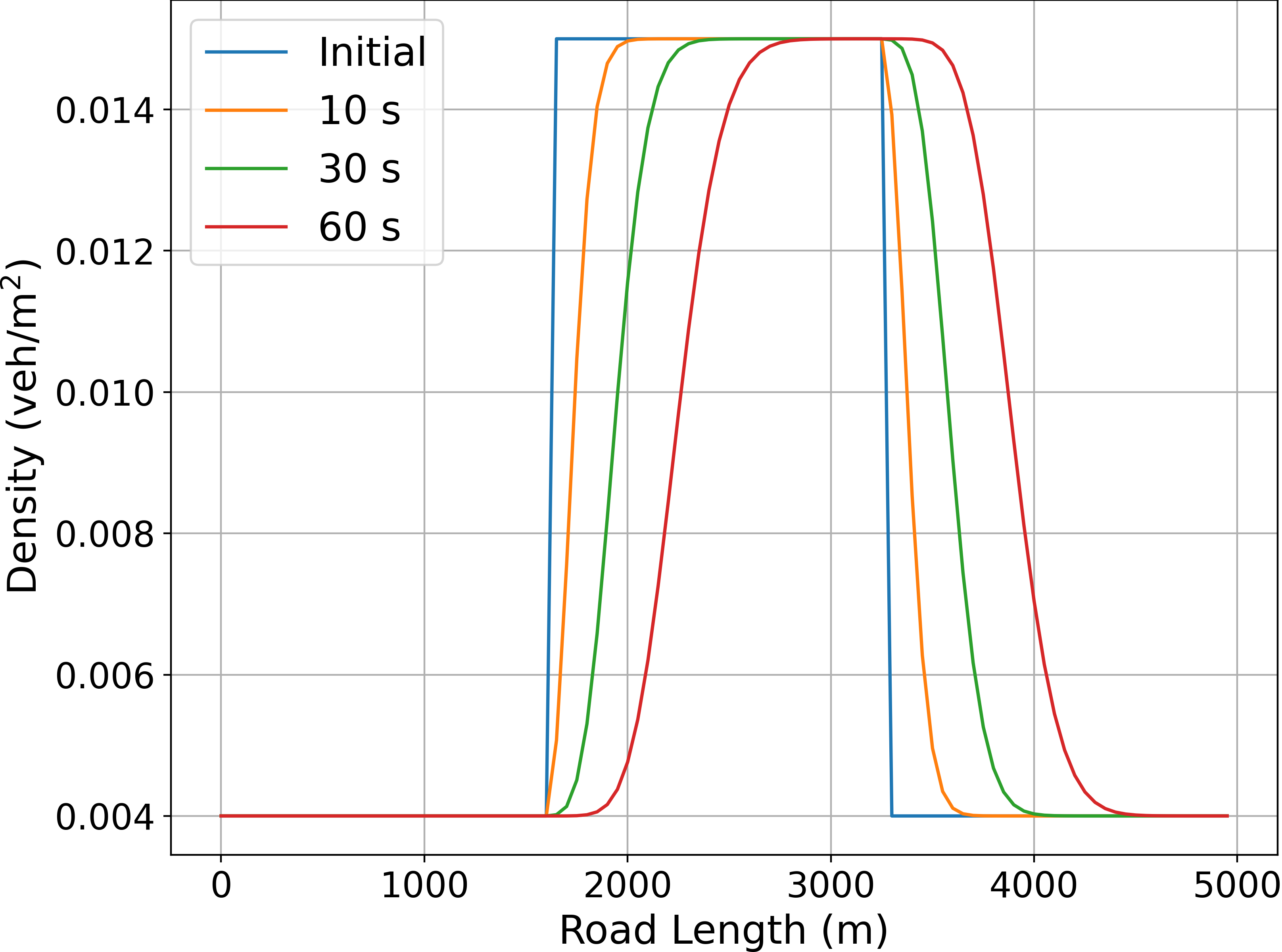}
        \caption{With Boundary Forces \martinc{at which $y$?} at $y = b$} 
        \label{density_with_bundary}
    \end{subfigure}
    
    \caption{Comparison of longitudinal density for Case 1.}
    \label{fig:density1d_1_lines}
\end{figure}

\begin{figure}[!htbp]
    \centering
    
    \begin{subfigure}{0.48\textwidth}
        \centering
        \includegraphics[width=\textwidth]{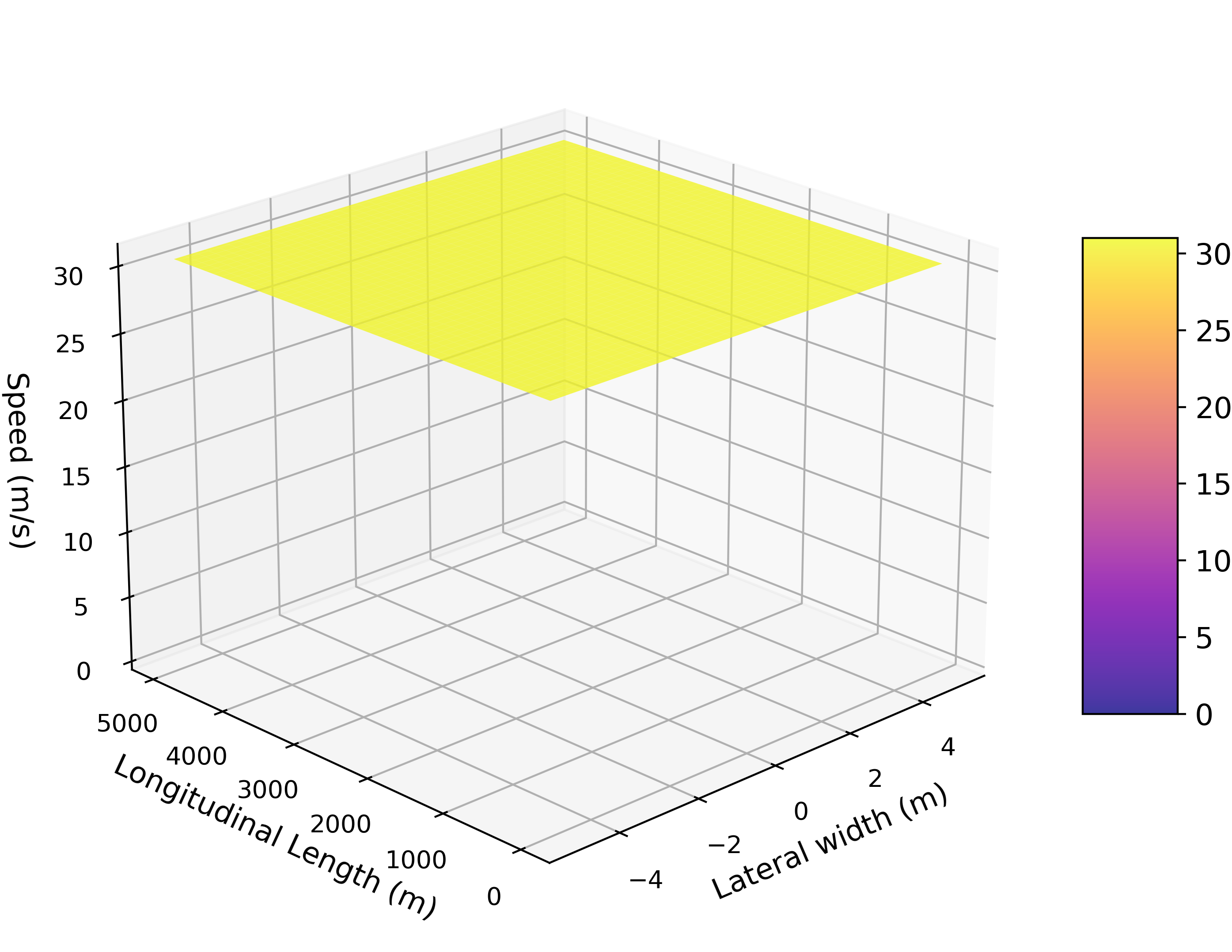}
        \caption{Without Boundary Forces}
        \label{speed_without_boundary}
    \end{subfigure}
    \hspace{0.02\textwidth}
    \begin{subfigure}{0.48\textwidth}
        \centering
        \includegraphics[width=\textwidth]{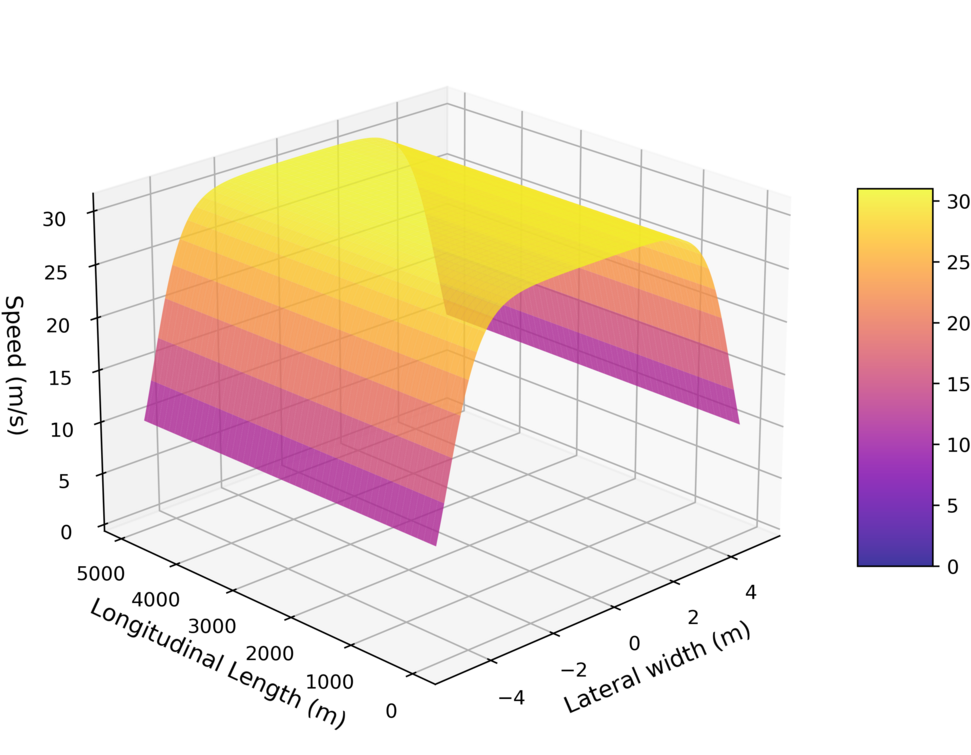}
        \caption{With Boundary Forces}
        \label{speed_with_boundary}
    \end{subfigure}
    
    \caption{Comparison of longitudinal speed profiles for Case 1.}
    \label{fig:speed1d_1_steady}
\end{figure}

\noindent The density discontinuity propagates downstream over time while largely preserving its structure. \martin{The observed smoothing is due to numerical diffusion which, for the upwind scheme and local speed $V$is given by \citealp{treiber2025traffic}
\begin{equation}
D_{\mathrm{num}}
=
\frac{V(y)\,\Delta x}{2}
\left(
1-
\frac{V(y)\,\Delta t}{\Delta x}
\right).
\label{Dnum}
\end{equation}
For the situation in \autoref{fig:density1d_1_lines}, we have a constant $V=v_f$ and a $y$ dependent speed $V(y)$ with $V(\pm b)=\unit[10.18]{m/s}$ for the cases (a) and (b), respectively. In both cases, we do not have an $x$ dependence of the speed (notice that we do not include lateral forces in this section) and an analytical solution of the linear diffusion-transport equation is available,

\begin{equation}
\rho(x,t,y)=\rho_0+\Delta\rho_0\left[\Phi\!\left(\frac{x-V(y)t-x_1}{\sqrt{2D_{\mathrm{num}}t}}\right)-\Phi\!\left(\frac{x-V(y)t-x_2}{\sqrt{2D_{\mathrm{num}}t}}\right)\right].
\label{eq:diffTransport}
\end{equation}

\noindent where $\Phi(z)$ denotes the standard normal distribution and $\rho_0=\unit[0.004]{veh/m^2}$, $\Delta \rho_0=\unit[0.011]{veh/m^2}$, $x_1=\unit[2\,000]{m}$, and $x_2=\unit[4\,000]{m}$ come from the initial condition \autoref{eq:init_case1}. We verified that this analytical solution 
agrees with the simulation as shown in \autoref{fig:density1d_3}. \martinc{see images rho?DT*.png}}
    \begin{figure}[H]
    \centering
    
    \begin{subfigure}{0.48\textwidth}
        \centering
        \includegraphics[width=\textwidth]{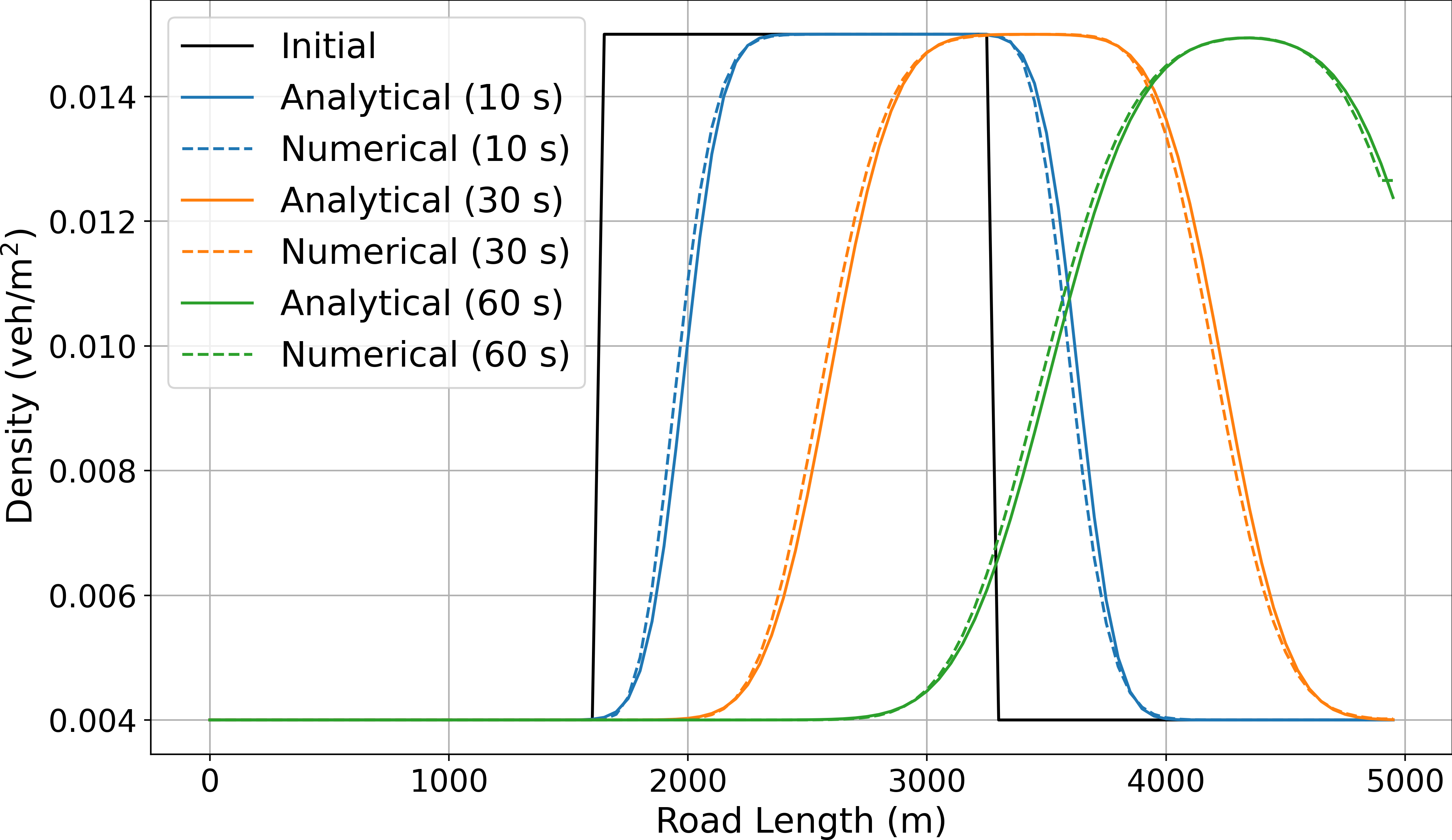}
        \caption{Without boundary forces}
    \end{subfigure}
    \hspace{0.02\textwidth}
    \begin{subfigure}{0.48\textwidth}
        \centering
        \includegraphics[width=\textwidth]{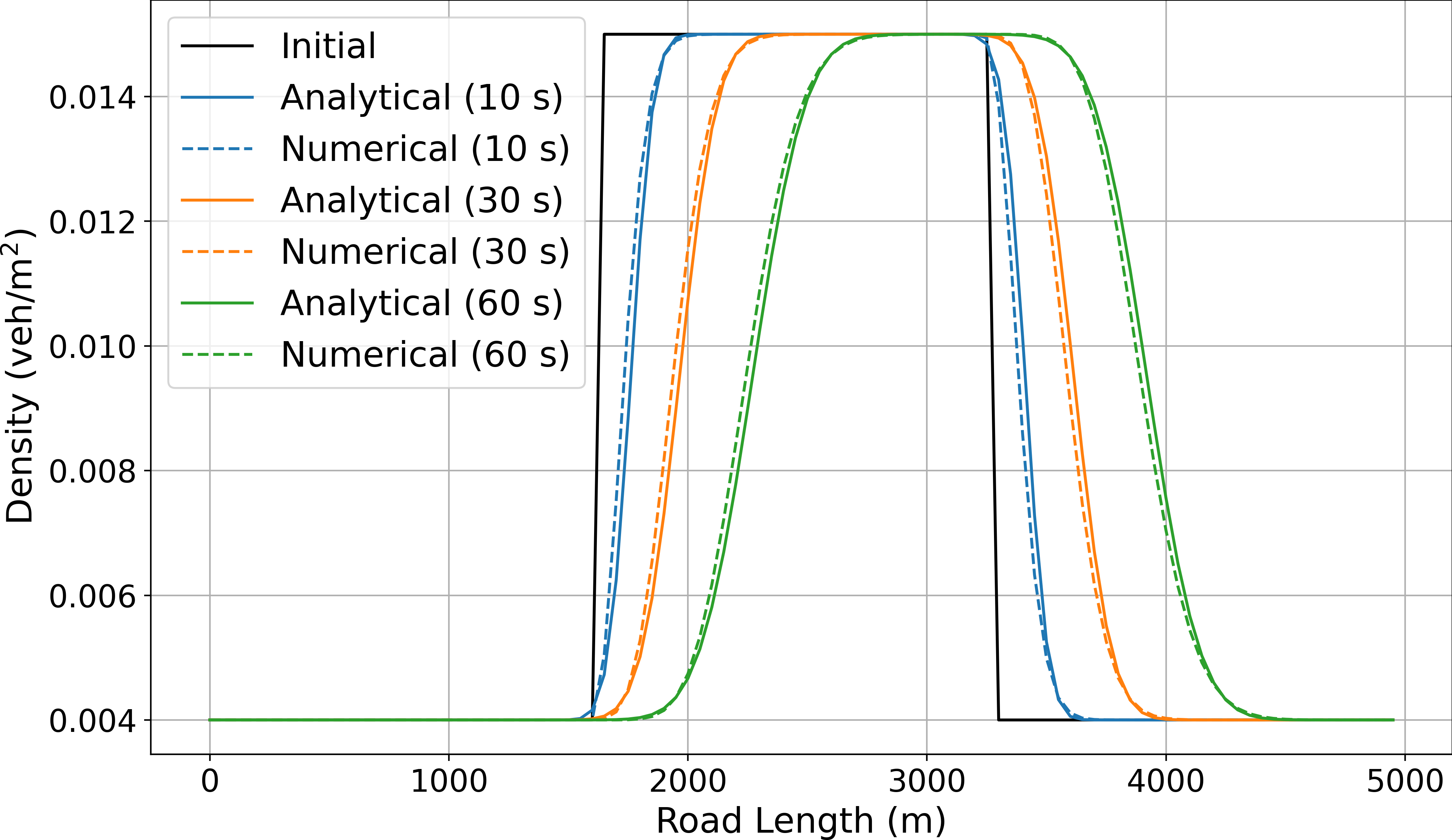}
        \caption{With Boundary forces at y = b}
    \end{subfigure}
    
    \caption{\martin{Comparison of the analytical solution for Case~1 with the numerical solutions using the standard discretisation of Table~\ref{tab:simulationparams} (dashed), and a coarser discretisation $\Delta x=\unit[50]{m}$, $\Delta y=\unit[0.5]{m}$, and $\Delta t=\unit[0.1]{s}$ (dotted)}}
    \label{fig:density1d_3}
\end{figure}
\begin{table}[htbp]
\centering
\caption{ Relative RMSE between numerical and analytical solutions.}
\label{tab:rmse_comparison}
\begin{tabular}{c c c}
\hline
\textbf{Time (s)} & {\textbf{Free Flow}} & {\textbf{Boundary}} \\
\hline
10  & 0.0309 & 0.0378 \\
30  & 0.0238 & 0.0286 \\
60  & 0.0150 & 0.0241 \\
\hline
\end{tabular}
\end{table}
\noindent
A quantitative comparison of density between the numerical and analytical solutions is performed using the relative root mean square error (RMSE), as shown in \autoref{rmse}, and the results are presented in \autoref{tab:rmse_comparison}. The low RMSE values indicate that the numerical solution closely matches the analytical solution.
\begin{equation}
 \mathrm{RMSE} = \sqrt{\frac{1}{N} \sum_{i=1}^{N} \left( \frac{\rho_{\text{num},i} - \rho_{\text{ana},i}}{\rho_{\text{ana},i}} \right)^2}   
 \label{rmse}
\end{equation}
 
\noindent \autoref{fig:density1d_1_without_005} presents the two-dimensional spatio-temporal evolution of traffic density in the absence of road-boundary forces, while \autoref{fig:density1d_1_with_005} illustrates the corresponding evolution when boundary effects are taken into account. The inclusion of boundary-induced resistance introduces an additional dissipative effect that slightly reduces the effective longitudinal velocity. Vehicles located near the road boundaries experience noticeable deceleration due to boundary interactions, whereas higher velocities are maintained near the road centre. 
\martin{This first case demonstrate that the numerical simulation produces valid results and the} proposed second-order macroscopic model consistently reproduces physically realistic free-flow behaviour while accounting for road-boundary effects in a stable and controlled manner.

\begin{figure}[!htbp]
    \centering

    \begin{subfigure}{0.48\textwidth}
        \centering
        \includegraphics[width=\linewidth]{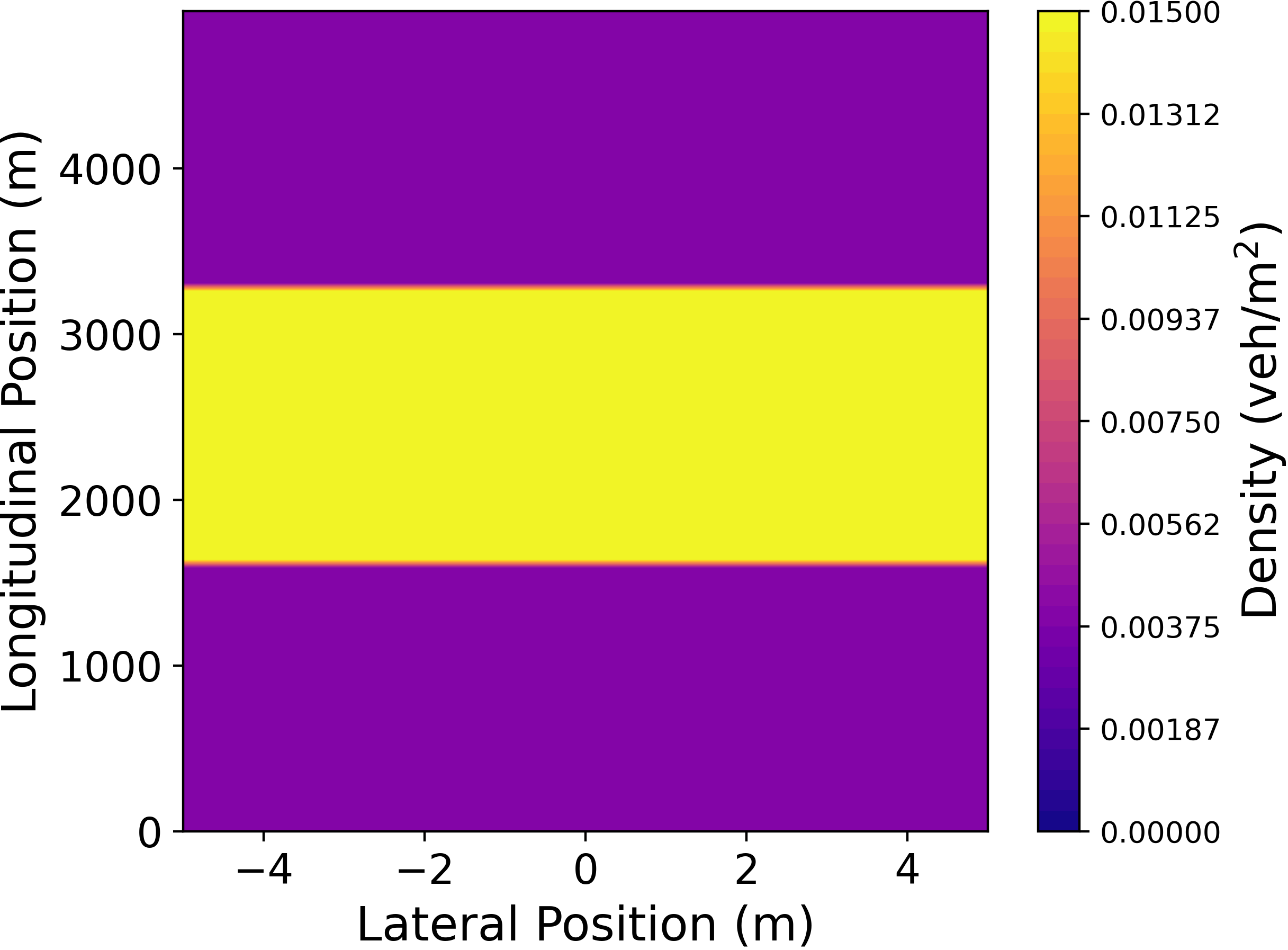}
        \caption{$t = 0$ s}
    \end{subfigure}
    \hfill
    \begin{subfigure}{0.48\textwidth}
        \centering
        \includegraphics[width=\linewidth]{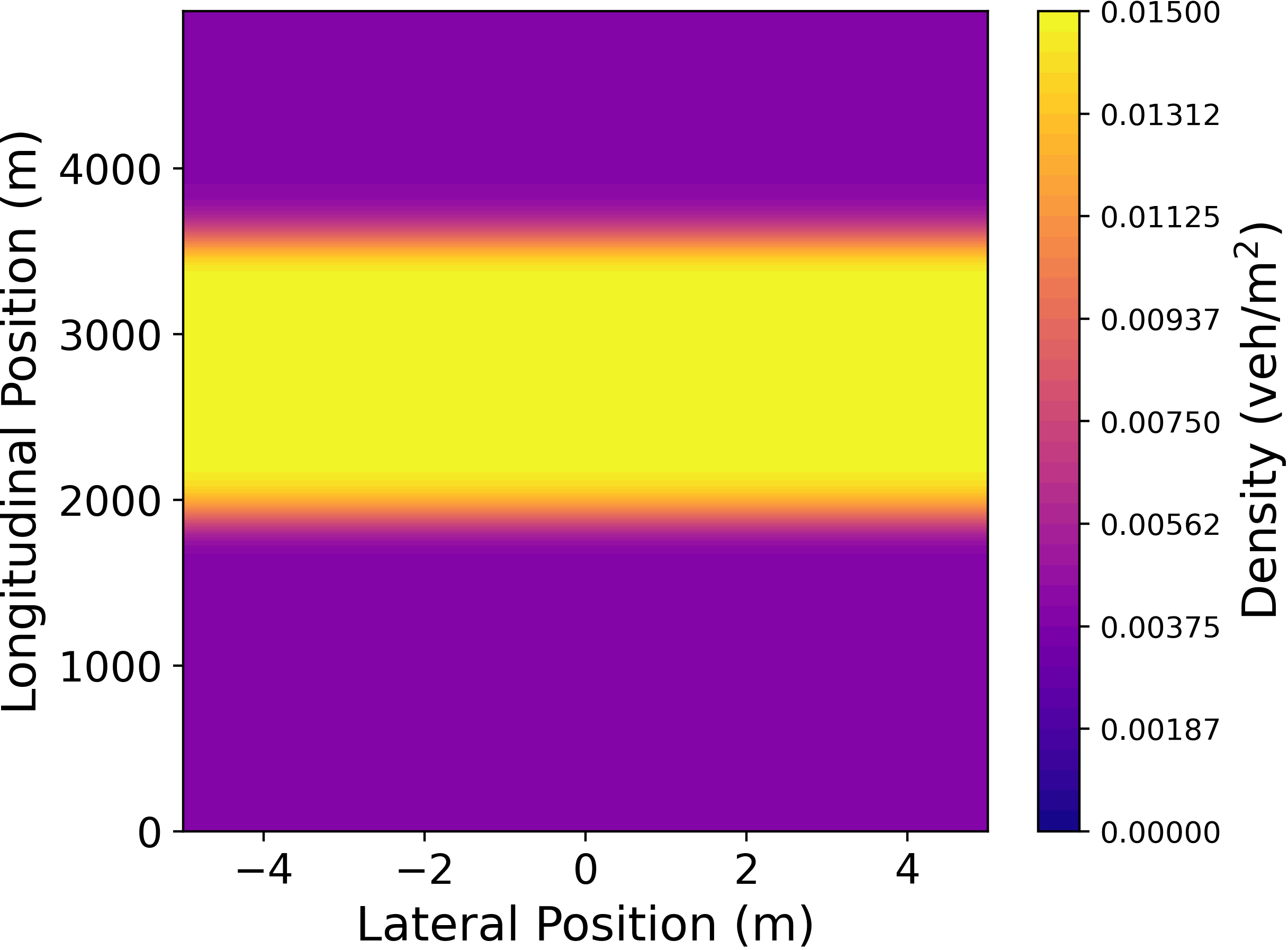}
        \caption{$t = 10$ s}
    \end{subfigure}

    \caption{Temporal evolution of traffic density without the road-boundary forces for Case~1.}
    \label{fig:density1d_1_without_005}
\end{figure}

\begin{figure}[!htbp]
\ContinuedFloat
    \centering

    \begin{subfigure}{0.48\textwidth}
        \centering
        \includegraphics[width=\linewidth]{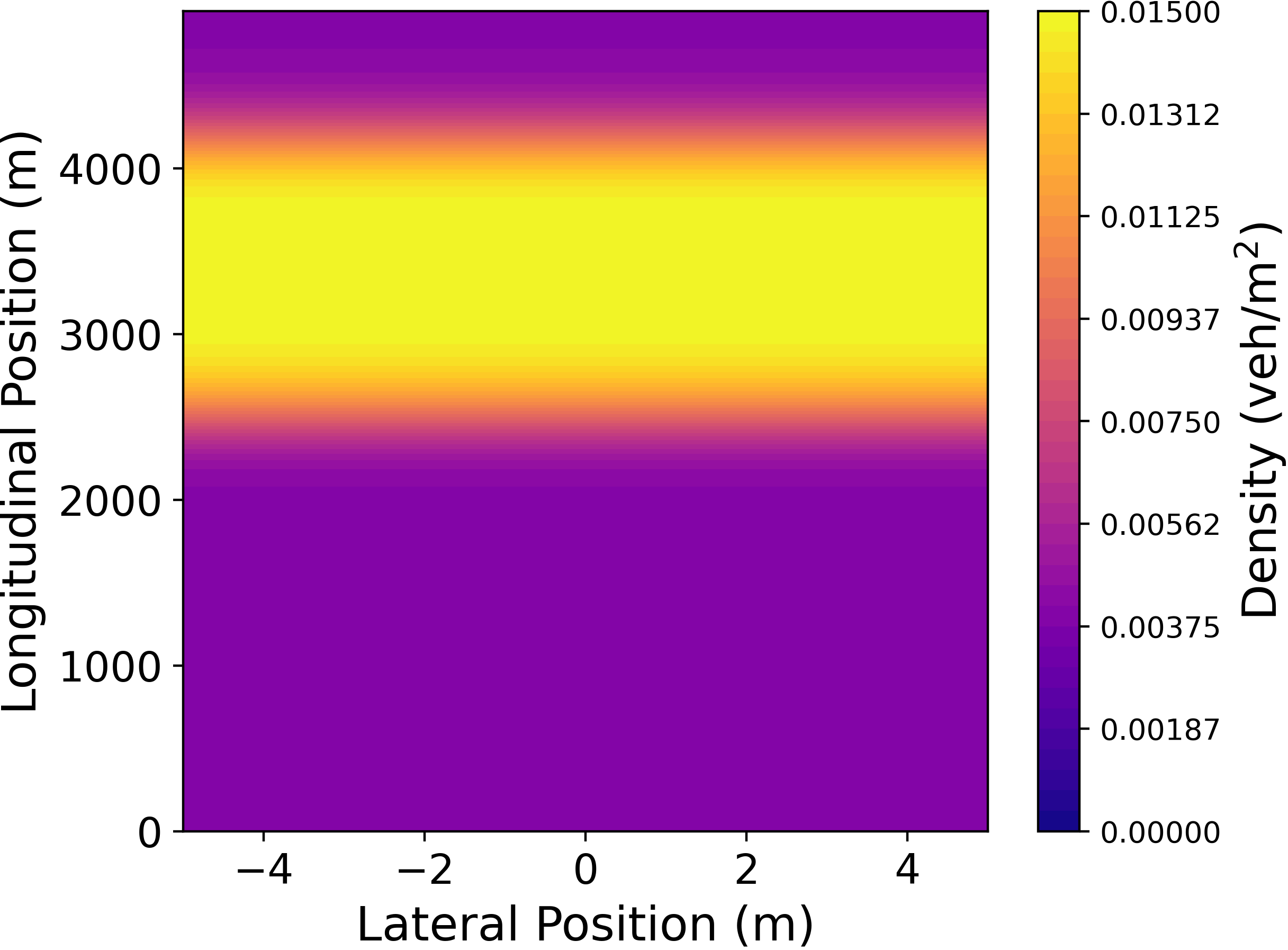}
        \caption{$t = 30$ s}
    \end{subfigure}
    \hfill
    \begin{subfigure}{0.48\textwidth}
        \centering
        \includegraphics[width=\linewidth]{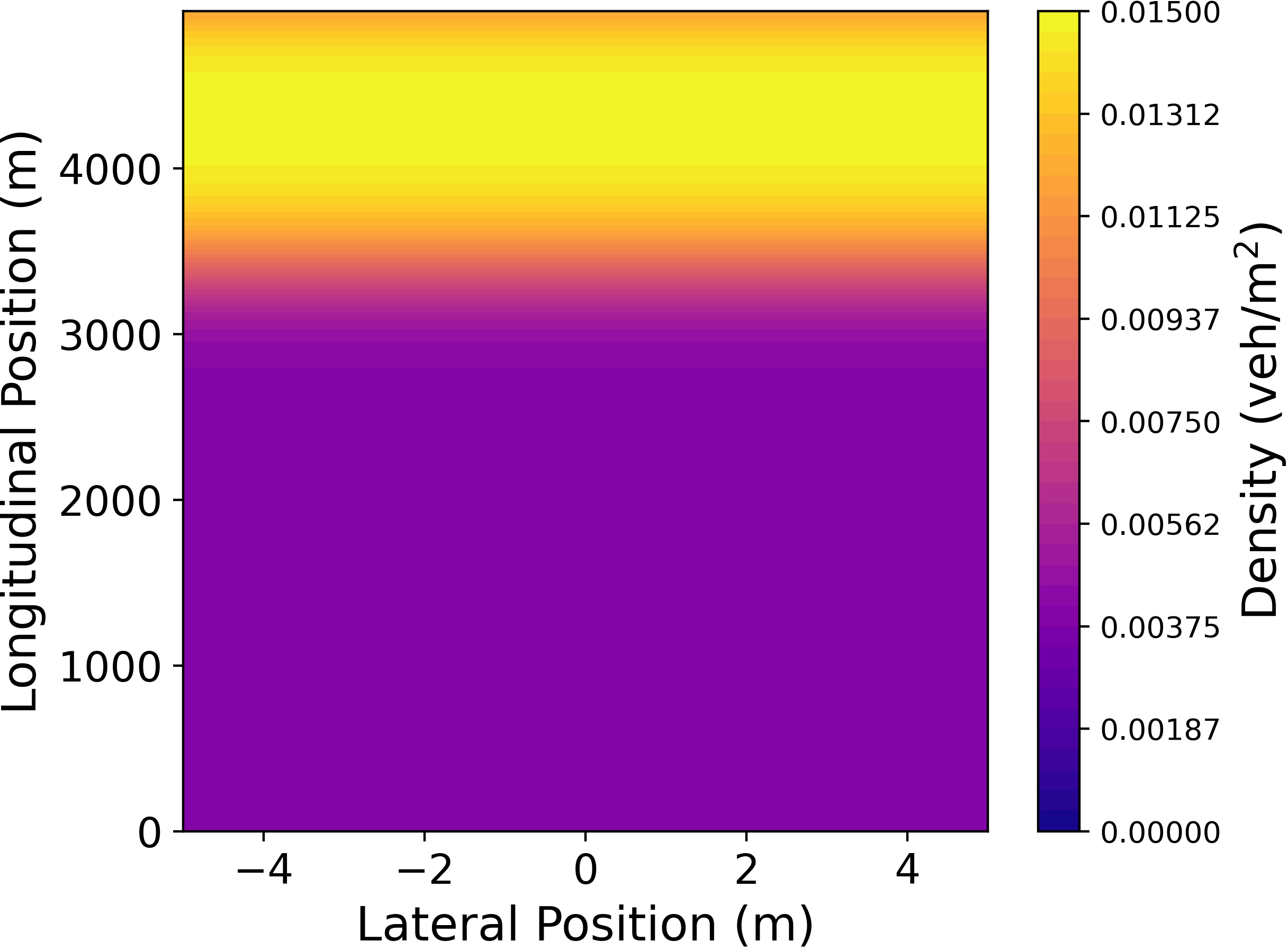}
        \caption{$t = 60$ s}
    \end{subfigure}

    \caption[]{Temporal evolution of traffic density without the road-boundary forces for Case~1 (continued).}
\end{figure}

    
    

\begin{figure}[H]
    \centering
    
    \begin{subfigure}{0.48\textwidth}
        \centering
        \includegraphics[width=\linewidth]{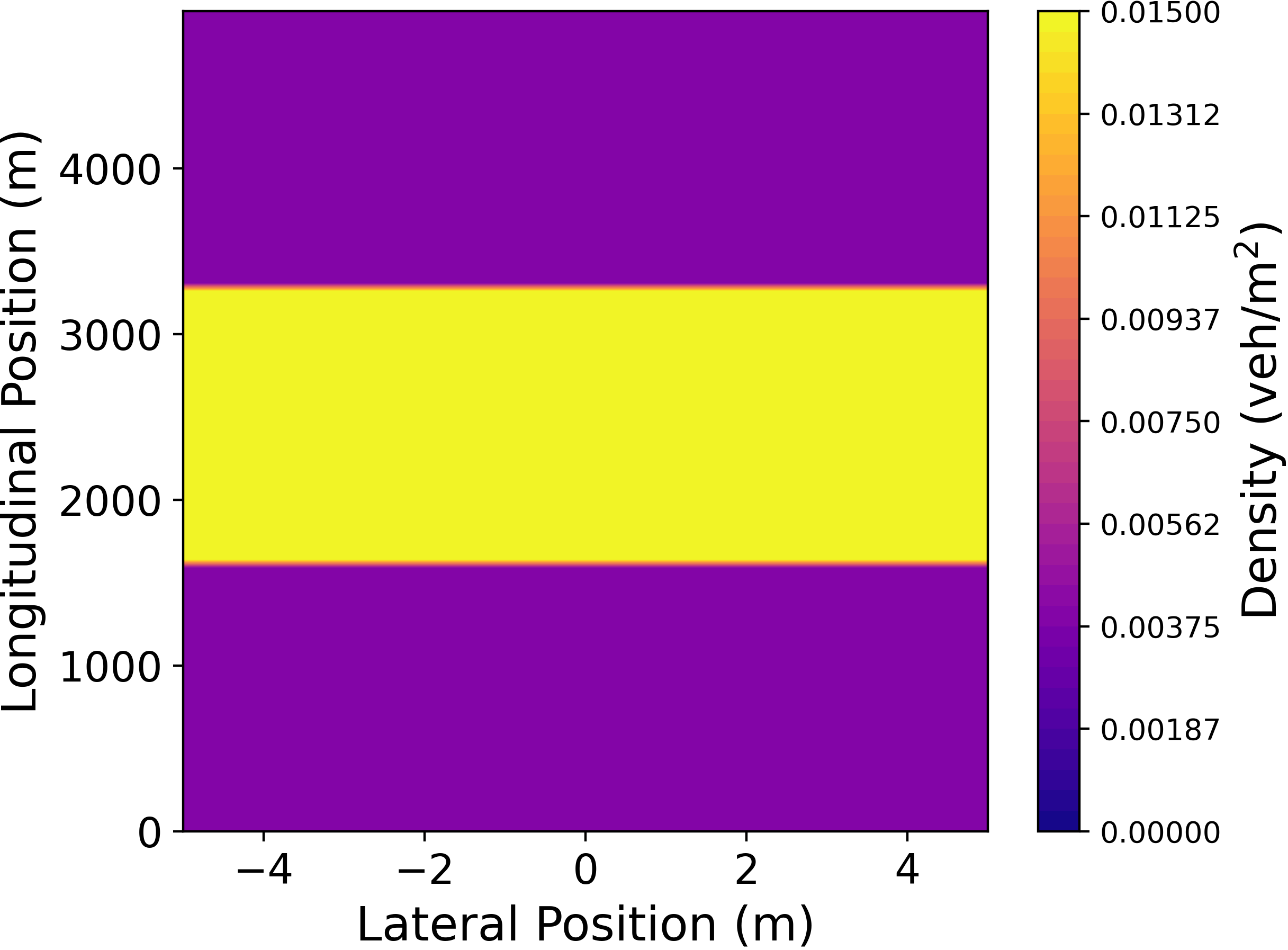}
        \caption{$t = 0$ s}
    \end{subfigure}
    \hspace{0.02\textwidth}
    \begin{subfigure}{0.48\textwidth}
        \centering
        \includegraphics[width=\linewidth]{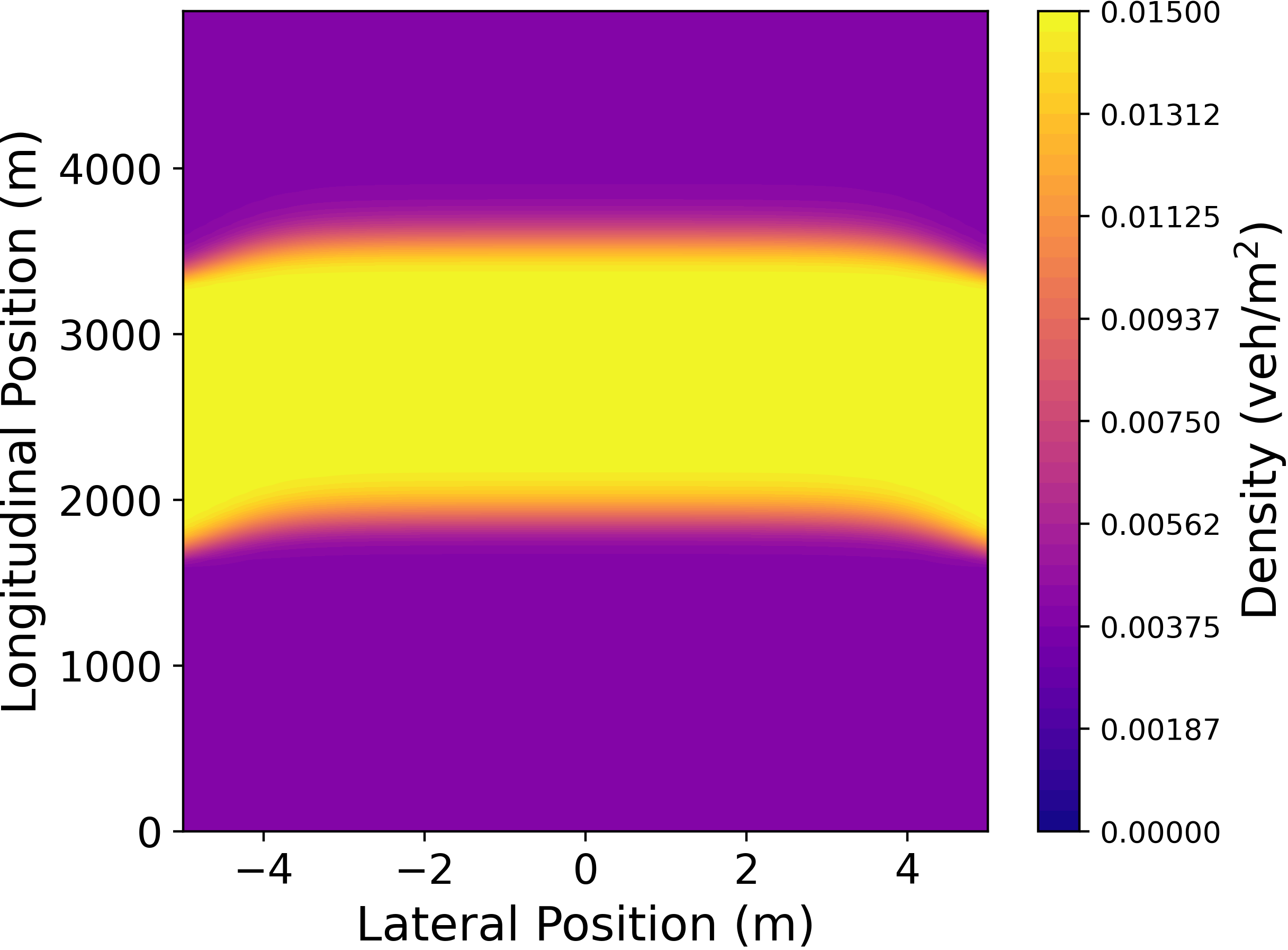}
        \caption{$t = 10$ s}
    \end{subfigure}
    
    \begin{subfigure}{0.48\textwidth}
        \centering
        \includegraphics[width=\linewidth]{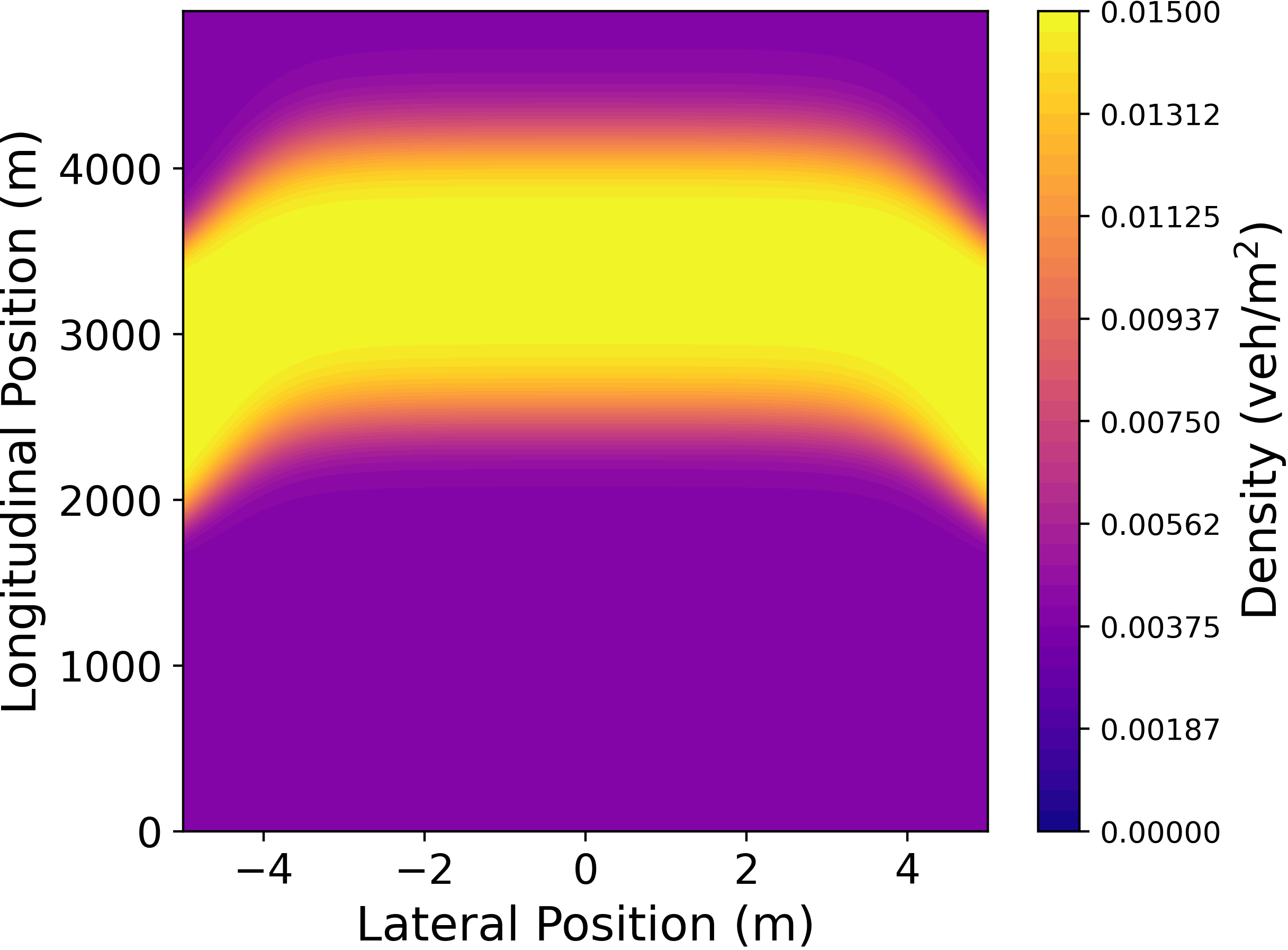}
        \caption{$t = 30$ s}
    \end{subfigure}
    \hspace{0.02\textwidth}
    \begin{subfigure}{0.48\textwidth}
        \centering
        \includegraphics[width=\linewidth]{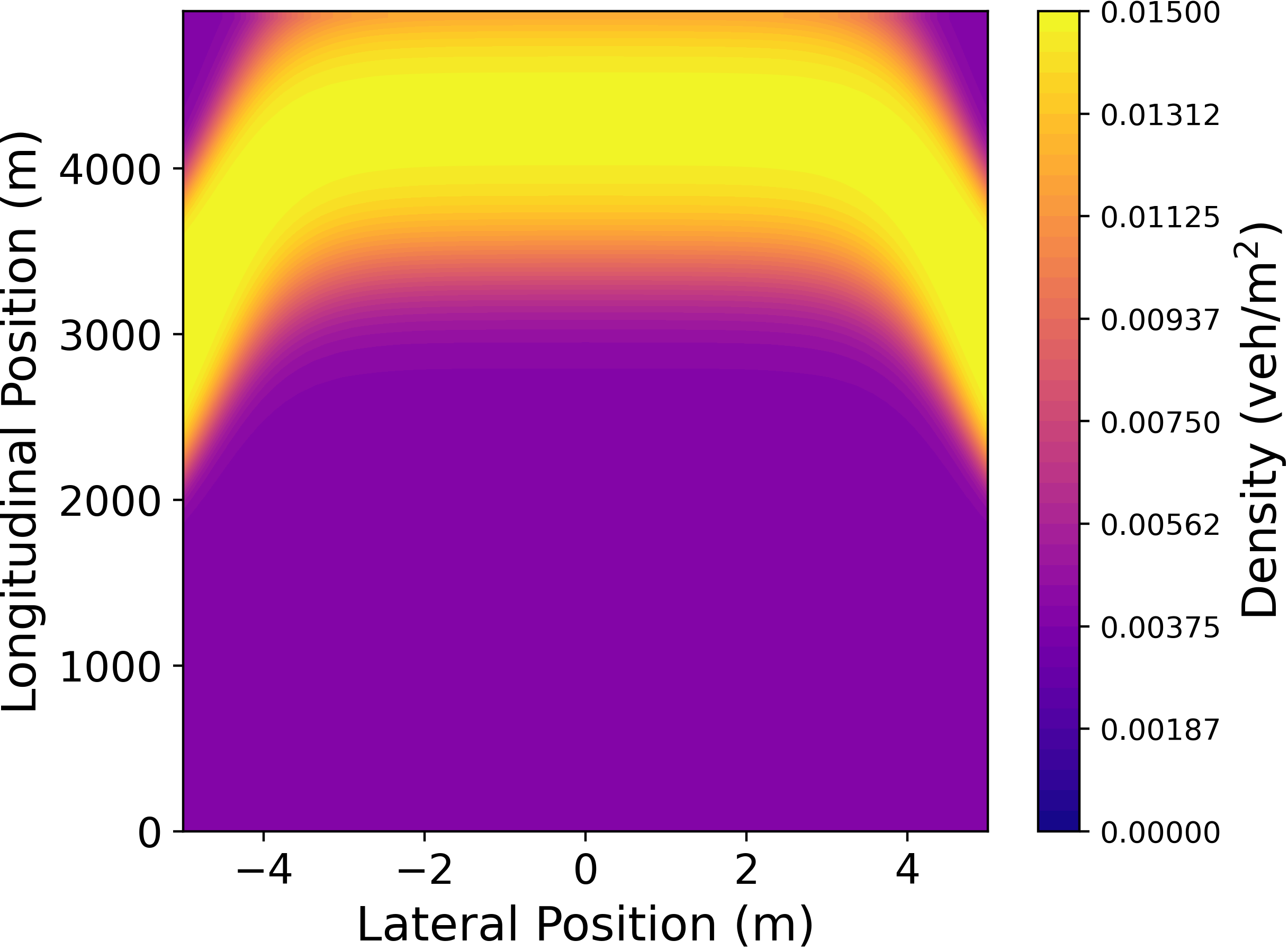}
        \caption{$t = 60$ s}
    \end{subfigure}
    
    \caption{Temporal evolution of traffic density with road-boundary forces for Case 1. \martinc{Here, a 2d contour plot is much clearer}}
    \label{fig:density1d_1_with_005}
\end{figure}

\subsubsection*{Case 2: Free-Flow and Congested Regions}

\noindent This case considers a more complex initial condition in which traffic alternates between free-flow and congested states. \martin{Free but high-flow traffic reaches a congested region with nearly standing traffic, while downstream the road is essentially empty. This can reflect a major bottleneck at the downstream end of the congestion that is removed at time $t=0$.} The initial density distribution is prescribed as
\begin{equation}
\rho(x\martin{,t}) =
\begin{cases}
0.015~\mathrm{veh/m^2}, & 0 \leq x < 750~\text{m}, \\
0.125~\mathrm{veh/m^2}, & 750~\text{m} \leq x < \martin{2\,500}~\text{m}, \\
0.001~\mathrm{veh/m^2}, & \martin{2\,500}~\text{m} \leq x < 5000~\text{m}
\end{cases}
\label{eq:case3_density}
\end{equation}
\martin{with flow initially at local steady state, $Q(x,0)=Q_e(\rho(x,0))$} \martinc{check}
The simulation reveals three distinct shock-wave like patterns. The first corresponds to the upstream-moving front between the free-flow and congested regions  near $x = 900~\text{m}$. \martin{The average propagation velocity is consistent with the LWR shockwave formula $c_{12}=(Q_1-Q_2)/\rho_1-\rho_2)$ which is expected by the Rankine-Hugoniot condition (\citealp{rankine1870xv,hugoniot1889propagation}).  However, due to the dynamic speed and numerical diffusion, the front is softened.}  The second pattern near $x = 3000~\text{m}$ represents the transition between  the congested and maximum-flow states \martin{at critical density $\rho_c=\unit[0.023]{veh/m^2}$ which is, again, expected, for theoretical reasons.} The third front separates the maximum-flow state from nearly empty traffic, and a rarefaction wave propagates downstream at a velocity with the desired free-flow speed $v_f$ as shown in \autoref{fig:density1d_2_lines_005} and \autoref{fig:speed1d_2}.
\begin{figure}[H]
\centering
    
    \begin{subfigure}{0.45\textwidth}
        \centering
        \includegraphics[width=\linewidth]{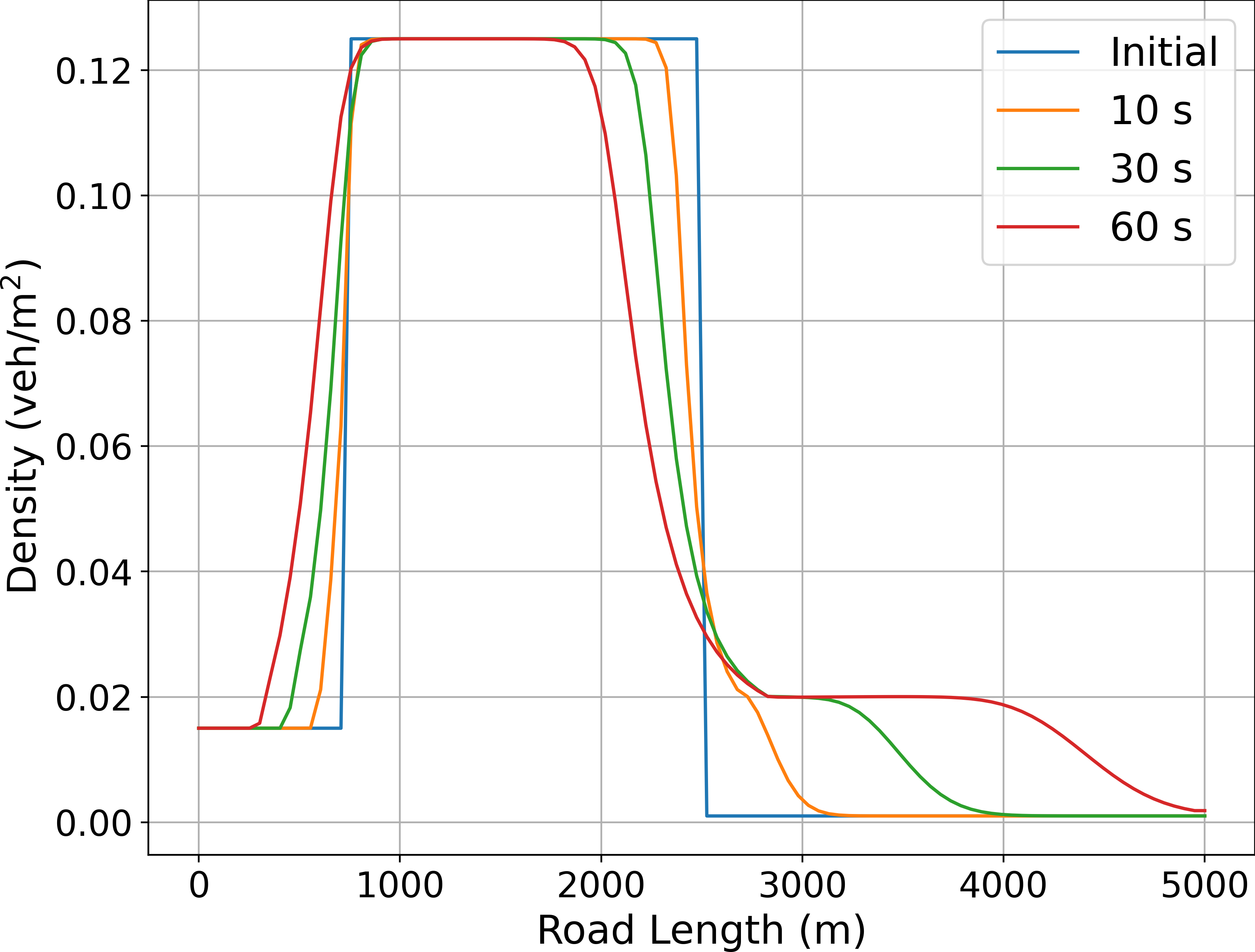}
        \caption{Without Boundary Forces}
    \end{subfigure}
    \hspace{0.04\textwidth}
    \begin{subfigure}{0.45\textwidth}
        \centering
        \includegraphics[width=\linewidth]{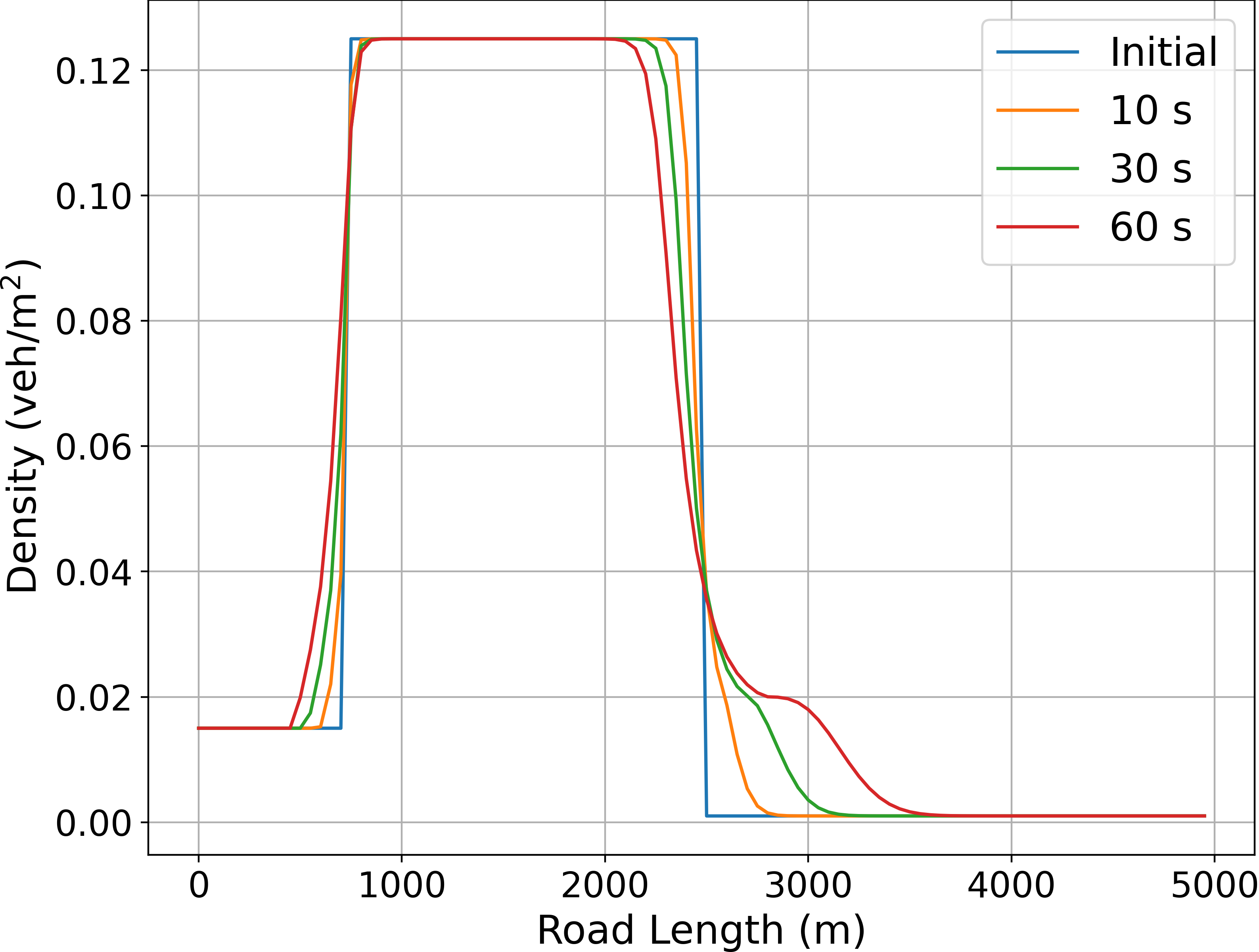}
        \caption{With Boundary Forces \martin{at $y=b$}\martinc{Check}}
    \end{subfigure}
    
    \caption{Comparison of longitudinal density for Case 2.}
    \label{fig:density1d_2_lines_005}
\end{figure}

\begin{figure}[!htbp]
    \centering
    
    \begin{subfigure}{0.48\textwidth}
        \centering
        \includegraphics[width=\textwidth]{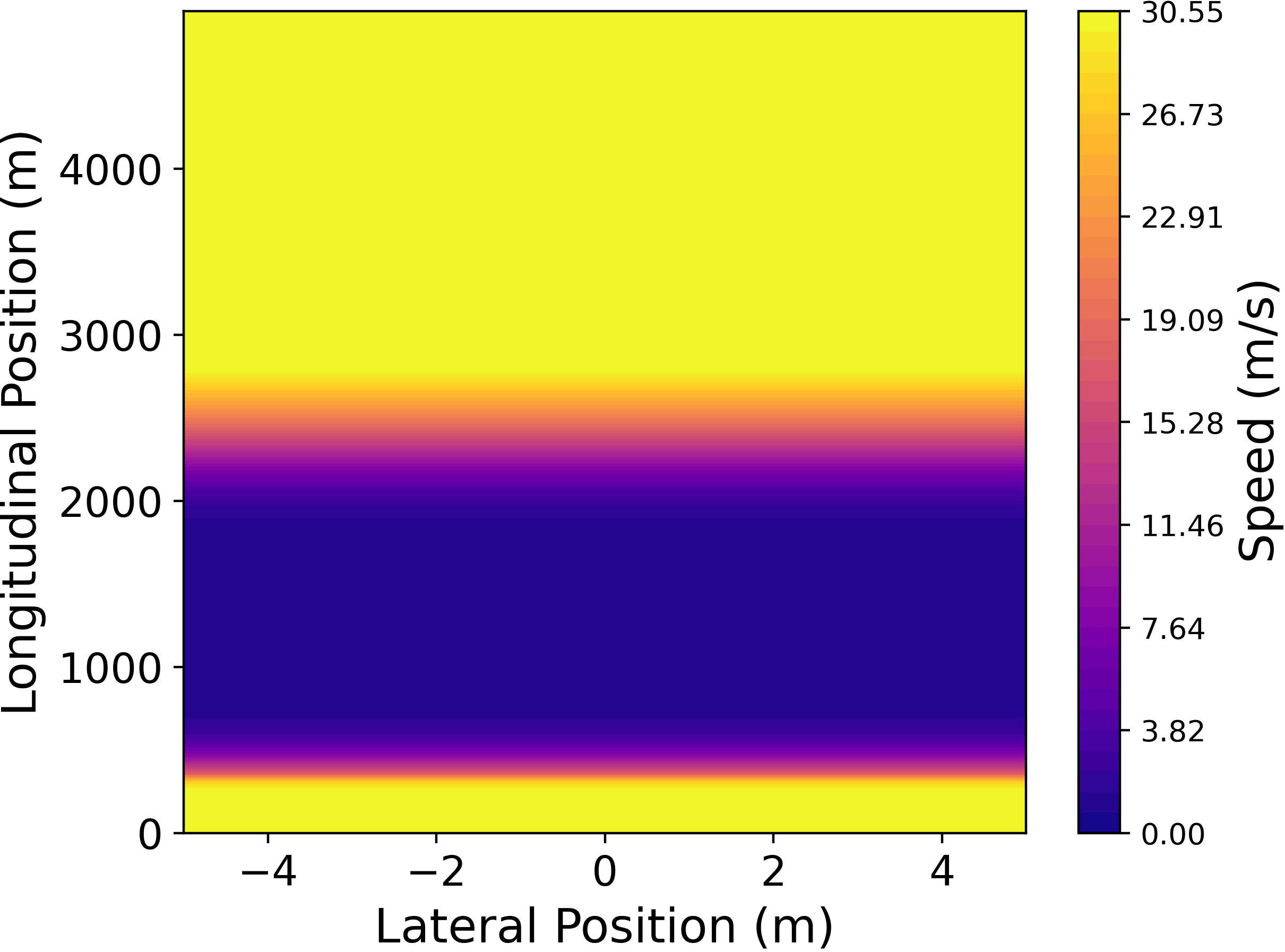}
        \caption{Without Boundary Forces}
    \end{subfigure}
    \hspace{0.02\textwidth}
    \begin{subfigure}{0.48\textwidth}
        \centering
        \includegraphics[width=\textwidth]{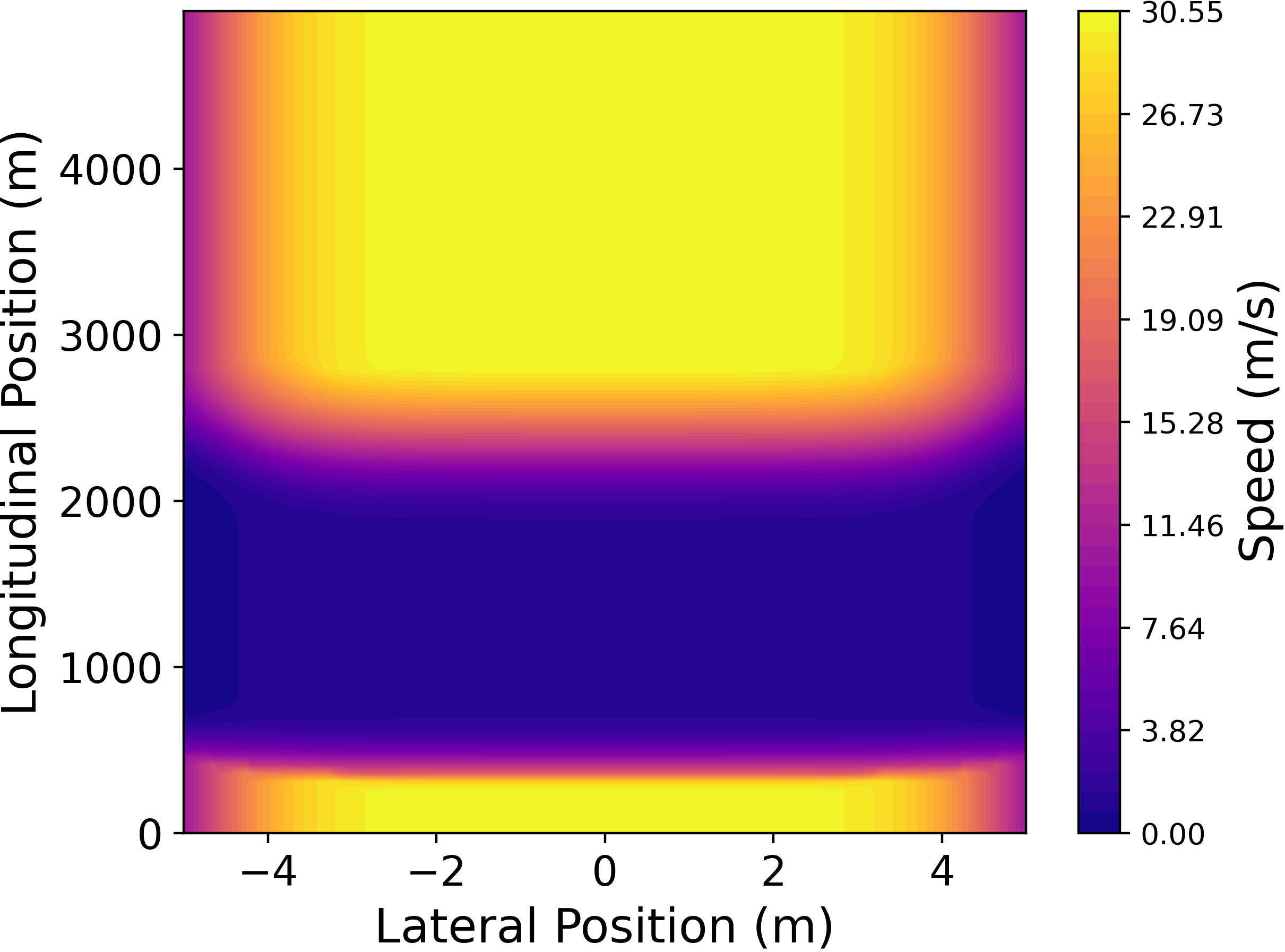}
        \caption{With Boundary Forces}
    \end{subfigure}
    
    \caption{Comparison of longitudinal speed profiles for Case 2 \martin{at $t=\unit[60]{s}$} \martinc{Check!}.}
    \label{fig:speed1d_2}
\end{figure}
\noindent The resulting traffic evolution gives rise to multiple moving interfaces whose propagation directions and relative speeds are governed by the local balance between self-driven acceleration and interaction-induced deceleration. The spatiotemporal density surfaces reveal the coexistence of upstream- and downstream-moving interfaces, reflecting the coupled evolution of free-flow and congested traffic states within a unified second-order framework as shown in \autoref{fig:density1d_2_lines_005}. 

%
    
    
\vspace{0.2cm}
\noindent
The influence of road-boundary forces is clearly visible
\autoref{fig:density1d_2_with}, where the density profiles exhibit distinct curvature across the road width. This lateral variation in density directly reflects the action of boundary-induced resistance, which selectively suppresses longitudinal motion near the edges while allowing relatively higher speeds in the central region. In this mixed traffic configuration, the upstream and downstream regions remain in a free-flow state, whereas the central region is congested. Despite this contrast, the density profiles remain smooth and well-resolved across the entire road width, with no evidence of spurious oscillations or numerical artifacts. In the congested central region, boundary effects continue to regulate vehicle motion, leading to a pronounced parabolic-like density structure, while in the free-flow upstream region the influence of boundary forces is weaker but still observable.

\begin{figure}[H]
\centering
    
    \begin{subfigure}{0.48\textwidth}
        \centering
        \includegraphics[width=\linewidth]{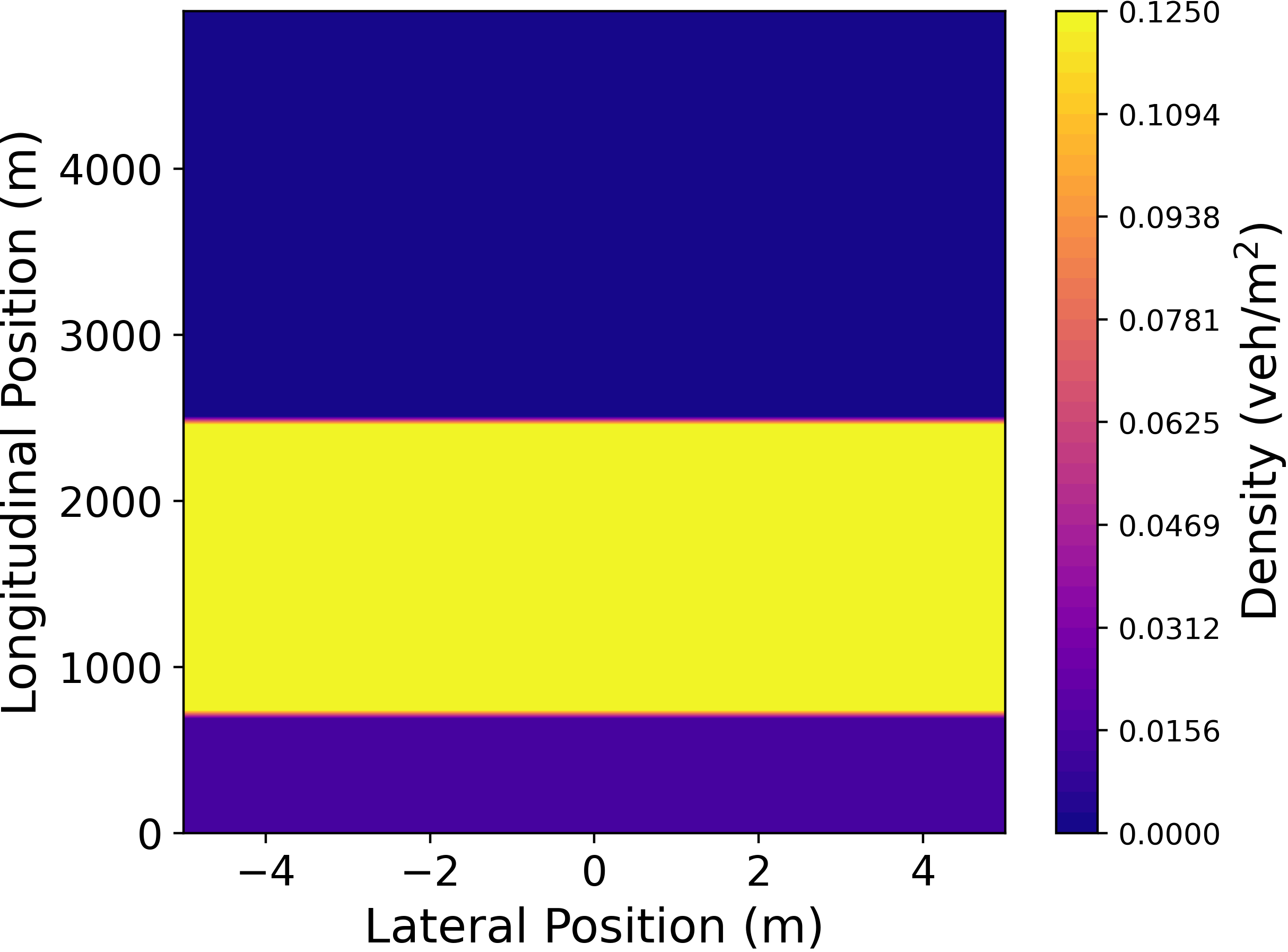}
        \caption{$t = 0$ s}
    \end{subfigure}
    \hspace{0.02\textwidth}
    \begin{subfigure}{0.48\textwidth}
        \centering
        \includegraphics[width=\linewidth]{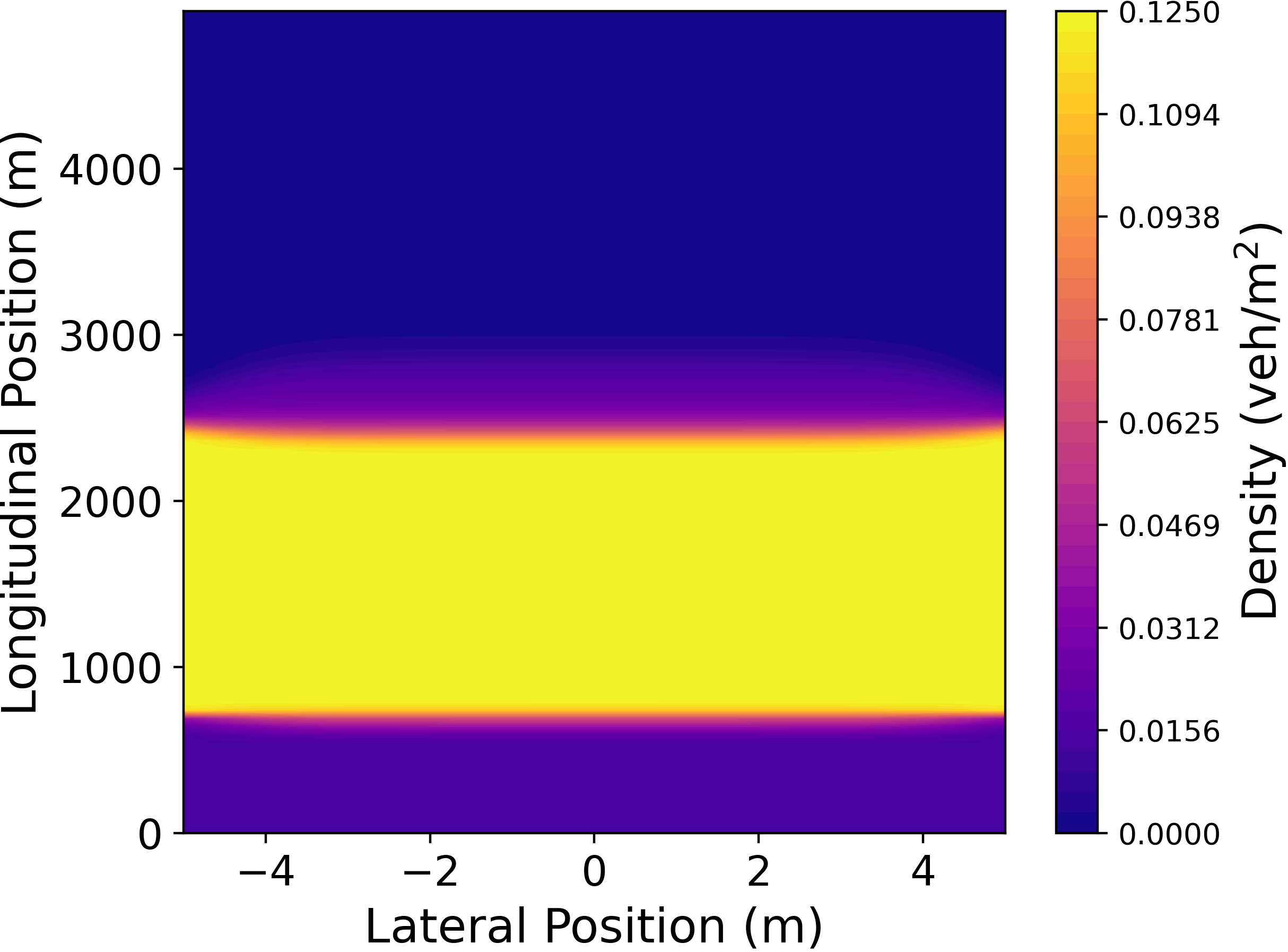}
        \caption{$t = 10$ s}
    \end{subfigure}
    
    \begin{subfigure}{0.48\textwidth}
        \centering
        \includegraphics[width=\linewidth]{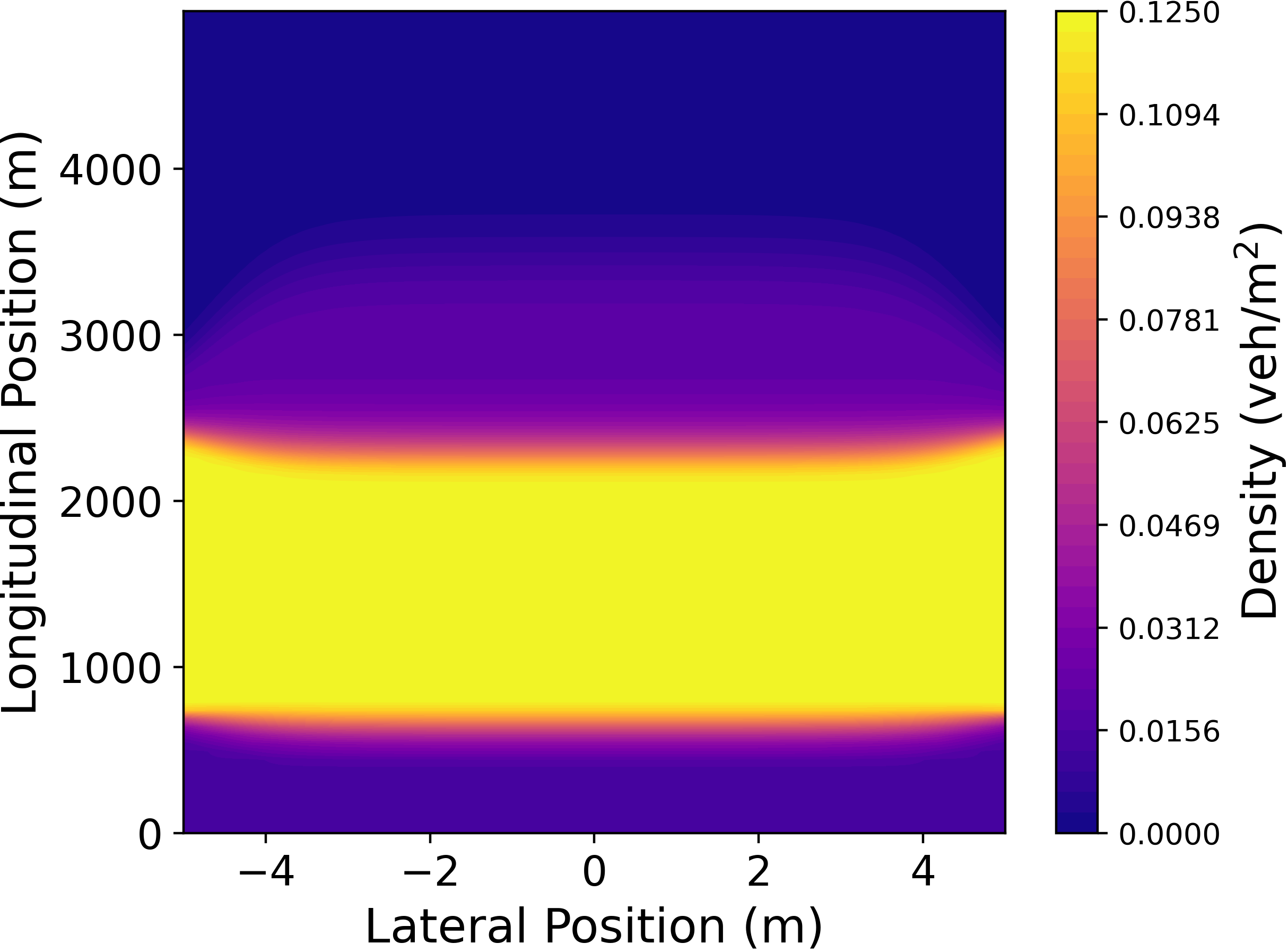}
        \caption{$t = 30$ s}
    \end{subfigure}
    \hspace{0.02\textwidth}
    \begin{subfigure}{0.48\textwidth}
        \centering
        \includegraphics[width=\linewidth]{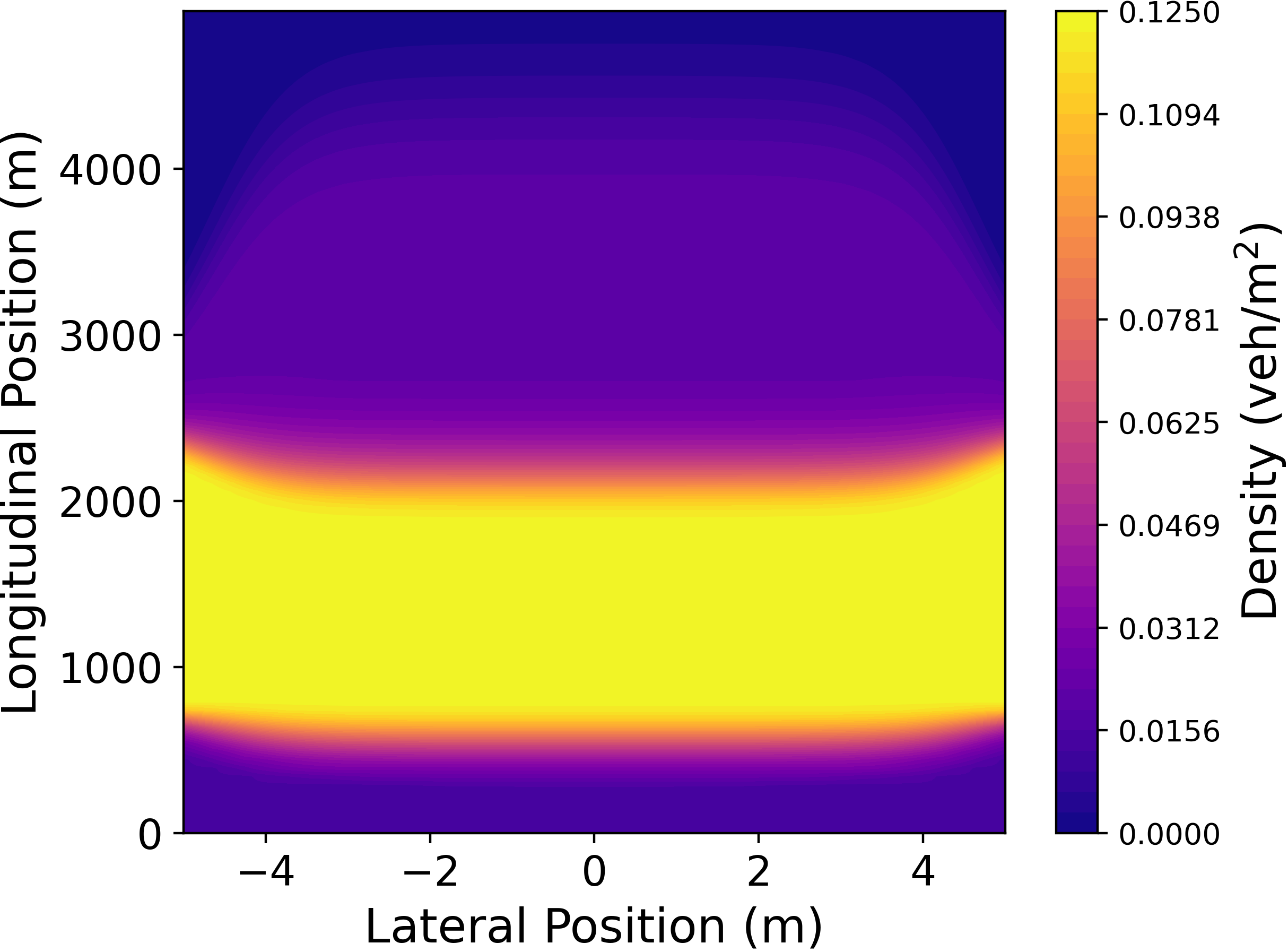}
        \caption{$t = 60$ s}
    \end{subfigure}
    
    \caption{Temporal evolution of traffic density with the road-boundary forces for Case 2}
    \label{fig:density1d_2_with}
\end{figure}

\noindent The propagation of all density interfaces is uniformly slowed due to the additional resistive effect introduced by the boundary force. Both the upstream-moving congestion front and the downstream-moving free-flow interface exhibit reduced propagation speeds. 

\subsubsection*{Case 3: Flow Instability}
\martinc{Thoroughly revised}
\noindent To examine the ability of the proposed longitudinal model to reproduce traffic instabilities, an initial density profile with a localized high-density perturbation in a homogeneous traffic stream is considered. The initial condition is defined as
\begin{equation}
\rho(x,0)=
\begin{cases}
0.025~\mathrm{veh/m^2}, & 0 \le x < 3300~\mathrm{m}, \\[4pt]
0.030~\mathrm{veh/m^2}, & 3300~\mathrm{m} \le x < 4150~\mathrm{m}, \\[4pt]
0.025~\mathrm{veh/m^2}, & 4150~\mathrm{m} \le x \le 5000~\mathrm{m}
\end{cases}
\label{eq:case3_ic}
\end{equation}
with the initial flow prescribed at local equilibrium, $Q_x(x,0)=Q_e(\rho(x,0))$.
\noindent
Following \citealp{treiber2025traffic}, the condition for linear stability of the longitudinal dynamics can be expressed as
\begin{equation}
\label{eq:long_stab_general}
\left(Q_e'(\rho)\right)^2 - i p_1\,Q_e'(\rho) - q_2 \le 0 ,
\end{equation}
where $i=\sqrt{-1}$, $Q_e(\rho)$ denotes the steady--state  flow--density relation. The coefficients $ip_1$ and $q_2$ originating from the linearization of the longitudinal acceleration function $\tilde{f}_x$  are are given by
\begin{equation}
\label{eq:p1_q2_gen}
\begin{aligned}
ip_1
&=-A_{V_x} - s_a\,A_{V_a} + 2 U_e,\\
q_2
&=U_e\!\left( A_{V_x} + s_a A_{V_a} \right)
-
\rho_e \!\left( A_{\rho_x} + s_a A_{\rho_a} \right) - U_e^{2},
\end{aligned}
\end{equation}
where $U_e$ and $\rho_e$ denote the steady--state speed and density, respectively, $s_a=u_xT$ is the anticipation distance, 
$A_{V_a}$ denotes the partial derivative of $\tilde{f}_x$, \autoref{eq:fx}, with respect to $u_x(x_a)$, $A_{V_x}$ denotes the partial derivative with respect to $\partial u_x/\partial x$, and $A_{\rho_a}$ and $A_{\rho_x}$ denote the corresponding partial derivatives with respect to density gradient and anticipated density, respectively.\footnote{Notice that the partial derivatives with respect to $u_x$ and $\rho$ itself are already absorbed in $Q'_e(\rho)$.} 
\vspace{0.2cm}

\noindent After a lengthy calculation, we get the (still general) stability criterion in terms of the derivatives of the acceleration function,
\begin{equation}
\label{eq:long_stab_A}
    (\rho_e U'_e)^2
    + \rho\left(U'_eA_{v_x}+s_aU'_eA_{v_a}
    +A_{\rho_x}+s_aA_{\rho_a}\right) \le 0
\end{equation}
Ignoring the boundary term of $\tilde{f}_x$, we obtain for our model
\begin{equation}
\label{eq:coefficients}
A_{V_x} = -\frac{\beta\sup{antic}}{\tau}\,\rho\,U_e'(\rho), 
\qquad
A_{V_a} = 0, 
\qquad
A_{\rho_x} = 0,
\qquad
A_{\rho_a} = \frac{U_e'(\rho)}{\tau} .
\end{equation}
\noindent
Substituting these coefficients into \autoref{eq:long_stab_A} results in the final condition
\begin{equation}
\rho_eU'_e\left(1-\frac{\beta\sup{anti}}{\tau}\right)+\frac{l_{\rm eff}+U_eT}{\tau} \ge 0
\end{equation}
For the triangular fundamental diagram \autoref{eq:triangularFD} with

\begin{equation}
U_e'(\rho) =
\begin{cases}
0, & \rho \le \rho_c, \\[6pt]
-\dfrac{Q_{\max}}{\rho^{2}}
\left(1 - \dfrac{w}{v_f}\right),
& \rho > \rho_c,
\end{cases}
\label{eq:U_e'}
\end{equation}
For the free-flow regime, we obtain unconditional stability while, for the congested regime, the flow stability criterion is given by
\begin{equation}
-\frac{Q_{\rm max}}{\rho}\left(1-\frac{\beta\sup{anti}}{\tau}\right)\left(1 - \frac{w}{v_f}\right)+\frac{l_{\rm eff}+U_eT}{\tau} \ge 0
\label{eq:stab_cond}
\end{equation}
This is an explicit stability condition that depends on the equilibrium speed--density relationship, the anticipation parameter $\beta\sup{anti}$, the longitudinal speed adaptation time $\tau$, and the anticipation distance $s_a=l_{\rm eff}+u_xT$. This condition provides a direct criterion for identifying stable and unstable traffic regimes in the longitudinal dynamics. To deliberately place the system in an unstable regime, the relative-speed anticipation parameter is reduced from $\beta\sup{anti}=\unit[0.65]{s}$ to 
\unit[0.6]{s} while the speed-adaptation time is increased from $\tau=0.65$ to $\tau=3$. Since, for the simulation of \autoref{fig:density1d_instab_lines} (baseline density $\rho=\unit[0.025]{veh/m^2}$), the critical value for the speed relaxation time $\tau_c=\unit[1.6]{s}$, this simulation runs in the flow unstable range $\tau>\tau_c$.  The resulting temporal evolution of the density profile is shown in \autoref{fig:density1d_instab_lines}. 
\begin{figure}[H]
        \centering
        \includegraphics[width=0.51\linewidth]{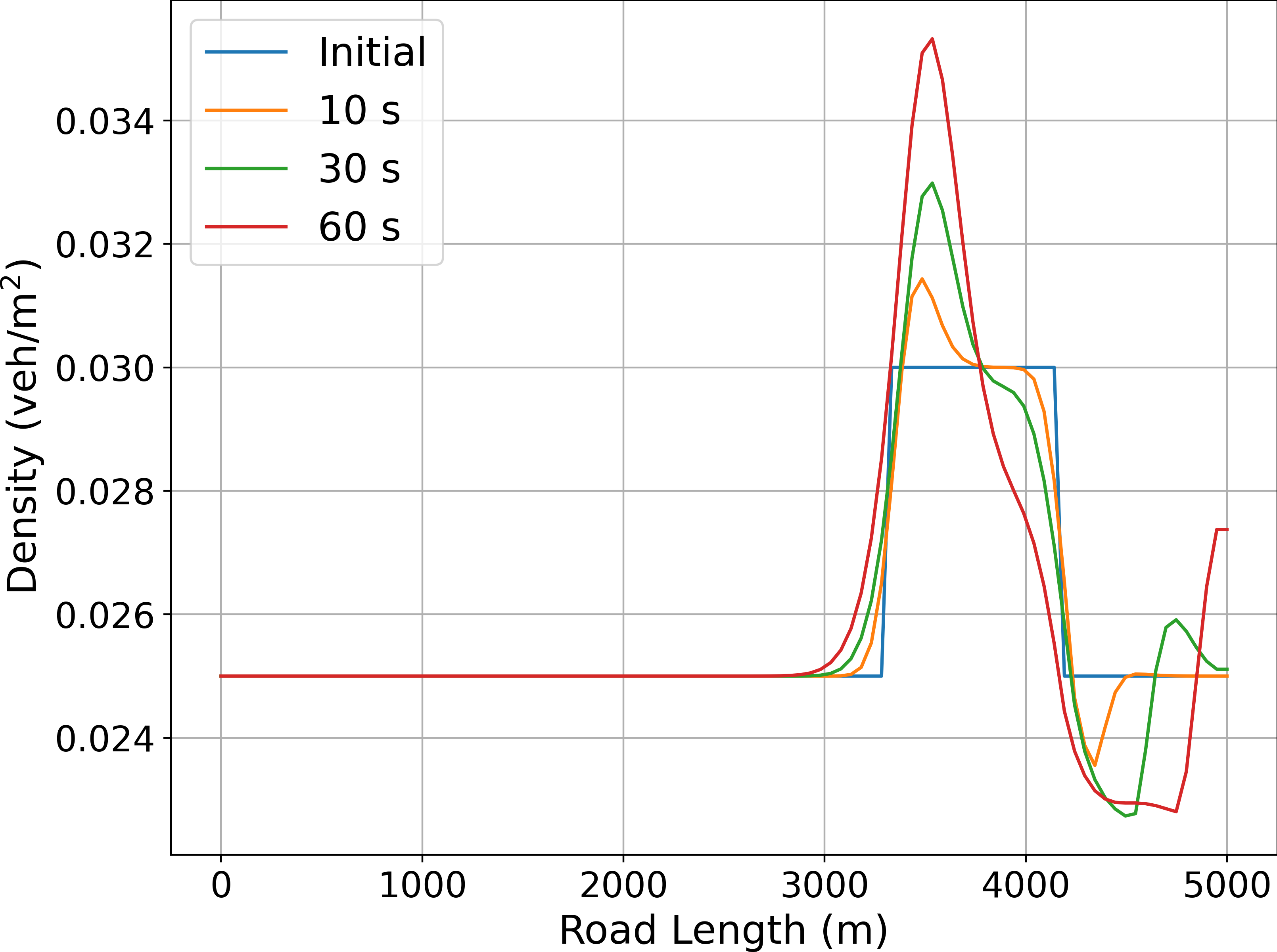}
        \caption{Traffic Instability \martin{without boundary effects}}
        \label{fig:density1d_instab_lines}
\end{figure}
\noindent The initial perturbation remains localized at early times, small-amplitude fluctuations progressively amplify and evolve into pronounced nonlinear waves. The wave amplitude increases over time and propagates upstream relative to the traffic flow, a characteristic of traffic instability and stop-and-go wave formation. These results confirm that the proposed longitudinal second-order macroscopic model can reproduce emerging and growing traffic waves when operated in the unstable regime, in agreement with theoretical stability predictions and established observations in macroscopic traffic flow theory.

\subsection{Lateral Dynamics}
\martinc{Needs to be completely redone since the model has been found to be inconsistent laterally and has now been corrected (no changes in the longitudinal dimension apart from parameter changes). Particularly, I did an analytic derivation of the longitudinal-hpomogeneous stationary state (includes this full subsection) on which the numerical solution can be tested, as in the longitudinal case. I attached the latex file and some images by mail (May 14, 2026).}

\noindent The lateral dynamics of the model are analyzed under the assumption of stationary and longitudinal homogeneity, i.e., $\frac{\partial}{\partial t}=0$ and 
$\frac{\partial}{\partial x}=0$ respectively. From \autoref{full_continuty}, it follows that $Q_y=\text{const}$. By virtue of the non-penetrable boundary conditions, we obtain
\begin{equation}
\rho=\rho(y), \qquad
Q_x=Q_x(y), \qquad
Q_y=0.    
\end{equation}

\noindent Using the momentum equations, \autoref{eq:fx} and \autoref{eq:fy}, we obtain for
\(\rho>0\) the conditions
\begin{equation}
 \tilde{f}_x
=
\tilde{f}_x^{\mathrm{OVM}}
+
\tilde{f}_x^{\mathrm{antic}}
+
\tilde{f}_x^{\mathrm{road}}
=0,   
\label{long_stationary}
\end{equation}

\noindent and
\begin{equation}
f_y
=
f_y^{\mathrm{traffic}}
+
f_y^{\mathrm{road}}
=0.    
\label{lat_stationary}
\end{equation}
 
\noindent Using \autoref{long_stationary} in \autoref{eq:fx} for the longitudinal momentum balance, a static relation between the longitudinal velocity component $u_x = Q_x/\rho$ as follows (notice $\rho(x_a,y,t) = \rho(x,y,t) = \rho(y)$ and that the anticipation term vanishes). 
\begin{equation}
   0 = \frac{1}{\tau}\left(
    U_e (\rho(x_a, y,t)) - u_x 
    -0
    -\beta_b\sup{long} u_x\, g(y)
    \right)
    \label{long_momentum}
\end{equation}
Hence, we obtain the steady-state longitudinal speed as a function of the optimal-speed $U_e(\rho(y))$:
\begin{equation}
    u_x = u_x(\rho(y),y) = \frac{U_e(\rho(y))}{1 + \beta_b\sup{long}g(y)}
    \label{longitudinally homogenous}
\end{equation}
where $(U_e(\rho(y)))$ denotes the optimal speed corresponding to the local density and $(g(y))$ represents the symmetric road-boundary function. This expression indicates that the actual longitudinal speed is reduced from its optimal value due to the influence of the road boundaries. Since $(g(y))$ increases as the road edges are approached, the denominator becomes larger near the boundaries, leading to a reduction in longitudinal speed. Conversely, near the road centre, where the influence of the boundaries is weaker, $(g(y))$ attains smaller values, and the vehicle speed approaches the optimal speed $U_e$. Therefore, the steady-state longitudinal velocity profile is governed by the combined effects of local traffic density through $(U_e(\rho)$ and boundary-induced deceleration through $(g(y))$. This relationship predicts lower speeds near the road edges and comparatively higher speeds toward the centre of the road, as shown in the \autoref{fig:steady-state longitudinal}.

\noindent Using \autoref{lat_stationary} in \autoref{eq:fy} for the transversal momentum balance, and for the longitudinally-homogenous steady state ($u_y = 0, u_y \sup{self} = 0$) we have
\begin{equation}
    0 = \frac{1}{\tau_y}\left(
    0
    +\beta\sup{lat} u_x \frac{\partial U_e(\rho)}{\partial y} - 0
    -
    \beta_b\sup{lat} u_x g'(y)\right)
        \label{lat_momentum}
\end{equation}

\noindent For a non-zero longitudinal speed $u_x$, the speed drops out and only a dynamical equation for the optimal speed function as a function of $y$ remains which can be solved analytically:

\begin{equation}
U_e(y)
=
U_e(0)
+
\alpha\left[g(y)-g(0)\right],
\qquad
\alpha
=
\frac{\beta_b^{\mathrm{lat}}}
     {\beta_b^{\mathrm{long}}}.
     \label{dynamical_speed}
\end{equation} 
where $g(y)$ is given from \autoref{eq:fy}. Once $U_e(y)$ is given, the density can be calculated by the inverse function
\begin{equation}
    \rho(y) = \rho_e (U_e(y))
    \label{inverse_rho}
\end{equation}
From the triangular fundamental diagram, we have the inverse speed-density relation given by \autoref{eq:rho_triang}.
\begin{equation}
\rho_e^{\mathrm{triang}}(U_e)
=
\begin{cases}
\dfrac{Q_{\max}\left(1-\dfrac{w}{v_f}\right)}
      {U_e-w},
& U_e \leq v_f, \\[10pt]
0,
& \text{otherwise}.
\end{cases}
\label{eq:rho_triang}
\end{equation}
In the free-flow regime of the triangular FD, the model yields somewhat unexpected results, since lateral density gradients have no effect in this regime (which, by definition, is completely interaction-free). So, in the presence of lateral repulsion effects, one always has a density very slightly above the critical density or
a corresponding optimal speed slightly below $v_f$
(\autoref{inverse_rho}). Near the boundaries, \autoref{dynamical_speed} will lead to $U_e(y)$ crossing the line $(v_f=\text{const.})$ corresponding to a sudden drop of the density to zero. The $y$ coordinates where this happens depend on the controlling 1d density $\rho_{\text{1d}}=\int_{-b}^b \rho(x,y) dy$. If it is below the critical 1d density, the repulsive boundary force has no counteracting gradient force in the free regime and will drive traffic towards the center until densities at least slightly above the critical density are reached. 
\vspace{0.2cm}

\noindent \martin{When replacing the triangular fundamental with a continuous fundamental diagram reflecting small interactions also in the "free" density regime $\rho<\rho_c$, the sharp transition from an empty region near the boundaries to densities above the critical densities near the center  for 1d densities $\rho_{\text{1d}}<\rho_{\text{1dc}}$ is replaced by a more natural smooth lateral profile. \autoref{fig:steady-state longitudinal-IDM} shows this for a fundamental diagram derived from the IDM steady-state gaps for $W_y=\unit[1]{m}$ wide stripes and $T_{\text{IDM}}=\unit[1.2]{s}$}
\martin{\begin{equation}
    \rho_e\sup{IDM}(U_e)=\frac{1}{W_y}\left(\frac{1}{\rho_{\text{max}}}+\frac{U_eT_{\text{IDM}}}{\sqrt{1-\left(\frac{U_e}{v_f}\right)^4}}\right)^{-1}
    \label{eq:funddiaIDM}
\end{equation}
}

   \begin{figure}[H]
    \centering
    
    \begin{subfigure}{0.47\textwidth}
        \centering
        \includegraphics[width=\textwidth]{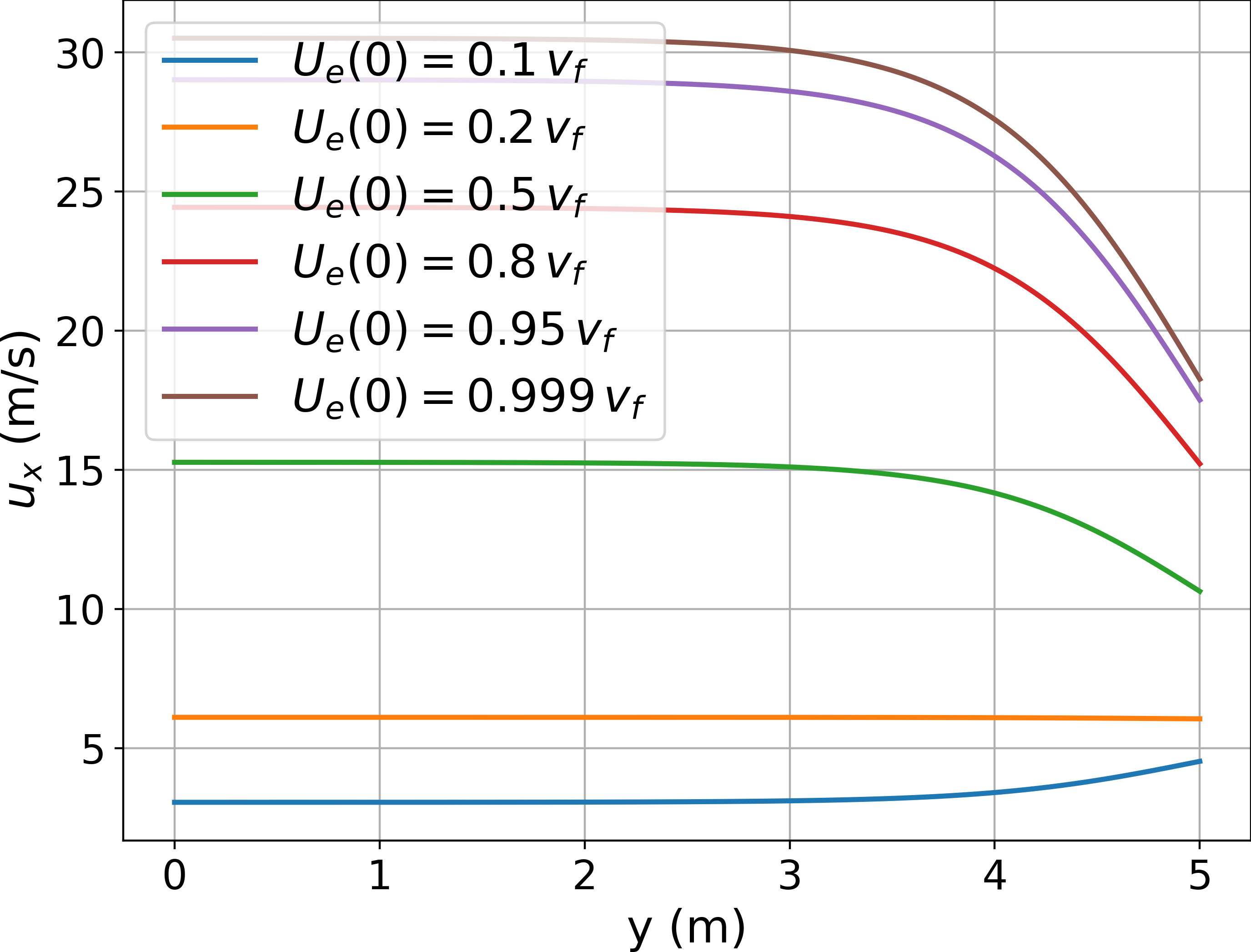}
        \caption{Steady-state speed $(u_x)$}
    \end{subfigure}
    \hspace{0.02\textwidth}
    \begin{subfigure}{0.48\textwidth}
        \centering
        \includegraphics[width=\textwidth]{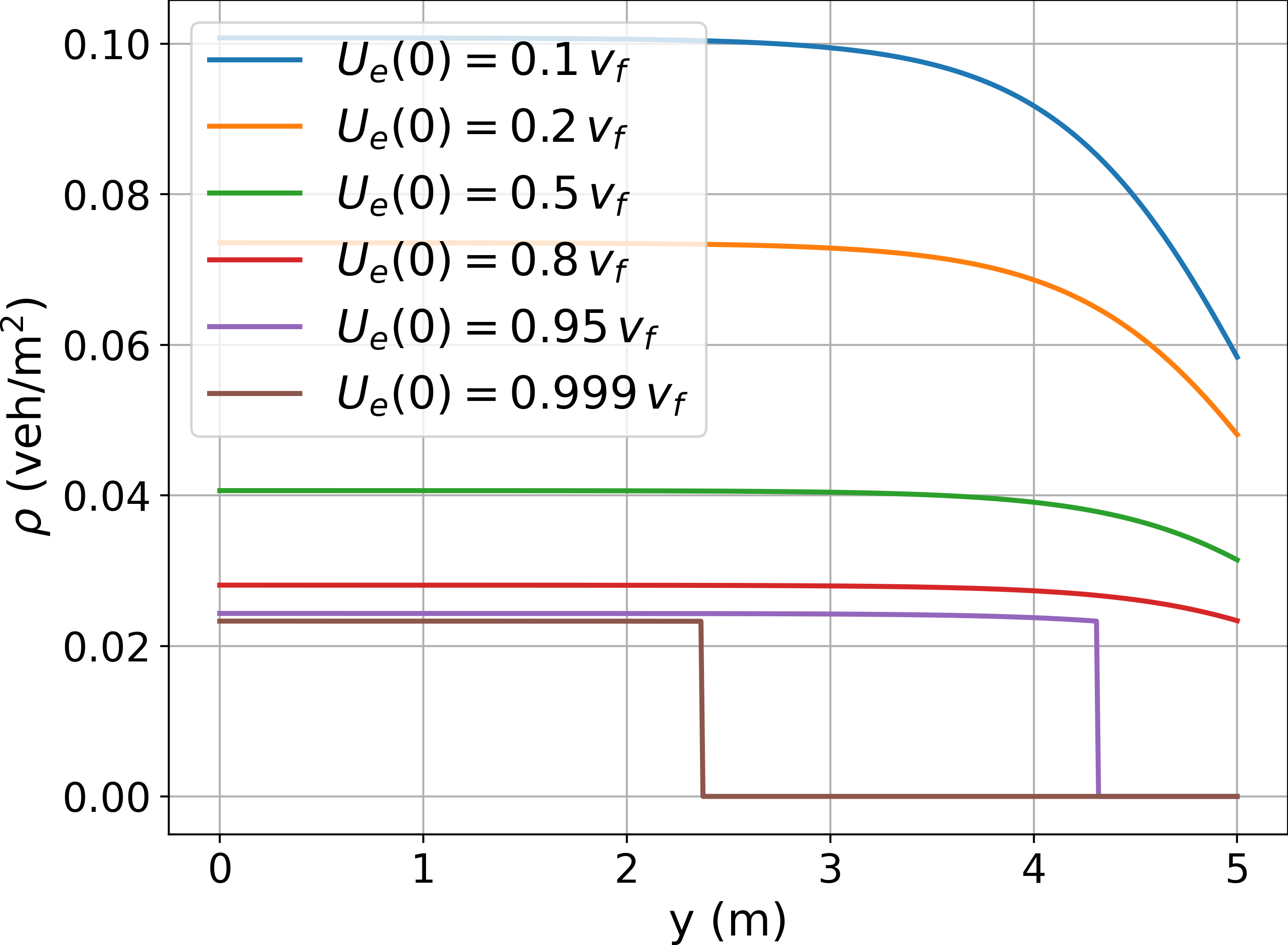}
        \caption{Density from inverse triangular speed-density relation}
    \end{subfigure}
    
    \caption{\martin{Stationary transversal profile for the triangular fundamental diagram \autoref{eq:triangularFD} and a longitudinal-homogeneous steady state characterized by the longitudinal speed $U_e(0)$ at the center}}
    \label{fig:steady-state longitudinal}
\end{figure}

 \begin{figure}[H]
     \centering
    
    \begin{subfigure}{0.47\textwidth}
        \centering
        \includegraphics[width=\textwidth]{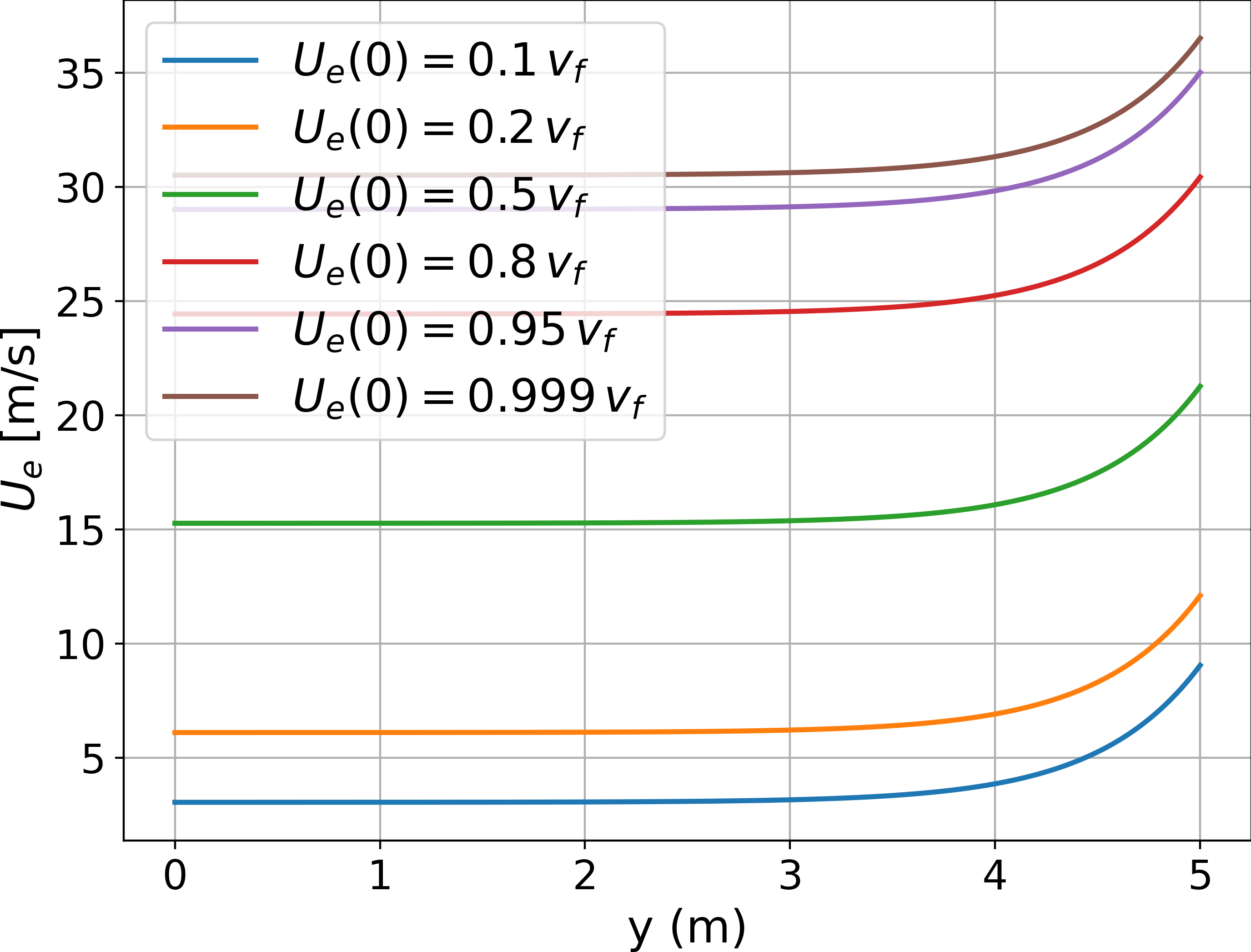}
        \caption{Equilibrium speed $(U_e)$}
    \end{subfigure}
    \hspace{0.02\textwidth}
    \begin{subfigure}{0.47\textwidth}
        \centering
        \includegraphics[width=\textwidth]{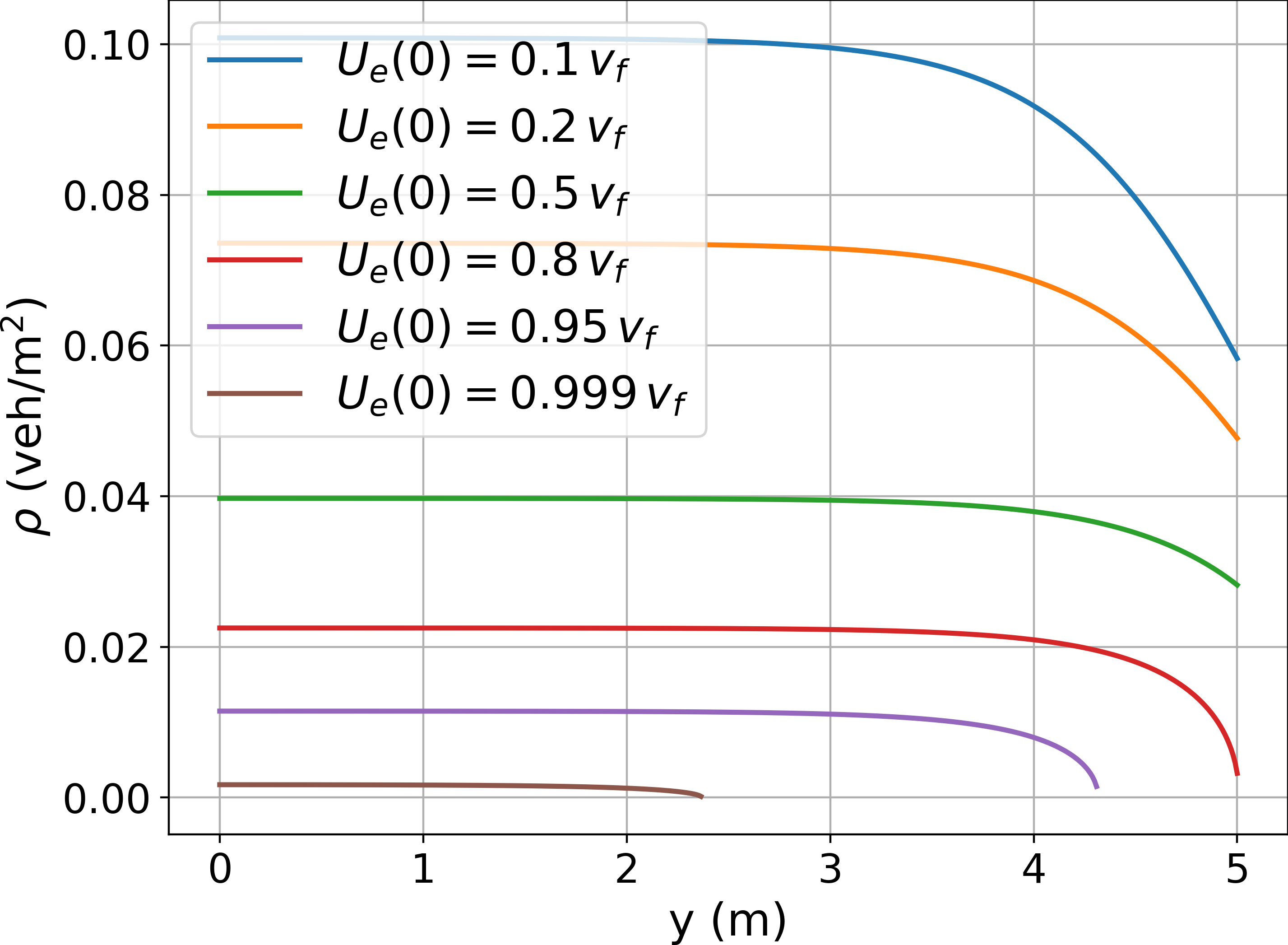}
        \caption{Density from inverse triangular speed-density relation}
    \end{subfigure}
    
 \caption{\martin{Stationary transversal profile for the IDM-like fundamental diagram \autoref{eq:funddiaIDM} and a longitudinal-homogeneous steady state characterized by the longitudinal speed $U_e(0)$ at the center} \martinc{Use my pdf and gnuplot files from my mail of 14.5.2026}}
    \label{fig:steady-state longitudinal-IDM}
\end{figure}

\noindent To verify the implementation of the Lax--Friedrichs scheme for the lateral direction, the numerical solution is compared with the analytical solution. \autoref{fig:rho_analytical_comparison} presents a comparison between the numerical and analytical steady-state, longitudinally homogeneous density profiles for three representative centerline equilibrium speeds, namely, $U_e(0)=0.1v_f$, $0.5v_f$, and $0.8v_f$. The simulations are initialized with different density profiles while ensuring that the total density in each initial profile is equal to the total density of the corresponding analytical solution. 
\begin{figure}[H]
\centering

\begin{subfigure}{0.48\linewidth}
    \centering
    \includegraphics[width=\linewidth]{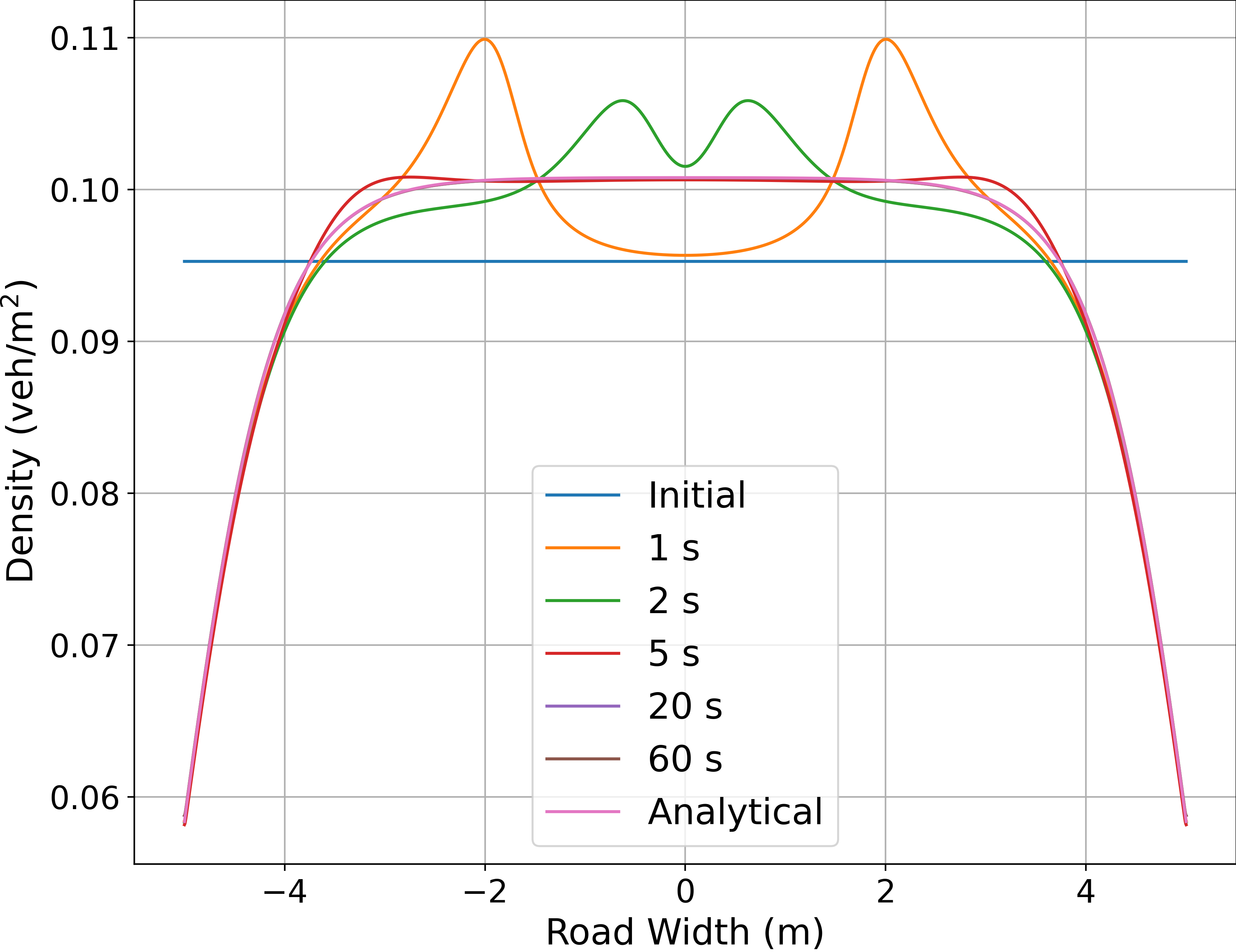}
    \caption{$U_e(0)=0.1\,v_f$}
    \label{fig:rho_analytical_01vf}
\end{subfigure}
\hspace{0.2cm}
\begin{subfigure}{0.48\linewidth}
    \centering
    \includegraphics[width=\linewidth]{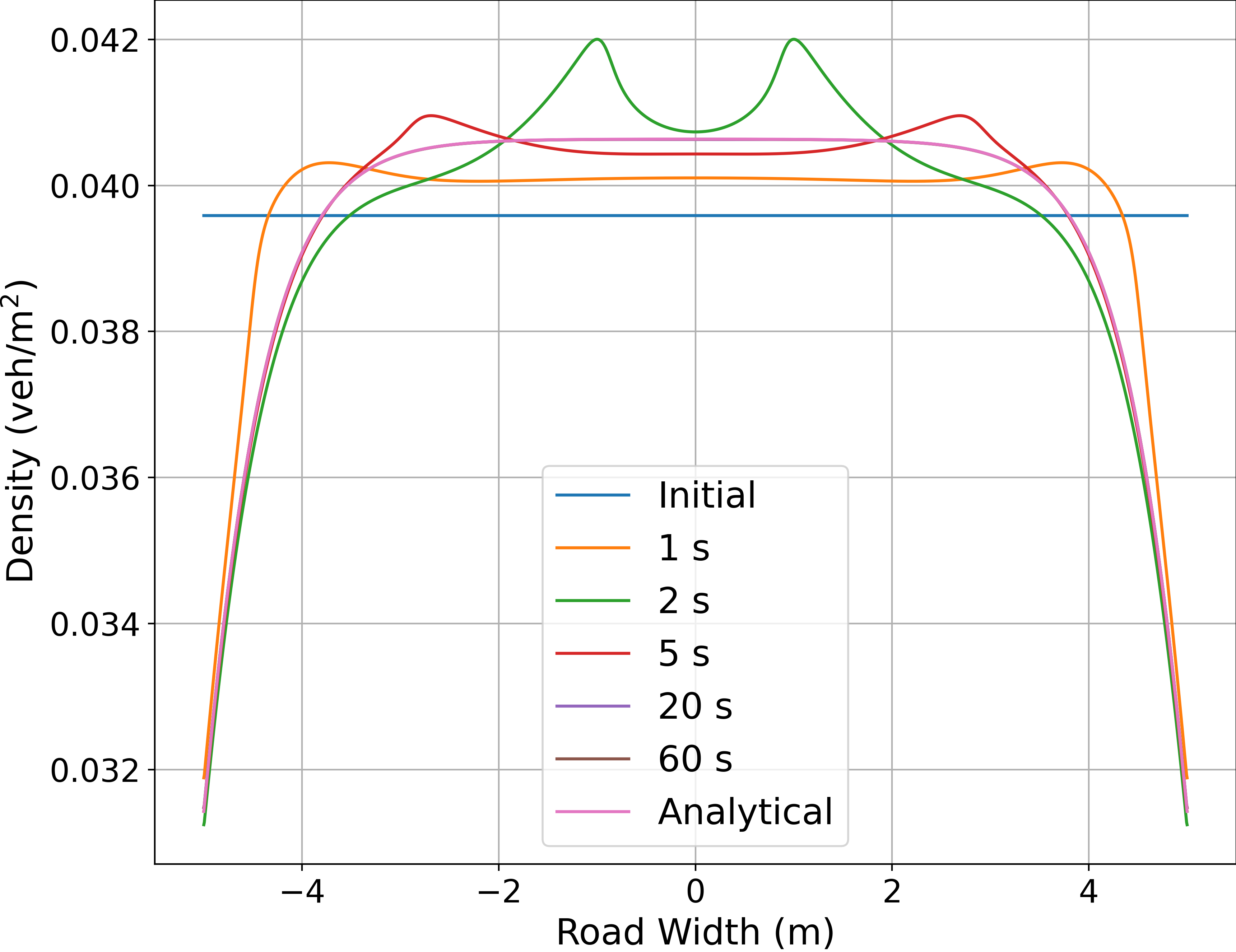}
    \caption{$U_e(0)=0.5\,v_f$}
    \label{fig:rho_analytical_02vf}
\end{subfigure}

\vspace{0.2cm}

\begin{subfigure}{0.48\linewidth}
    \centering
    \includegraphics[width=\linewidth]{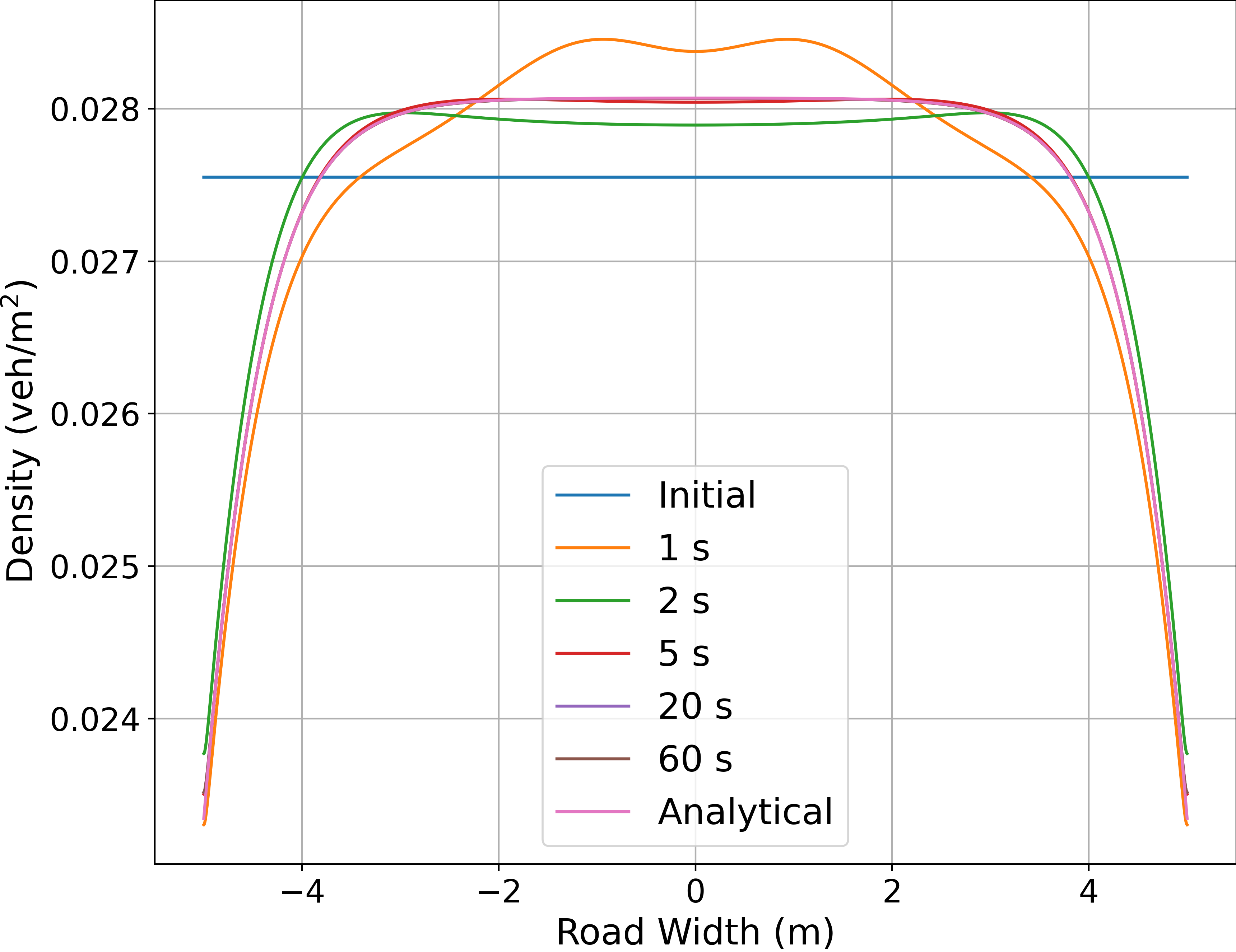}
    \caption{$U_e(0)=0.8\,v_f$}
    \label{fig:rho_analytical_05vf}
\end{subfigure}

\caption{Comparison of the numerical and analytical steady-state density profiles for different centerline equilibrium speeds, $U_e(0)$. \martinc{Since I cannot see any differences between 30s, 60s, and analytical, it would be better to show the profile at smaller time instances, e.g., 1 s, 2 s, 5 s, and 20 s. This also applies to Fig. 13(b)}}
\label{fig:rho_analytical_comparison}

\end{figure}

\noindent As the simulation progresses, the lateral interaction and road-boundary forces redistribute the traffic density across the road width. After a sufficient period, the numerical solution converges to the analytical solution corresponding to the steady-state and longitudinally homogeneous condition for all three cases. This agreement demonstrates the consistency and correctness of the implemented Lax--Friedrichs numerical scheme. subsequently, the lateral dynamics of the model are explored through three distinct cases, with initial conditions adopted from \citealp{agrawal2023two}. These cases are designed to provide a comprehensive understanding of lateral interactions, including lane-changing behavior, vehicle distribution across the road section, and the role of lateral driving forces. The three cases considered for the analysis of lateral dynamics are described as follows:

\subsubsection*{Case 1: Concave Lateral Density}

\noindent \martinc{I would skip \emph{all} the 3d plots of Cases 1 and 2 because it only makes the paper unnecessarily long and data-intensive without giving new information. Just Fig. 13(b) and 15(b) which both can become the new Fig. 13 are enough}

\noindent In this scenario, the initial lateral density distribution is concave, with the maximum density concentrated at the center of the road. The mathematical representation of the initial condition is given by
\begin{equation}
\rho(x,y) = \rho_{\text{max}}
\left(1 - \left(\frac{y}{b}\right)^2\right),
\label{eq:case_lat_concave}
\end{equation}
where $\rho_{\text{max}}$ denotes the maximum density (complete standstill) and $b$ represents half the road width. The evolution of lateral density over space and time is illustrated in \autoref{Case 1}. \autoref{Case 1} illustrates the influence of road-boundary forces on the lateral redistribution of traffic density, highlighting the differences in the density profiles with boundary effects.
\vspace{0.2cm}

\noindent The early evolution during \martin{the first seconds} is characterized by rapid lateral dispersion, \martin{driven by a desire of the drivers to occupy the initially nearly empty space away from the road center.} During this phase, vehicle--vehicle interactions remain relatively weak, allowing vehicles to redistribute efficiently across the road width. \martin{Afterwards}, the rate of lateral movement decreases as the available lateral space becomes progressively occupied and interaction effects intensify, limiting further dispersion. \martin{For $t>\unit[5]{s}$}, only minimal lateral rearrangement is observed, indicating that the system has approached a stable lateral configuration. The resulting density profile is not perfectly uniform, with a persistent concentration near the road center due to the repulsive road-boundary forces that discourage vehicle accumulation near the road edges. 
This lateral redistribution of density is accompanied by a competing effect in the longitudinal momentum balance. The same lateral equilibrium profile that redistributes density away from the road boundaries also introduces a direct longitudinal braking term near the edges through the boundary interaction force (\autoref{longitudinally homogenous}), independent of the local density. Since the longitudinal boundary-braking coefficient is larger than the lateral-repulsion coefficient in this configuration ($\beta_b\sup{long}$ > $\beta_b\sup{lat}$), the braking effect dominates the density-induced acceleration expected from the reduced edge density. Consequently, the longitudinal speed attains its maximum at the road center and decreases toward both boundaries. This demonstrates that the steady-state longitudinal speed profile is governed not solely by the density distribution, but by the balance between the density-induced acceleration and the direct boundary-induced braking effect.}
\vspace{0.2cm}

\noindent This equilibrium reflects a balance between interaction-driven lateral dispersion and boundary-induced confinement, demonstrating that the proposed second-order model captures both the transient and steady-state characteristic. These observations are consistent with the reported average lane change durations ranging from $4$ to $17~\text{s}$ as reported by \citealp{toledo2007modeling}, thus supporting the model’s fidelity in representing realistic traffic behaviors.
\begin{figure}[!htbp]
\centering  
    \begin{subfigure}{0.48\textwidth}
        \centering
        \includegraphics[width=\linewidth]{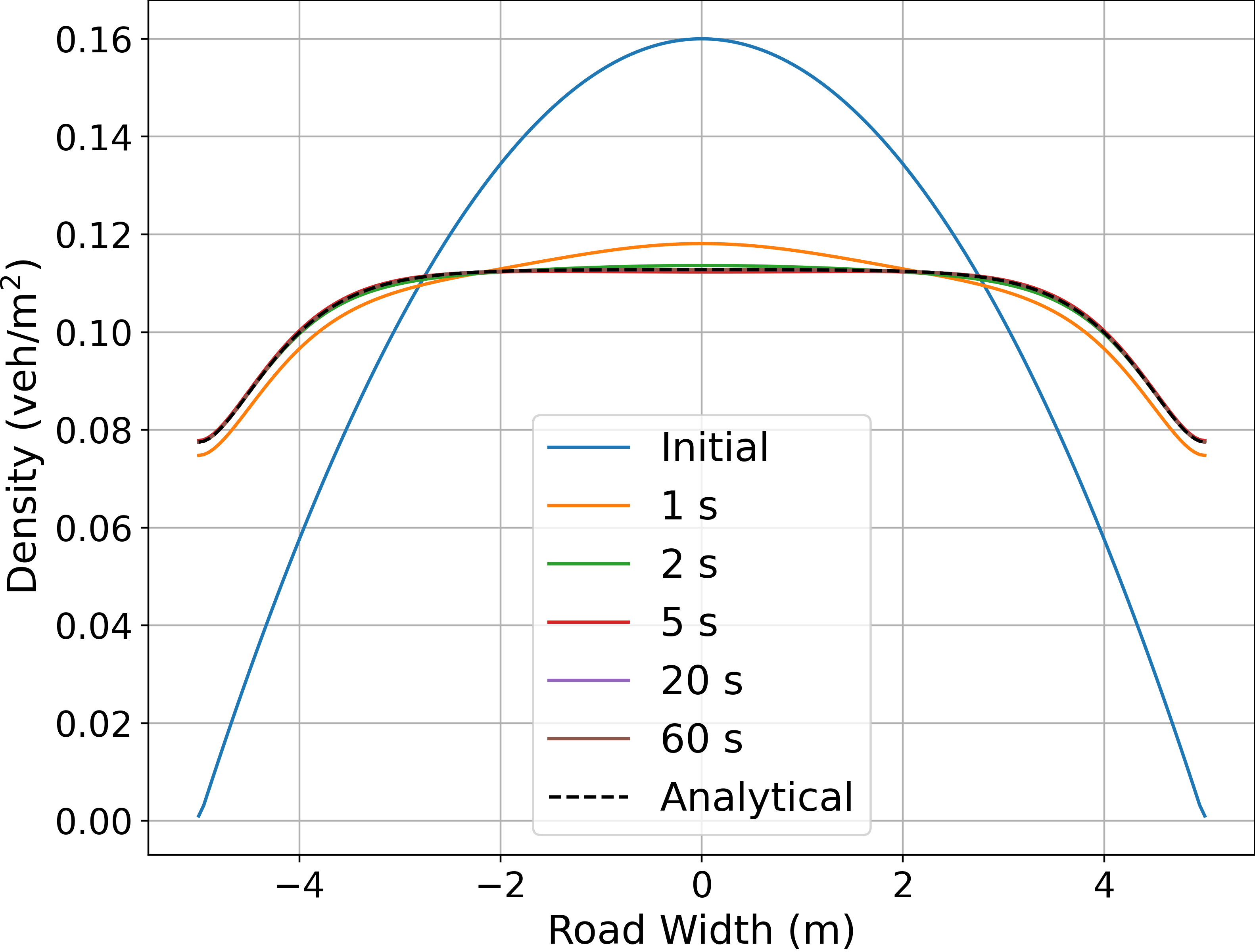}
        \caption{Case 1.}
        \label{Case 1}
    \end{subfigure}
 \hspace{0.02\textwidth}
        \begin{subfigure}{0.48\textwidth}
        \centering
        \includegraphics[width=\linewidth]{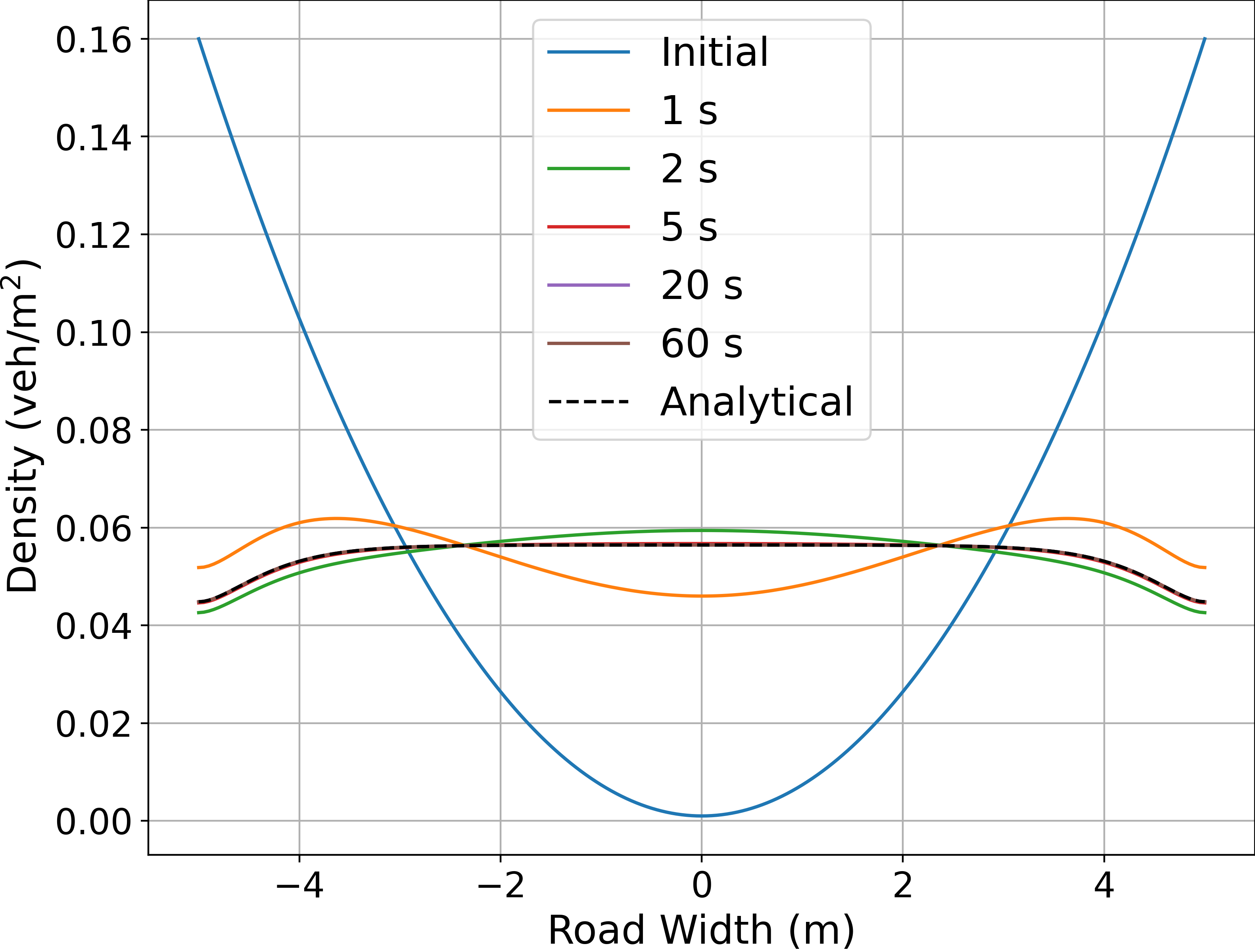}
        \caption{Case 2.}
        \label{Case 2}
    \end{subfigure}
    
    \caption{Evolution of lateral density distributions for Case 1 and Case 2 at $ y = b$, highlighting the influence of road-boundary forces.}
    \label{fig:density1d_2_lines_lat0}
\end{figure}









\subsubsection*{Case 2: Convex Lateral Density}
\martinc{remove all 3d figures as specified below the caption ''Case 1''}

\noindent In this case, the initial lateral density distribution is convex, with the maximum density located near the road boundaries and the minimum density at the center. The initial condition is defined as
\begin{equation}
\rho(x,y) = \rho_{\text{max}}
\left(\frac{y}{b}\right)^2,
\label{eq:case_lat_convex}
\end{equation}
which represents a higher concentration of vehicles close to the road edges. As shown in \autoref{Case 2}, the early evolution up to approximately \unit[5]{s} is characterized by a pronounced inward lateral movement, driven by strong interaction-induced forces arising from high-density accumulation near the road edges \martin{and boundary effects acting in the same direction.}  Between 5--10~s, the rate of lateral redistribution decreases as density gradients weaken and vehicle--vehicle interactions become more balanced across the road width. The speed is lower near the road boundaries and higher toward the center, consistent with the same boundary-braking mechanism discussed in Case 1 ($\beta_b^{\text{long}} > \beta_b^{\text{lat}}$). At later times ($t>\unit[20]{s}$), the system approaches \martin{essentially the same stable lateral configuration as in the concave initial configuration, Case~1, with minimal differences only due to the lower number of vehicles of this case. This reflects the interplay between the boundary repulsion and the gradient-driven traffic component.}

\subsubsection*{Case 3: Concave Lateral Density over Half Road Width}
\martinc{The same applies: The 3d figures give no new information and Fig. 17(b) is enough. It also seems that the simulation does not approach exactly the analytical solution. Check by plotting also the analytical solution in the plots Fig. 13(b), 15(b) and 17(b) becoming the new Figures 13 and 14}
\noindent In this scenario, only one half of the road width initially exhibits a concave lateral density distribution, while the other half remains nearly empty with negligible traffic. The initial density distribution is prescribed as
\begin{equation}
\rho(x,y) =
\begin{cases}
\rho_{\text{max}}
\left(
1 -
\left(
\frac{y - b/2}{\,b/2\,}
\right)^2
\right),
& y \leq 0, \\[6pt]
0.001, & \text{otherwise},
\end{cases}
\label{eq:case_lat_half_concave}
\end{equation}

\noindent As a result of this highly asymmetric initial condition, vehicles migrate toward the initially empty half of the road, as shown in \autoref{Case 3}. This redistribution is further illustrated in \autoref{Case 3}, highlighting the lateral spreading from the occupied to the vacant region. \martin{As in the other cases,} the early evolution is characterized by rapid lateral migration toward the initially unoccupied half of the road, driven by large lateral density gradients and the availability of substantial free space. However, \martin{due to the initial asymmetry, the total lateral movement to reach a steady state is much larger as in the fully concave or convex cases, so the eventual relaxation to the steady state takes much longer as in the other cases.} In summary, the observed dynamics demonstrate that the proposed second-order model accurately captures both transient crowding effects and the prolonged equilibration process associated with highly uneven lateral density distributions. In summary, the observed dynamics demonstrate that the proposed second-order model accurately captures both transient crowding effects and the prolonged equilibration process associated with highly uneven lateral density distributions.
\begin{figure}[H]
\centering
    \includegraphics[width= 0.5\textwidth]{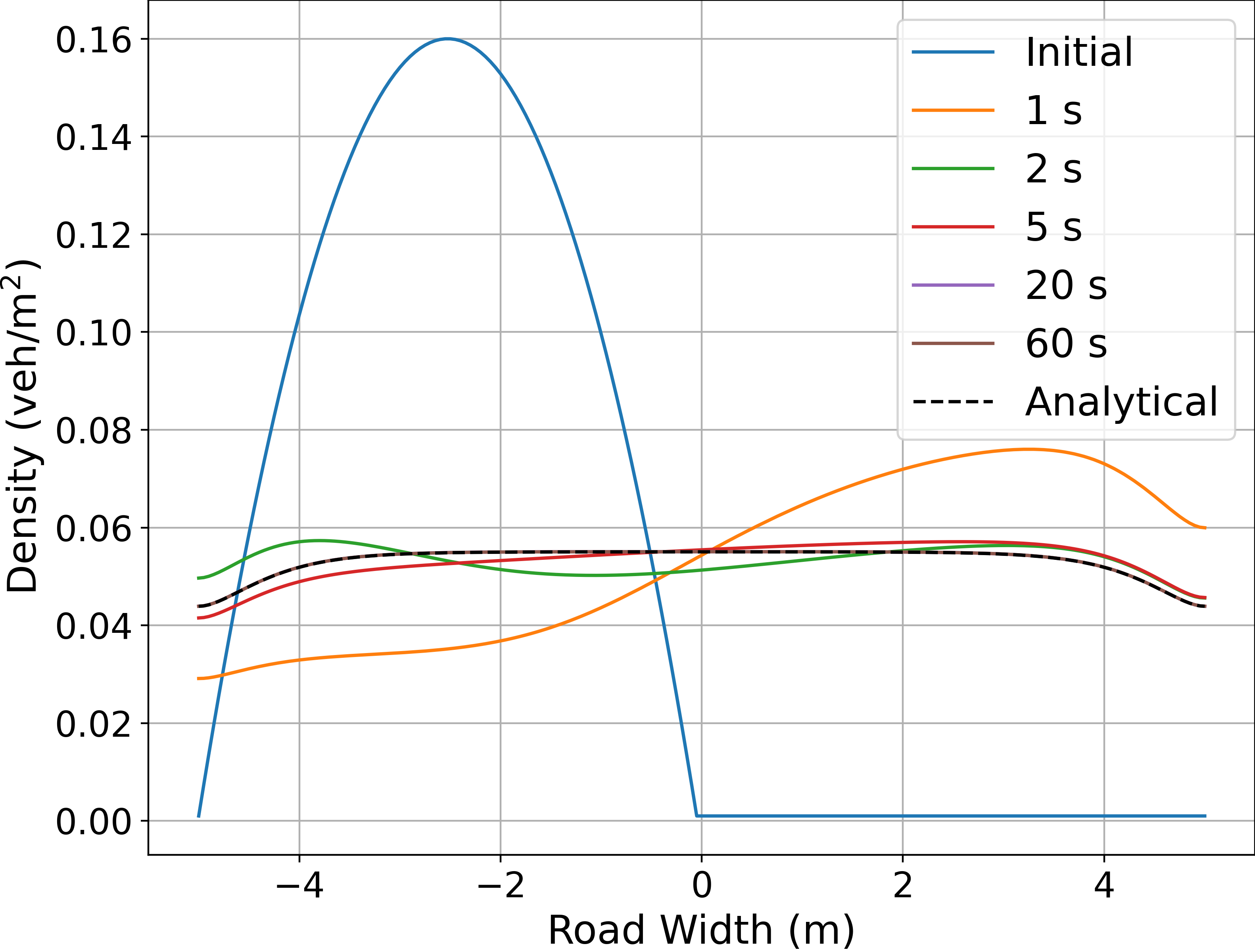}
    \label{Case 3}
    \caption{Evolution of density distributions for Case 3 at $y = b$, highlighting the influence of road-boundary forces.}
    \label{Case 3}
\end{figure}

\subsection{Longitudinal and Lateral dynamics}
\martinc{Here, $(x,y,t)$ information is essential. However, I would replace the 3d figures (which more look like modern art than scientific) by the color-coded contour plots you have used from Fig. 4 onwards.}

\noindent In this section, we will simulate lateral and longitudinal dynamics together under various initial conditions to assess whether the model can effectively replicate real-world macroscopic traffic behavior. Studying the combined behavior is essential because real-world traffic dynamics are usually \martin{two}-dimensional, where longitudinal flow and lateral interactions coexist and influence each other. For instance, lateral lane-changing behavior often impacts longitudinal flow characteristics. Capturing these coupled dynamics allows for a more comprehensive understanding of traffic phenomena and ensures the model's robustness in representing complex scenarios like lane-changing maneuvers, merging traffic, and the effect of road boundaries on vehicular movement.

\subsubsection*{Case 1: Free-Flow with Medium Congestion}
\begin{figure}[H]
\centering
\begin{subfigure}{0.48\textwidth}
    \centering
    \includegraphics[width=\linewidth]{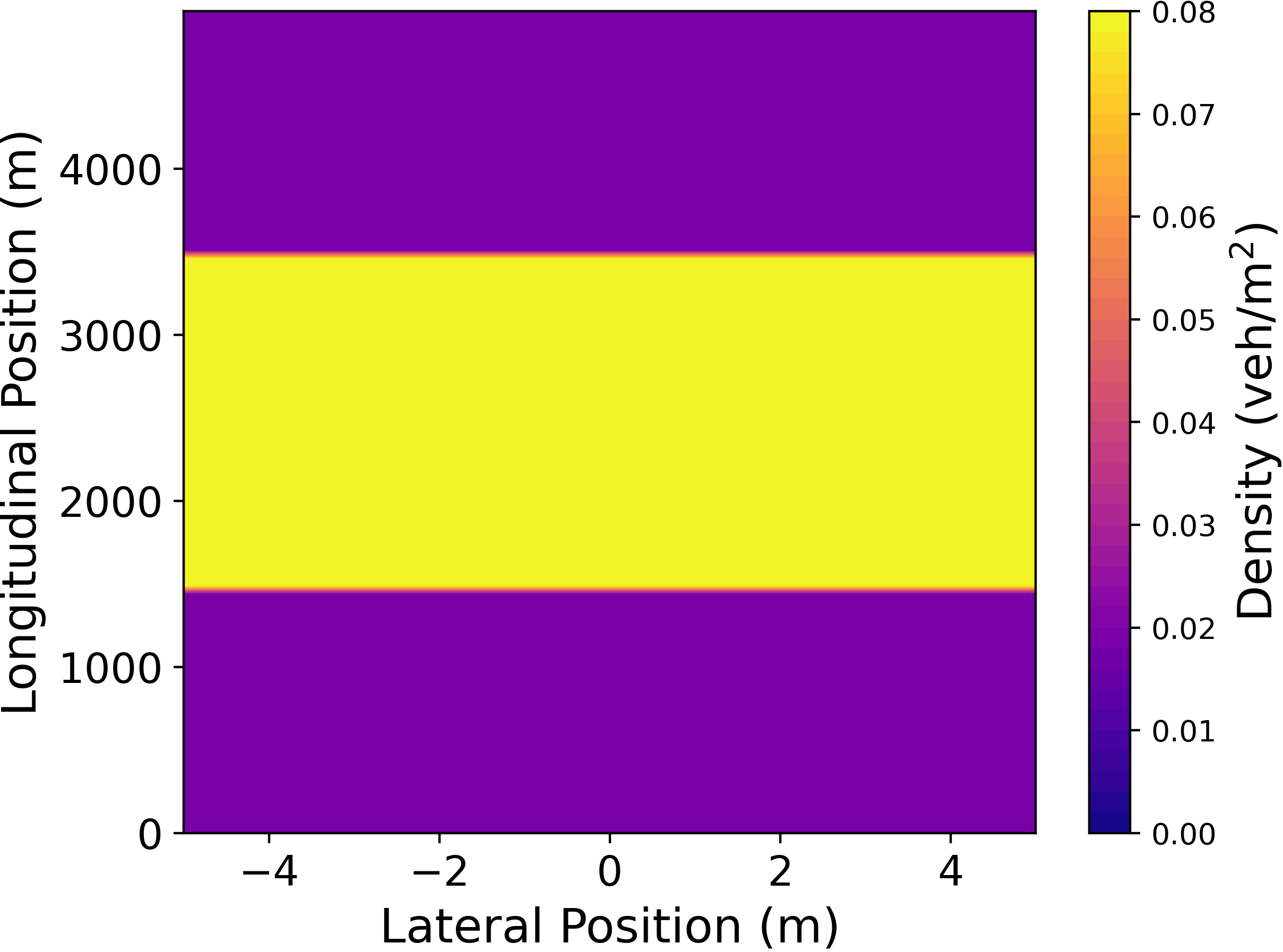}
    \caption{$t = 0$ s}
\end{subfigure}
\hspace{0.02\textwidth}
\begin{subfigure}{0.48\textwidth}
    \centering
    \includegraphics[width=\linewidth]{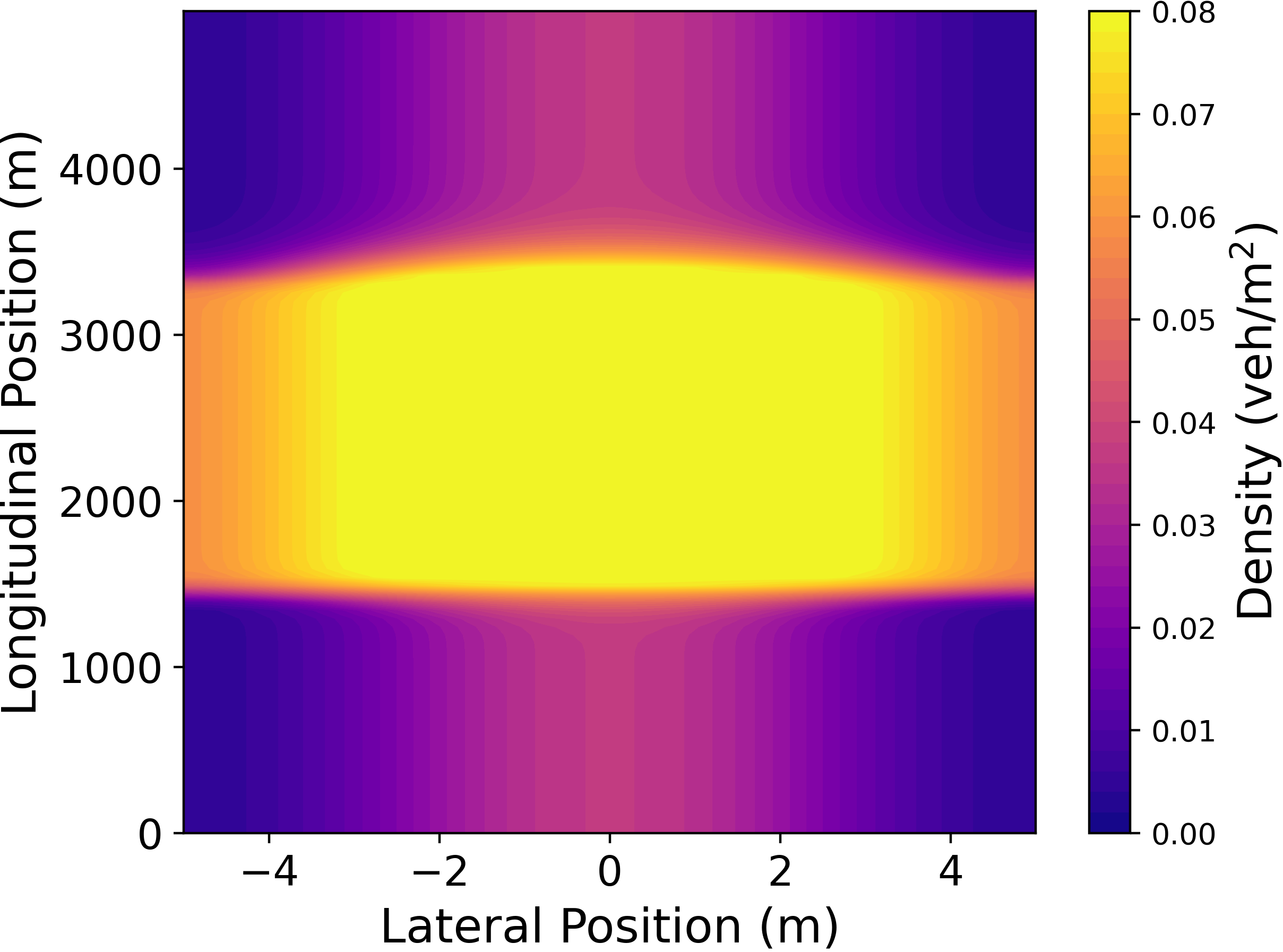}
    \caption{$t = 10$ s}
\end{subfigure}

\vspace{0.2cm}
\begin{subfigure}{0.48\textwidth}
    \centering
    \includegraphics[width=\linewidth]{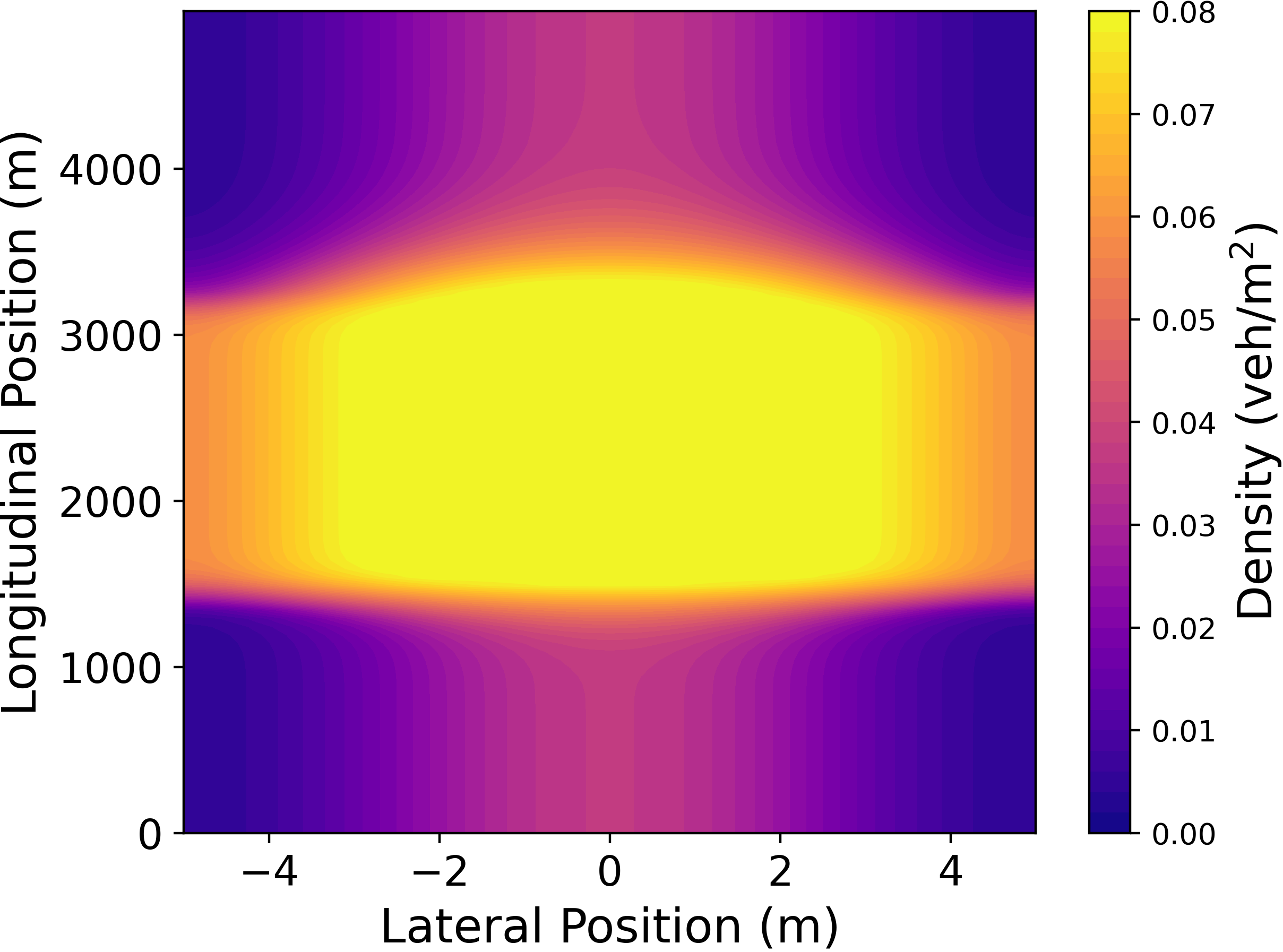}
    \caption{$t = 30$ s}
\end{subfigure}
\hspace{0.02\textwidth}
\begin{subfigure}{0.48\textwidth}
    \centering
    \includegraphics[width=\linewidth]{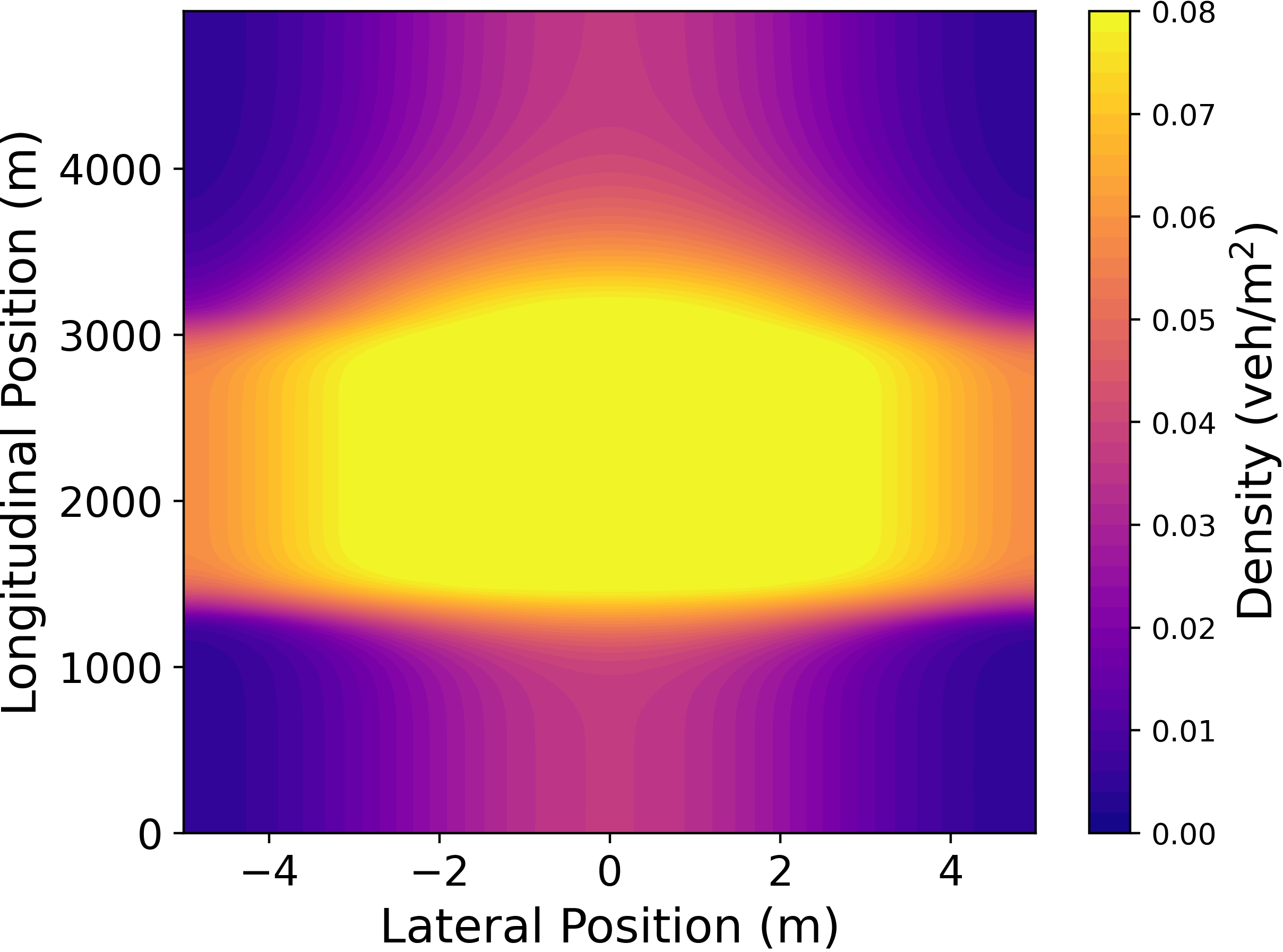}
    \caption{$t = 60$ s}
\end{subfigure}

\caption{Temporal evolution of traffic density for the proposed two-dimensional model for Case~1.}
\label{fig:fulldynamics_case0}
\end{figure}

\noindent In this scenario, the central section of the road is initialized in a medium-congested state, while the upstream and downstream sections are in free-flow conditions. The initial density distribution is prescribed as
\martinc{I redefined the initial conditions. Please re-simulate accordingly (see my mail)}
\martin{\begin{equation}
\rho(x) =
\begin{cases}
\unit[0.080]{veh/m^2}, & \unit[1\,500]{m} \leq x < \unit[3\,500]{m}, \\
\unit[0.020]{veh/m^2} & \text{otherwise},
\end{cases}
\label{eq:case_latlong_medium}
\end{equation}
}
\martin{and periodic longitudinal boundary conditions have been used.} \autoref{fig:fulldynamics_case0} illustrates the temporal evolution of traffic density predicted by the proposed two-dimensional second-order macroscopic traffic flow model, which incorporates both longitudinal and lateral dynamics. The density evolution is presented at times ranging from the initial condition ($T = 0~\text{s}$) to $T = 60~\text{s}$. The selected initial condition is designed to examine whether vehicles perform lateral manoeuvres under medium congestion levels, where the traffic density remains significantly below the maximum density and sufficient lateral space is available for vehicle movement.

\vspace{0.2cm}
\noindent For $T > 10~\text{s}$, the initially rectangular, medium-congested block gives rise to two distinct shock waves, one forming at its upstream boundary ($x=\unit[1\,500]{m}$) and the other at its downstream boundary ($x=\unit[3\,500]{m}$), as shown in \autoref{fig:fulldynamics_case0}. As the simulation progresses, both boundaries propagate in the upstream direction. The downstream boundary propagates upstream due to the free-flow traffic downstream, while the upstream boundary also propagates upstream due to the medium congested traffic conditions. Simultaneously, vehicles begin to rearrange laterally under the combined influence of lateral interactions and road-boundary forces. During this stage, lateral interactions become increasingly prominent, resulting in a redistribution of traffic density across the road width, and the density field gradually evolves into a fully two-dimensional structure. Within the congested core bounded by the two shock fronts, the density remains high but below the maximum (jam) density, allowing vehicles to continue rearranging laterally across the road width. Outside this core, where the density has decreased, the lower traffic density provides greater lateral freedom, resulting in a more uniform vehicle distribution across the roadway. The influence of the road-boundary effects becomes evident through the accumulation of vehicles toward the centre of the road and the corresponding reduction in density near the road boundaries on both the leading and trailing sides of the congested region. Throughout the simulation, the density gradients remain well resolved, and no spurious oscillations are observed in regions of steep density variation, including at both shock fronts, as shown in \autoref{fig:fulldynamics_case0}.

\vspace{0.2cm}

\noindent Overall, these results demonstrate that the proposed two-dimensional second-order macroscopic model successfully captures the coupled longitudinal and lateral evolution of traffic density, including nonlinear wave propagation, lateral redistribution under boundary influences, and upstream shock wave propagation. This highlights the importance of incorporating lateral dynamics for the realistic representation of disordered traffic flow.

\subsubsection*{Case 2: Free-Flow with Heavy Congestion}
\begin{figure}[H]
    \centering
    
    \begin{subfigure}{0.48\textwidth}
        \centering
        \includegraphics[width=\linewidth]{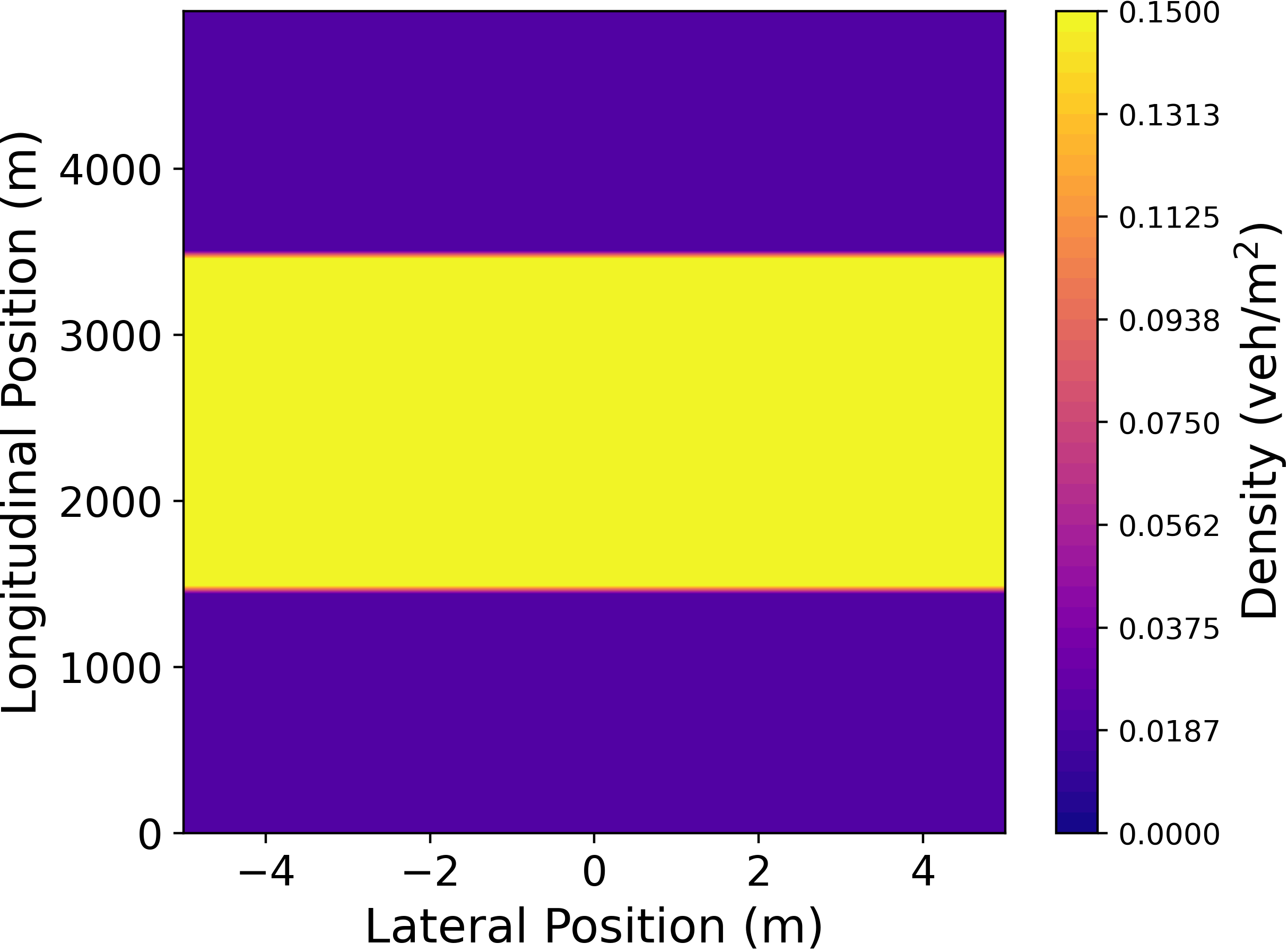}
        \caption{$t = 0$ s}
    \end{subfigure}
    \hspace{0.02\textwidth}
    \begin{subfigure}{0.48\textwidth}
        \centering
        \includegraphics[width=\linewidth]{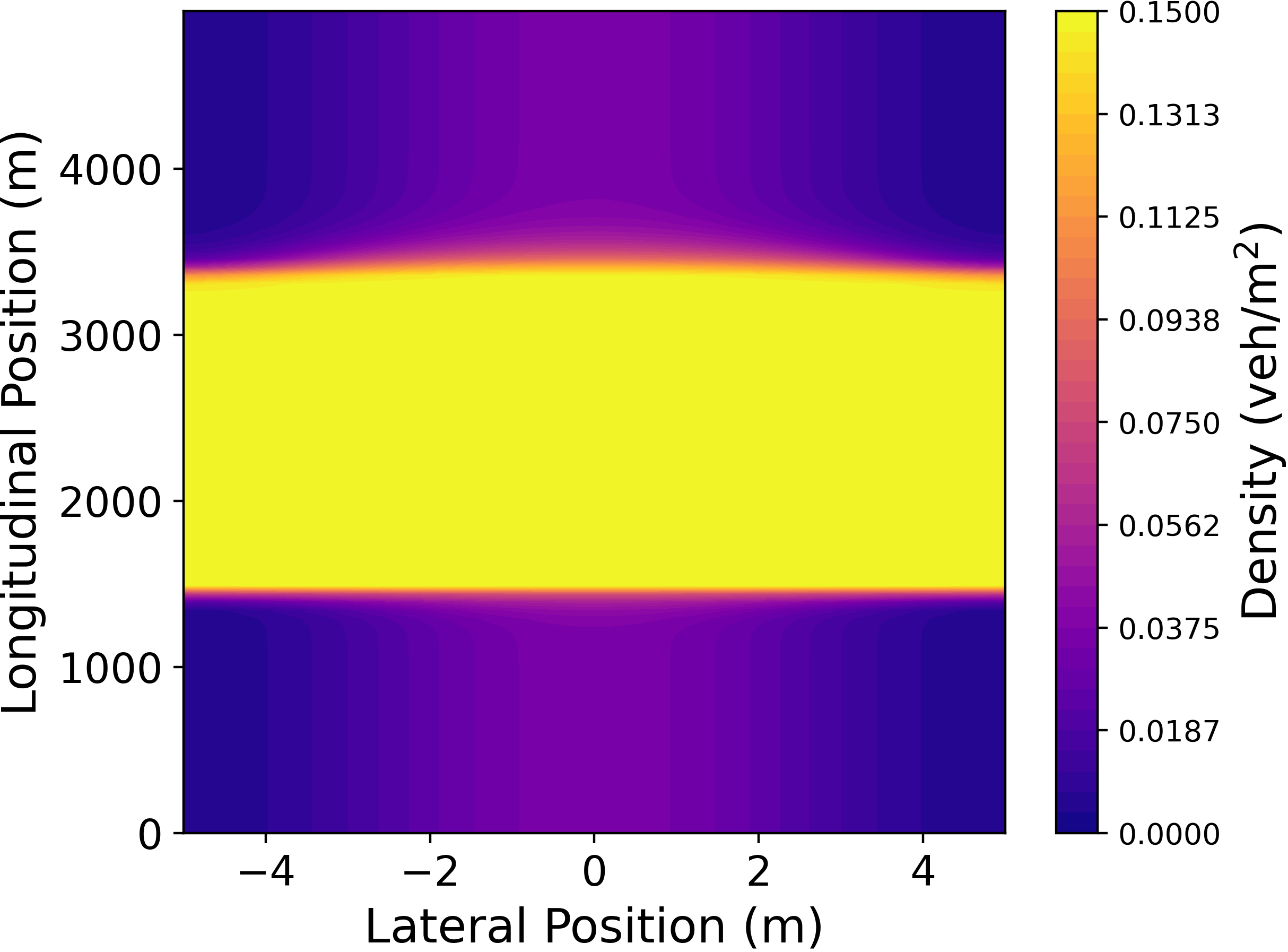}
        \caption{$t = 10$ s}
    \end{subfigure}
    
    \begin{subfigure}{0.48\textwidth}
        \centering
        \includegraphics[width=\linewidth]{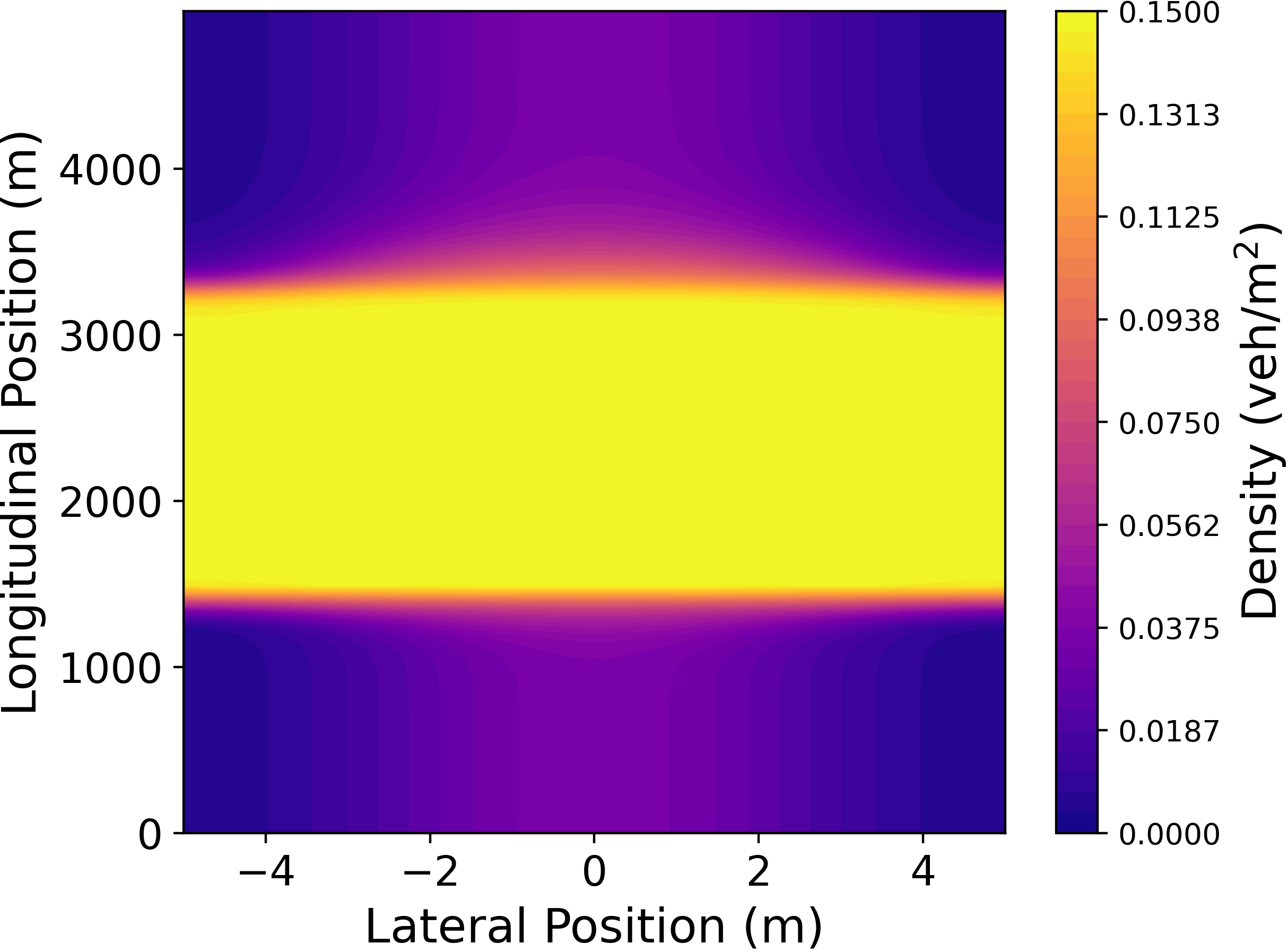}
        \caption{$t = 30$ s}
    \end{subfigure}
    \hspace{0.02\textwidth}
    \begin{subfigure}{0.48\textwidth}
        \centering
        \includegraphics[width=\linewidth]{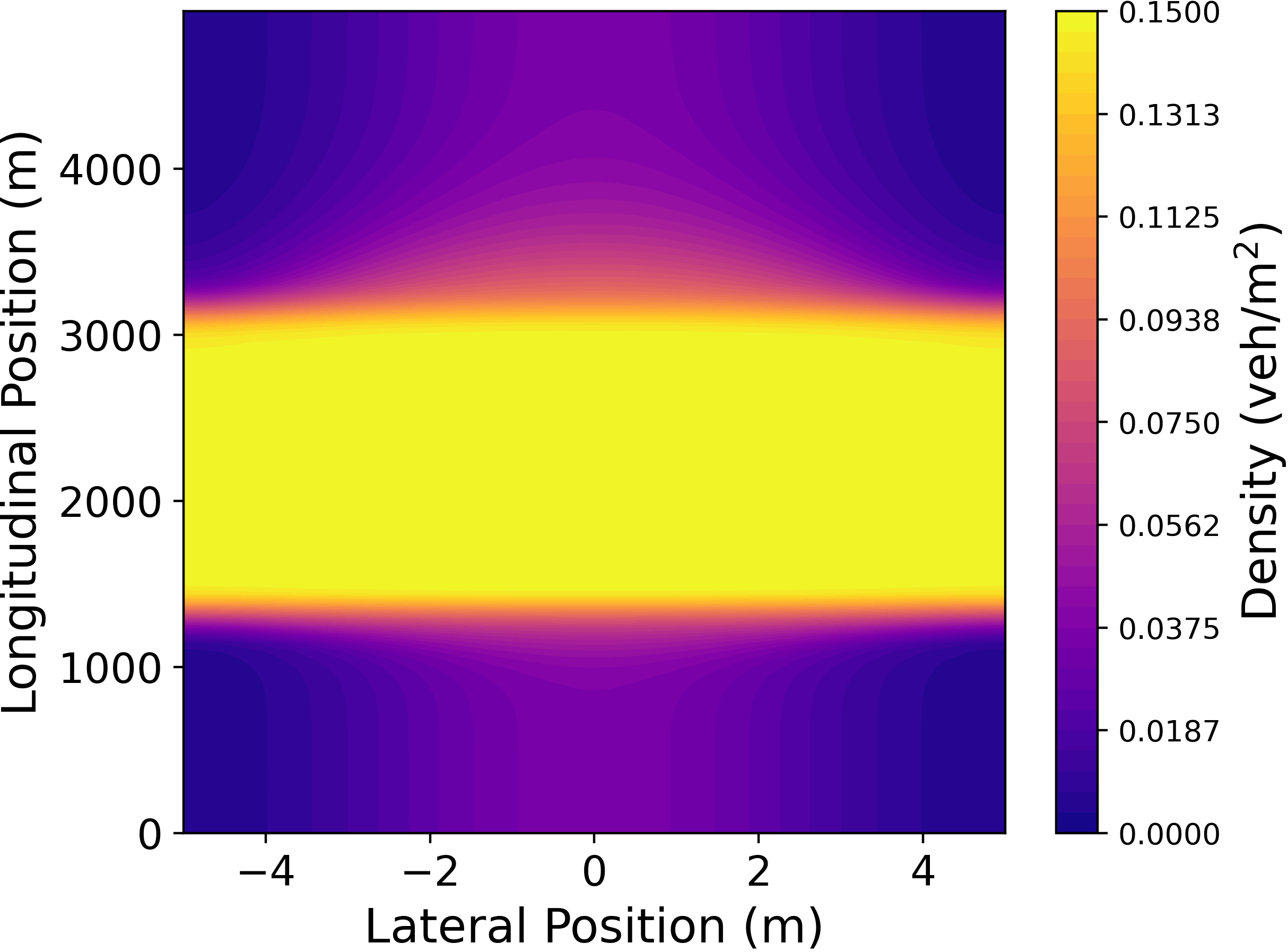}
        \caption{$t = 60$ s}
    \end{subfigure}
    
    \caption{Temporal evolution of traffic density for proposed two-dimensional model for Case 2.} 
    \label{fig:fulldynamics_case1}
\end{figure}

\noindent In this case, the road is initially divided into three longitudinal zones: a heavily congested region, close to the maximum density, occupying the central portion of the road, flanked by free-flow conditions both upstream and downstream. The corresponding initial density distribution is given by
\martinc{Please re-simulate according to the new initial condition \autoref{eq:case_latlong_heavy}. Notice that presently the simulations are inconsistent also with the past initial condition}
\martin{\begin{equation}
\rho(x) =
\begin{cases}
\unit[0.150]{veh/m^2}, & \unit[1\,500]{m} \leq x < \unit[3\,500]{m}, \\
\unit[0.020]{veh/m^2} & \text{otherwise}.
\end{cases}
\label{eq:case_latlong_heavy}
\end{equation}
}
\autoref{fig:fulldynamics_case1} presents the temporal evolution of traffic density obtained from the proposed two-dimensional macroscopic model. The density field is shown as a function of both longitudinal and lateral spatial coordinates, thereby highlighting the coupled dynamics characteristic of disordered traffic flow. At $T = 0~\text{s}$, the density distribution corresponds to the prescribed initial condition and exhibits a sharp discontinuity in the longitudinal direction. 
For $T > 10~\text{s}$, the initially rectangular, near-jam congested block gives rise to two distinct shock waves, one forming at its downstream interface ($x = \unit[3\,500]{m}$) and one at its upstream interface ($x = \unit[1\,500]{m}$), as shown in \autoref{fig:fulldynamics_case1}. The downstream boundary propagates upstream because of the free-flow traffic downstream, whereas the upstream boundary propagates upstream owing to the heavy congestion. During this phase, the influence of lateral interactions and road boundary forces becomes evident, resulting in a redistribution of traffic density across the road's width. At intermediate times, the density field develops into a fully two-dimensional structure. Because the interior of the block is initially close to the maximum density, lateral vehicle movement is strongly restricted there at early times; only as each shock front sweeps through and locally reduces the density below jam conditions does sufficient free space become available for lateral redistribution to develop. This effect is visible in \autoref{fig:fulldynamics_case1}(b)--(d): near both interfaces, where density has relaxed below jam, vehicles progressively accumulate toward the center of the road and are depleted near the road boundaries, whereas the still near-jam core in between remains comparatively unaffected by lateral dynamics. At later times ($T > 30~\text{s}$), this boundary-driven lateral accumulation is clearly visible along both the upstream and downstream interfaces, bulging symmetrically toward the center of the domain and following the corresponding shock fronts as they advance. This behaviour demonstrates the ability of the model to capture complex spatio-temporal traffic dynamics simultaneously at both interfaces of a congested region. Overall, the results confirm that the proposed two-dimensional second-order macroscopic model successfully captures nonlinear density wave propagation, lateral redistribution of traffic density, and smooth shock resolution without introducing numerical artefacts. These findings show the importance of incorporating both longitudinal and lateral dynamics for the realistic modeling of disordered traffic flow.


\section{Conclusion}
\noindent This study developed a two-dimensional second-order macroscopic traffic flow model to describe disordered traffic by explicitly coupling longitudinal and lateral vehicle dynamics within a unified continuum framework. By extending the continuity equation to two spatial dimensions and formulating acceleration equations in both the longitudinal and lateral directions, the proposed model overcomes fundamental limitations of existing one-dimensional second-order and two-dimensional first-order approaches. The longitudinal dynamics were derived from the Full Velocity Difference Model (FVDM) with an additional Lagrangian propagation term to improve speed adaptation under varying spacing conditions, while the lateral dynamics were formulated based on interaction-driven principles consistent with the Optimal Velocity (OV) framework. A distinguishing feature of the model is the explicit incorporation of road-boundary effects in both directions, allowing realistic representation of vehicle confinement, space-sharing behaviour, and boundary-induced resistance that are essential for modelling lane-free and weakly lane-disciplined traffic systems.
\vspace{0.2cm}

\noindent Extensive numerical experiments demonstrated that the proposed formulation consistently captures a wide spectrum of traffic phenomena. Under laterally homogeneous conditions, the model reproduced stable free-flow propagation, upstream-moving congestion waves, and controlled shock evolution across free-flow, congested, and mixed regimes. The inclusion of road-boundary forces was shown to systematically moderate wave propagation speeds while preserving density interfaces, confirming their role as regulating mechanisms. Simulations of purely lateral dynamics further revealed that the model accurately captures rapid dispersion, inward migration, and delayed equilibration under concave, convex, and asymmetric density configurations.
\vspace{0.2cm}

\noindent The coupled longitudinal--lateral simulations demonstrate that the proposed two-dimensional second-order macroscopic model consistently captures regime-dependent traffic dynamics under disordered conditions. In both medium and heavy congestion scenarios, upstream-propagating shock waves arise from longitudinal interactions, while lateral redistribution is governed by the interplay of interaction-induced forces and road-boundary effects. When density remains below the maximum density, available lateral space enables two-dimensional density evolution, whereas near jam conditions lateral motion in the downstream region becomes restricted and remains active primarily in lower-density upstream areas. In all cases, vehicle accumulation away from road boundaries and well-resolved density gradients are consistently observed, with no spurious oscillations near sharp interfaces. These results confirm that incorporating both longitudinal and lateral dynamics is essential for realistically representing disordered traffic flow and capturing nonlinear wave propagation and two-dimensional density evolution within a macroscopic modelling framework.
\vspace{0.2cm}

\noindent The present work provides a physically consistent and interpretable macroscopic framework for analysing disordered traffic flow. Several important extensions can be pursued in future research. Incorporating multiple vehicle classes with heterogeneous dynamic characteristics would allow the model to better represent mixed traffic commonly observed in developing regions. Calibration and validation using traffic data would further strengthen the model’s empirical grounding and enable systematic parameter estimation. From a modelling perspective, extensions incorporating stochastic driving behaviour, non-local anticipation, or dynamic road geometries could enhance realism. Finally, the proposed two-dimensional second-order framework offers a strong foundation for applications in traffic control, infrastructure design, and safety analysis, particularly in environments where lateral manoeuvring and space-sharing play a dominant role.

\noindent\textbf{Declaration of Generative AI and AI-assisted technologies in the writing process}\\
The authors declare that a chatbot was used to assist with grammar and language editing of this manuscript. All content generated by the tool was carefully reviewed, revised, and approved by the authors, who take full responsibility for the content of this publication.\\

\noindent\textbf{CRediT authorship contribution statement}\\
Shashank Rajput: Writing – original draft, Software, Formal analysis, Methodology. Venkatesan Kanagaraj : Writing – review \& editing, Methodology, Funding acquisition, Conceptualization.  Martin Treiber: Writing – review \& editing, Methodology, Conceptualization. Gowri Asaithambi: Writing – review \& editing, Methodology, Funding acquisition, Conceptualization. Ostap Okhrin: Writing – review \& editing, Methodology, Conceptualization. \\ 

\noindent\textbf{Acknowledgments}\\
The authors gratefully acknowledge the support received through the Scheme for Promotion of Academic and Research Collaboration (SPARC) project, funded by the Ministry of Education, Government of India (Project No. SPARC/2019–2020/P2384/SL). The second author acknowledges the fellowship support provided by the Alexander von Humboldt Foundation, Germany, during 2025, hosted at the Technical University of Dresden, Germany.

\noindent

\label{Bibliography}

\bibliographystyle{apalike}

\bibliography{Bibliography} 






\end{document}